\documentclass[twocolumn]{aastex63}

\newcommand{\cii}{[C{~\small{II}}]$_{158\, \mu\rm{m}}$}

\newcommand{\ci}{[C{~\small{I}}]$_{809\mu\rm{m}}$}
\newcommand{\kms}{\rm{km \,s}^{-1}}
\newcommand{\um}{\mu\rm{m}}

\received{October 29, 2021}
\revised{January 7, 2022}
\accepted{January 24, 2022}

\submitjournal{ApJ}

\shorttitle{Galaxies overdensities around three $z\sim6.5$ quasars}
\shortauthors{Meyer et al.}

\graphicspath{{./}{figures/}}

\begin{document}

\title{Constraining galaxy overdensities around three $z\sim 6.5$ quasars with ALMA and MUSE}

\correspondingauthor{Romain A. Meyer}
\email{meyer@mpia.de}

\author[0000-0001-5492-4522]{Romain A. Meyer}
\affiliation{Max Planck Institut f\"ur Astronomie, K\"onigstuhl 17, D-69117, Heidelberg, Germany}
\author[0000-0002-2662-8803]{Roberto Decarli}
\affiliation{INAF – Osservatorio di Astrofisica e Scienza dello Spazio di Bologna, via Gobetti 93/3, I-40129, Bologna, Italy}
\author[0000-0003-4793-7880]{Fabian Walter}
\affiliation{Max Planck Institut f\"ur Astronomie, K\"onigstuhl 17, D-69117, Heidelberg, Germany}
\author[0000-0002-3119-9003]{Qiong Li}
\affiliation{Department of Astronomy, School of Physics, Peking University, Beijing 100871, P. R. China}
\affiliation{Department of Astronomy, University of Michigan, 311 West Hall, 1085 S. University Ave, Ann Arbor, MI, 48109-1107, U.S.A.}
\affiliation{Kavli Institute for Astronomy and Astrophysics, Peking University, Beijing, 100871, P. R. China}
\author[0000-0003-4956-5742]{Ran Wang}
\affiliation{Department of Astronomy, School of Physics, Peking University, Beijing 100871, P. R. China}
\affiliation{Kavli Institute for Astronomy and Astrophysics, Peking University, Beijing, 100871, P. R. China}
\author[0000-0002-5941-5214]{Chiara Mazzucchelli}
\affiliation{European Southern Observatory, Alonso de Cordova 3107, Vitacura, Region Metropolitana, Chile}
\author[0000-0002-2931-7824]{Eduardo Ba\~nados}
\affiliation{{Max Planck Institut f\"ur Astronomie, K\"onigstuhl 17, D-69117, Heidelberg, Germany}}
\author[0000-0002-6822-2254]{Emanuele P. Farina}
\affiliation{Max Planck Institut fur Astrophysik, Karl–Schwarzschild–Straße 1, D-85748 Garching bei M\"unchen, Germany}
\author[0000-0001-9024-8322]{Bram Venemans}
\affiliation{Leiden Observatory, Leiden University, PO Box 9513, 2300 RA Leiden, The Netherlands}

\begin{abstract}
We quantify galaxy overdensities around three high--redshift quasars with known \cii\ companions: PJ231--20 ($z=6.59$), PJ308--21 ($z=6.24$) and J0305--3150 ($z=6.61$). Recent SCUBA2 imaging revealed the presence of $17$ submillimeter galaxies (SMG) with sky separations $0.7'
< \theta < 2.4'$ from these three quasars. We present ALMA Band 6 follow--up observations of these SCUBA2--selected SMGs to confirm their nature and redshift. We also search for continuum--undetected \cii\ emitters in the ALMA pointings and make use of archival MUSE observations to search for Lyman-$\alpha$ Emitters (LAE) associated with the quasars. While most of the SCUBA2--selected sources are detected
with ALMA in the continuum, no \cii\ line emission could be detected, indicating that they are not at the quasar redshifts. Based on the serendipitous detection of CO 7--6 and \ci\ emission lines, we find that four SMGs in the field of PJ231--20 are at $z\sim2.4$, which is coincident with the redshift of a Mg{~\small II} absorber in the quasar rest-frame UV spectrum. We report the discovery of $2$ LAEs within $<0.6\ \rm{cMpc}$ of PJ231--20 at the same redshift, indicating a LAE overdensity around this quasar. Taken together, these observations provide new constraints on the large--scale excess of Lyman--$\alpha$-- and \cii--emitting galaxies around $z>6$ quasars and suggest that only wide--field observations, such as MUSE, ALMA or \textit{JWST} mosaics, can reveal a comprehensive picture of large--scale structure around quasars in the first billion years of the Universe.
\end{abstract}


\section{Introduction}
\label{sec:intro}
Observations of $z>6$ quasars show that they are powered by supermassive black holes (SMBH) as massive as $10^{9} M_\odot$  \citep[e.g.,][]{DeRosa2011,DeRosa2014,Banados2018,Mazzucchelli2017,Yang2020,Wang2021}. Their surprisingly high masses, accumulated within less than a Gyr since the Big Bang, are a puzzle for galaxy evolution and black hole growth theories. One formation pathway is through the existence of massive seed black holes ($\gtrsim 10^{3} M_\odot$) at $z\sim 15-30$ created either by the collapse of massive gas clouds \citep[e.g.,][]{Oh2002, Bromm2003, Begelman2006, Ferrara2014, Inayoshi2014}, that of Population III stars \citep[e.g.,][]{Bond1984,Madau2001, Latif2013, Valiante2016} or the runaway collision of stars in compact clusters \citep[e.g.,][]{Omukai2008,Devecchi2009,Katz2015,Sakurai2017}. Radiatively inefficient accretion, close to or even above the Eddington limit is another possible scenario explaining the presence of SMBHs with masses of $\sim 10^9 M_\odot$ already at $z\sim 6$. Such extreme accretion histories are thought to be made possible by an abundance of gas--rich mergers or the presence of SMBH seeds in massive metal--poor gas halos \citep[e.g.,][]{Narayanan2008, Hopkins2008, Overzier2009, Angulo2012, Latif2015, Habouzit2019, Wise2019}. Either way, most models of black hole formation and growth postulate or find that luminous $z>6$ quasars should reside in the densest environments and effectively trace the emergence of the first large--scale structures in the Universe \citep[see, e.g.,][for comprehensive reviews]{Haiman2004,Overzier2009,Volonteri2010,Latif2016,Inayoshi2020}. 

A direct prediction of this hypothesis is the presence of galaxy overdensities around quasars in the first billion years. Since the first discoveries, more than $200$ quasars at $z>6$ have been detected \citep[e.g.][]{Fan2001,Fan2004, Fan2006,Venemans2007,Mortlock2011,Venemans2015a,Carnall2015,Banados2016,Jiang2016, Reed2015,Reed2017, Banados2018,Yang2018a,Wang2018a,Wang2021}. This large sample of early luminous quasars has enabled the possibility to probe their supposedly overdense environment. However, despite long and sustained efforts with optical/IR ground and space--based observatories, evidence for galaxy overdensities around these objects is mixed and contradictory \citep[e.g.][]{Willott2005,Ajiki2006,Kim2009,Utsumi2010,Banados2013,Simpson2014,Mazzucchelli2017a,Goto2017,Farina2017,Champagne2018,Mignoli2020,Miller2020}. Although most $z>6$ quasar fields have not been searched systematically and uniformly for overdensities, the current absence of clear evidence of galaxy overdensities around $z>6$ quasars is an outstanding challenge to our current paradigm of black hole growth and galaxy evolution.

Recently, ALMA and NOEMA observations of the \cii\ line in high--redshift quasars have revealed the presence of close ($<60$ proper kpc, $<1000\ \kms$) \cii--bright companions found around $\sim 30\%$ of luminous $z\sim 6$ quasars \citep{Decarli2017,Willott2017,Decarli2018,Decarli2019, Neeleman2019a,Venemans2020}. These objects tantalisingly hint at the long--predicted large--scale overdensity around early quasars, but the limited field of view of ALMA only constrains the smaller scales ($< 1$ comoving Mpc) of the galaxy--quasar correlation at $z>6$. 
Alternatively it is possible that these companion galaxies could simply be in the process of merging with the quasar \citep[e.g.,][]{Decarli2019, Neeleman2019a}, as is expected if SMBH growth is driven by mergers \citep[e.g.,][]{Hopkins2008}. Therefore, the overabundance of companions on small scales might not necessarily trace larger overdensities, but rather result from a selection bias towards ongoing or recent mergers (that fuel the SMBH gas accretion and increase its likelihood of being detected as a hyper--luminous quasar). Constraining the large--scale cross--correlation of galaxies and quasars at $z>6$ is thus necessary to distinguish these two competing hypotheses.

In this paper, we investigate the large--scale environment of three $z>6$ quasars with known bright \cii\ companions: J0305--3150  \citep[][]{Venemans2019}, PJ231--20 and PJ308--21  \citep[][]{Decarli2017}. J0305--3150, PJ231--20 and PJ308--21 have SMBH masses of  $2.00^{+0.22}_{-0.64}\times10^{9} M_\odot$, $1.89^{+0.34}_{-0.45}\times10^{9} M_\odot$ and $1.69^{+0.20}_{-0.35}\times10^{9} M_\odot$, respectively \citep[][ Farina et al., in prep.]{Mazzucchelli2017, Neeleman2021}. These quasars were observed with the Submillimeter Common--User Bolometer Array--2 (SCUBA2) on the James Clerk Maxwell Telescope \citep[][]{Holland2013} at $850\um$ and $450\um$ as part of a larger survey of quasar environments described in \citet[][]{Li2020_sherry} and  Li et al. (in prep.). The SCUBA2 images revealed numerous sub--mm galaxies (SMGs) detected at $0.7 \lesssim \theta  \lesssim 2.4\ \rm{arcmin}$ from each quasar (corresponding to $\sim 1.8-5.8$ comoving Mpc at the quasar redshifts). To test the possibility that these SMGs could be part of a large--scale overdensity associated with the $z>6$ quasars, we have observed the 17 brightest with ALMA to confirm their redshift. The spectral tunings were placed such that the \cii\ line would fall in the upper sideband if the SMGs were at the quasar redshift. 
Moreover, we also make use of archival MUSE observations of the quasars to probe potential overdensities of LAEs on smaller scales ($\lesssim 2$ comoving Mpc) than the SMG--ALMA pointings ($\sim 1.8-5.8$ comoving Mpc). This paper thus aims to present a comprehensive analysis of the galaxy under/over--density around three $z>6$ quasars probed by SCUBA2, ALMA and MUSE.

The structure of the paper is as follows. In Section \ref{sec:data}, we describe our ALMA observations of the selected SCUBA2 SMGs in three high--redshift quasar fields. We present in Section \ref{sec:cont_sources} the continuum sources detected with ALMA and discuss the ALMA and the SCUBA2 continuum fluxes. We assess the redshift of the detected continuum sources using emission lines and photometric redshifts capitalizing on ancillary \textit{HST} and \textit{Spitzer} imaging combined with the ALMA and SCUBA2 measurements. In Section \ref{sec:line_sources}, we present the results of a search for serendipitous \cii\ line emitters (undetected in the continuum) in the ALMA pointings. We present in Section \ref{sec:LAEs} the LAEs found in the archival MUSE observations of our three quasars. Finally, we conclude in Section \ref{sec:conclusion} and present updated constraints of the overdensity of galaxies around our quasars using our new observations as well as LAEs and \cii\ quasar companions from the literature. 

Throughout this paper, we assume a concordance cosmology with $H_0=70\, \rm{km\, s}^{-1}\ \rm{Mpc}^{-1}$, $\Omega_M = 0.3, \Omega_\Lambda=0.7$. All magnitudes are given in the AB system \citep[][]{Oke1983}. At the redshift of the quasars ($z\sim6.5$), $1"$ corresponds to $5.46$ proper kpc. 

\section{Observations and data reduction} \label{sec:data}
\subsection{Quasar fields studied in this work and existing archival data} 
This work focuses on three $z>6$ quasars: J0305--3150, PJ231--20 and PJ308--21. J0305--3150 was originally discovered in the VISTA Kilo--Degree Infrared Galaxy (VIKING) Survey \citep[][]{Venemans2013}, while PJ308--21 \citep[][]{Banados2016} and PJ231--20 \citep[][]{Mazzucchelli2017} were discovered in the Panoramic Survey Telescope and Rapid Response System (PAN--STARRS) quasar surveys. All three quasars were observed with ALMA to detect the redshifted \cii\ emission line, obtaining amongst other properties precise redshifts: $z_{\rm{J0305-3150}}=6.6145\pm 0.0001$ \citep[][]{Venemans2016}, $z_{PJ231-20}=6.58651\pm 0.00017$  and $z_{PJ308-21}=6.2342\pm0.0010$ \citep[][]{Decarli2017}. The \cii\ observations also revealed the presence of companions bright in \cii\ and in the dust continuum emission \citep[e.g.,][]{Decarli2017, Neeleman2019a,Venemans2020}. Further ALMA and MUSE observations uncovered three  close ($<40$ kpc) \cii\ emitters \citep[][]{Venemans2019} as well as a nearby Lyman--$\alpha$ emitter \citep[][]{Farina2017} in the vicinity of J0305--3150. These studies suggest that these three quasars could trace particularly overdense environments.

\subsection{ALMA observations of sub--mm SCUBA2 sources}
The three quasar fields have been observed with SCUBA2 as part of a larger sub--mm survey of $z\sim 6$ quasars to study their environments \citep[][, Li et al., in prep.]{Li2020_sherry}. In order to study the possible large--scale overdensities around the three $z\sim 6.5$ quasars, we have selected $>3.5\sigma$ ($\gtrsim 4 \,\rm{mJy}$) sub--mm galaxies (SMG) in their vicinity ($0.7 \lesssim  \theta \lesssim 2.4\ \rm{arcmin}$) based on the SCUBA2 850$\mu\rm{m}$ maps. No cut--off was imposed on the maximum distance of SMGs from the quasar, which simply results from the SCUBA2 Field of View and the depth of the data. The resulting sample contains 4 SCUBA2--selected SMGs around both J0305 and PJ231, and 9 around PJ308 for a total of $17$. The reduction and analysis of the SCUBA2 data is detailed extensively in \citet[][see also Li. et al, in prep. for the SMG results]{Li2020_sherry}, to which we direct the interested reader. We reproduce in Appendix \ref{app:scuba_images} the SCUBA2 imaging to make this paper self--contained.

The SCUBA2 sources were each observed in Band 6 for $\simeq10$ minutes on source with a single ALMA pointing (program 2019.1.01003.S, PI: R. Decarli). The spectral setup was chosen such that the center of the upper sideband was at the redshifted frequency of \cii\ of the central quasar ($\sim 250-270$ GHz). The array configuration were chosen to have a relatively low spatial resolution between $0.7"$ and  $1.4"$ as the aim is to detect the \cii\ line in the SMGs and thus try to confirm whether they are at the quasar redshift.

\begin{figure*}
    \centering
    \includegraphics[width=0.32\textwidth]{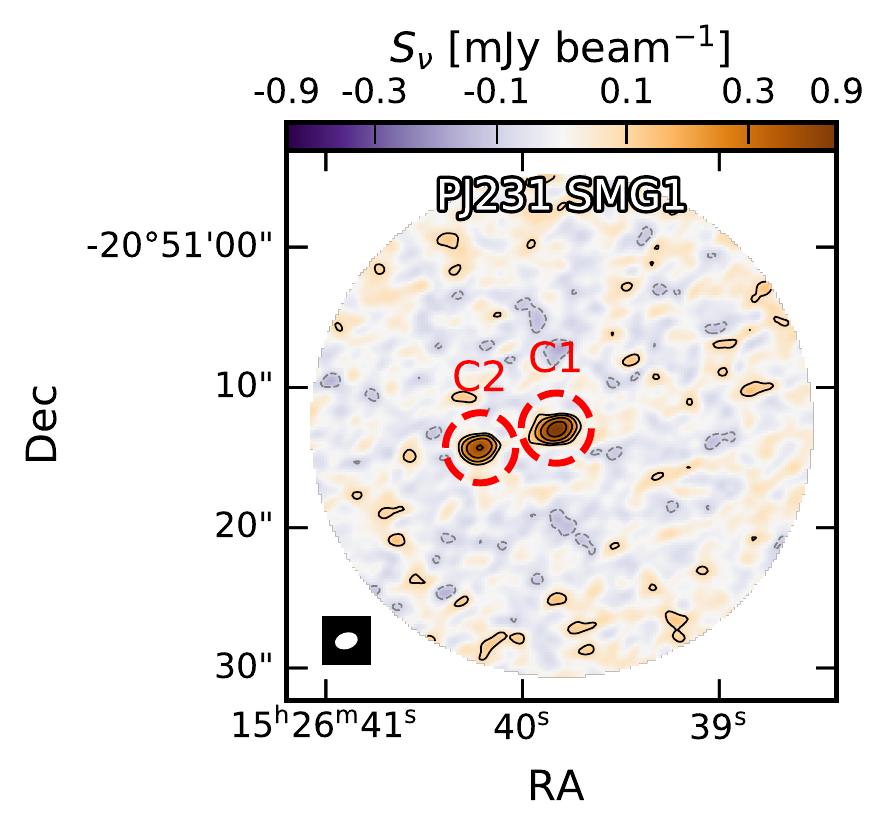}
    \includegraphics[width=0.32\textwidth]{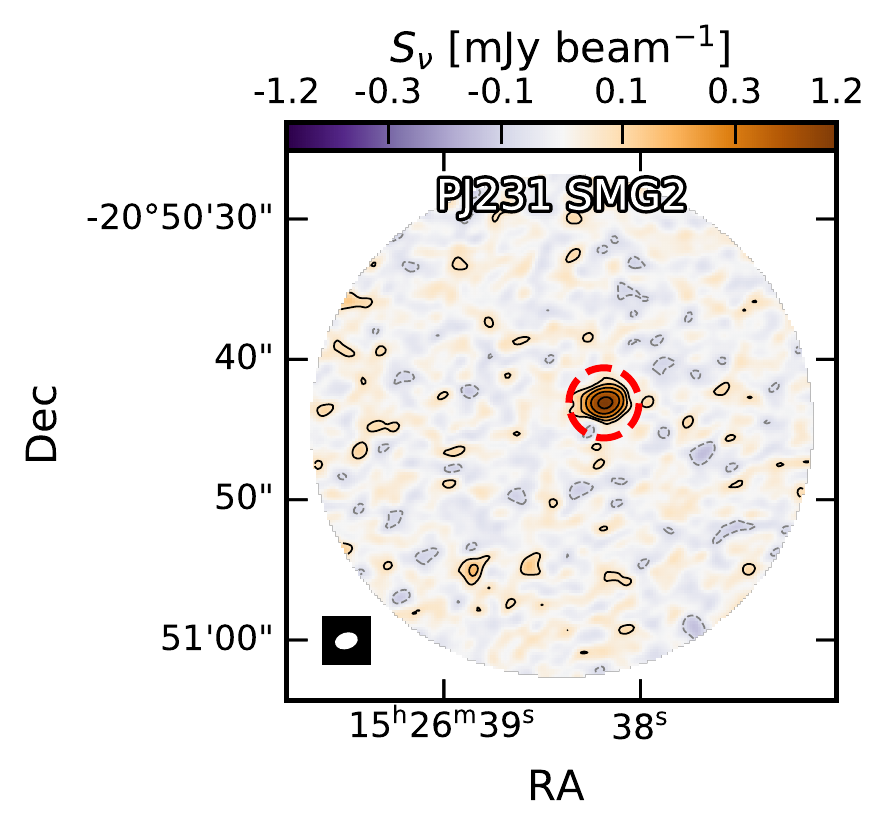}
    \includegraphics[width=0.32\textwidth]{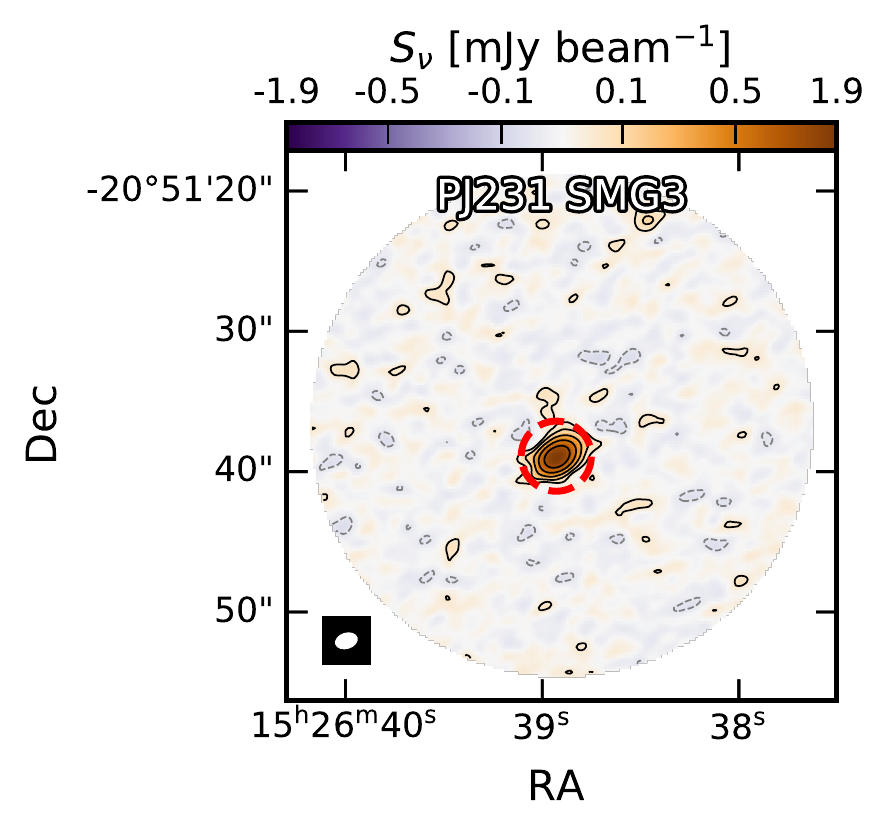}   \\  
    \includegraphics[width=0.32\textwidth]{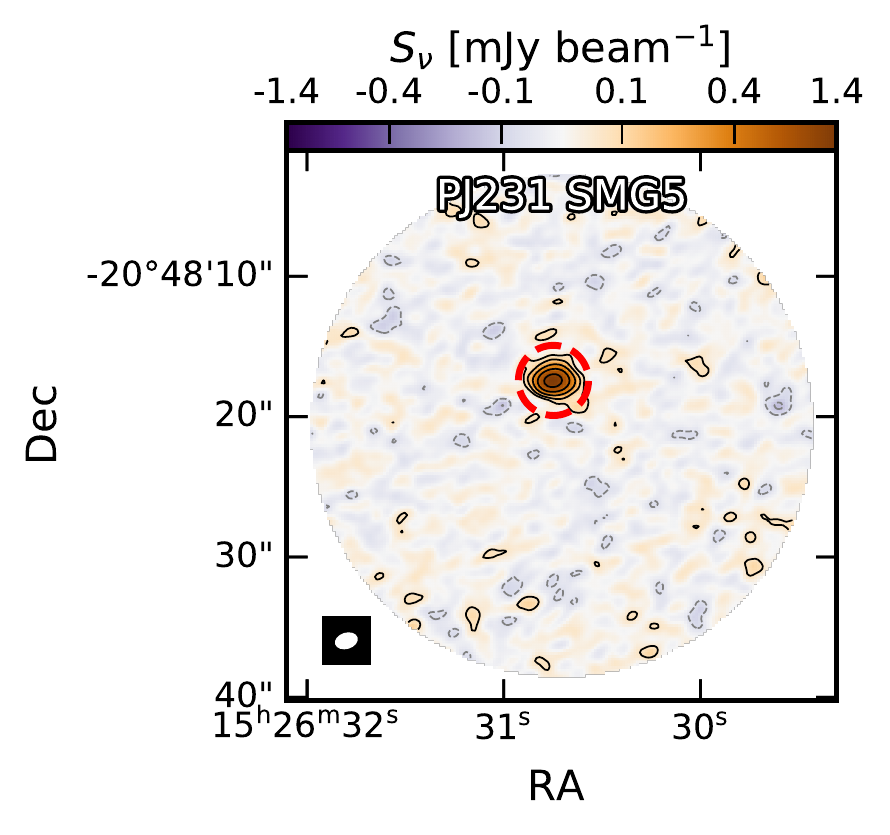}  
    \includegraphics[width=0.32\textwidth]{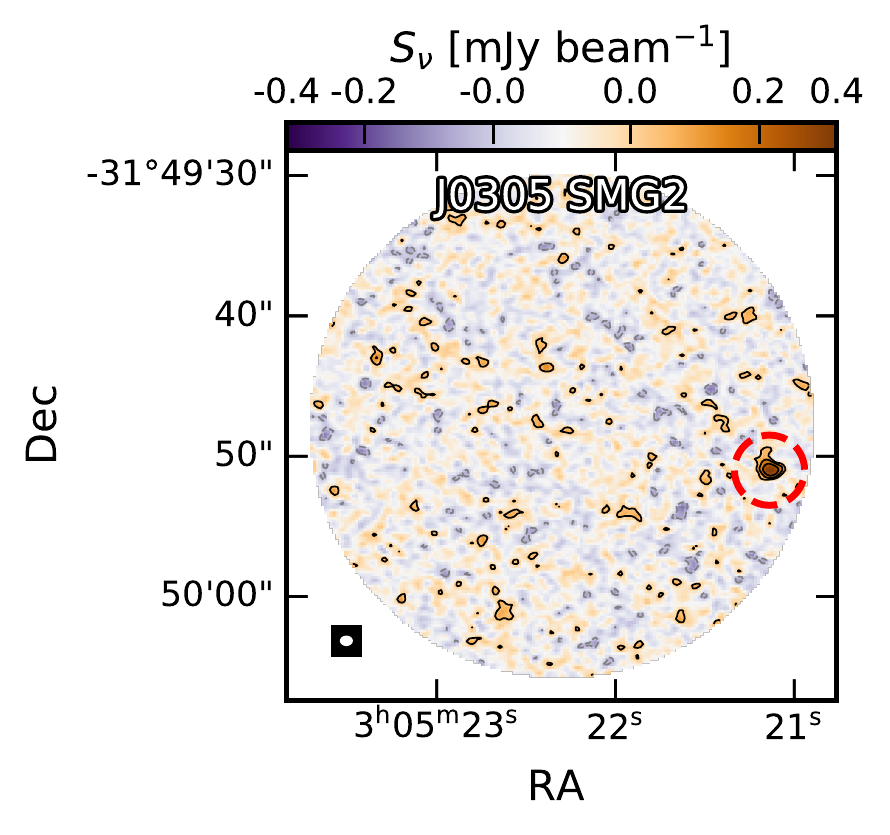} 
    \includegraphics[width=0.32\textwidth]{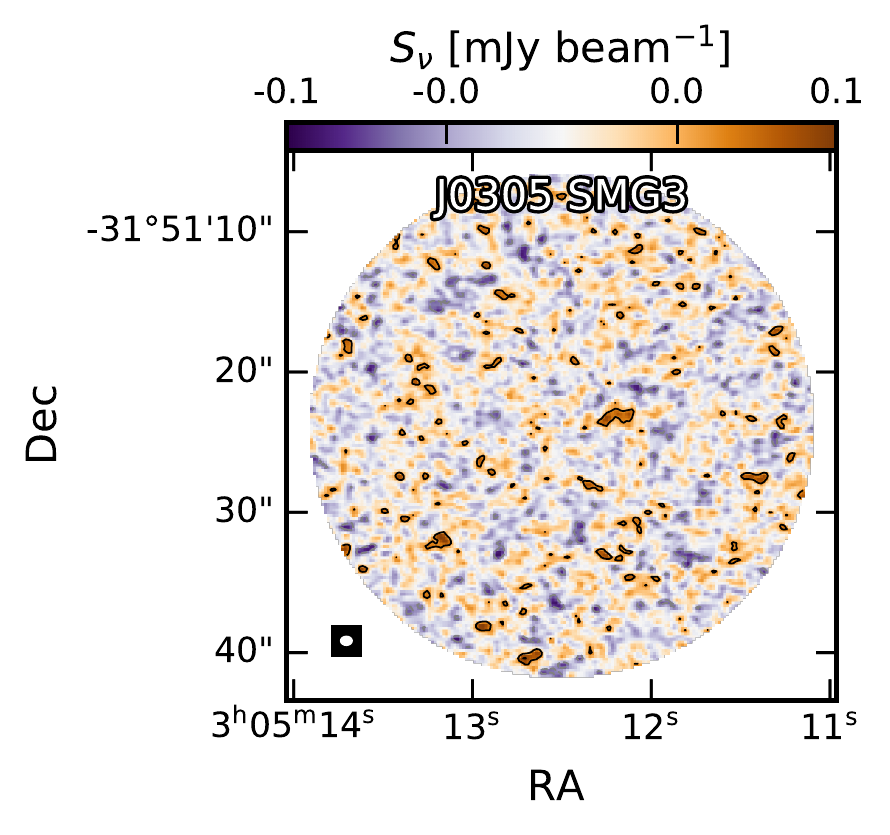} \\    
    \includegraphics[width=0.32\textwidth]{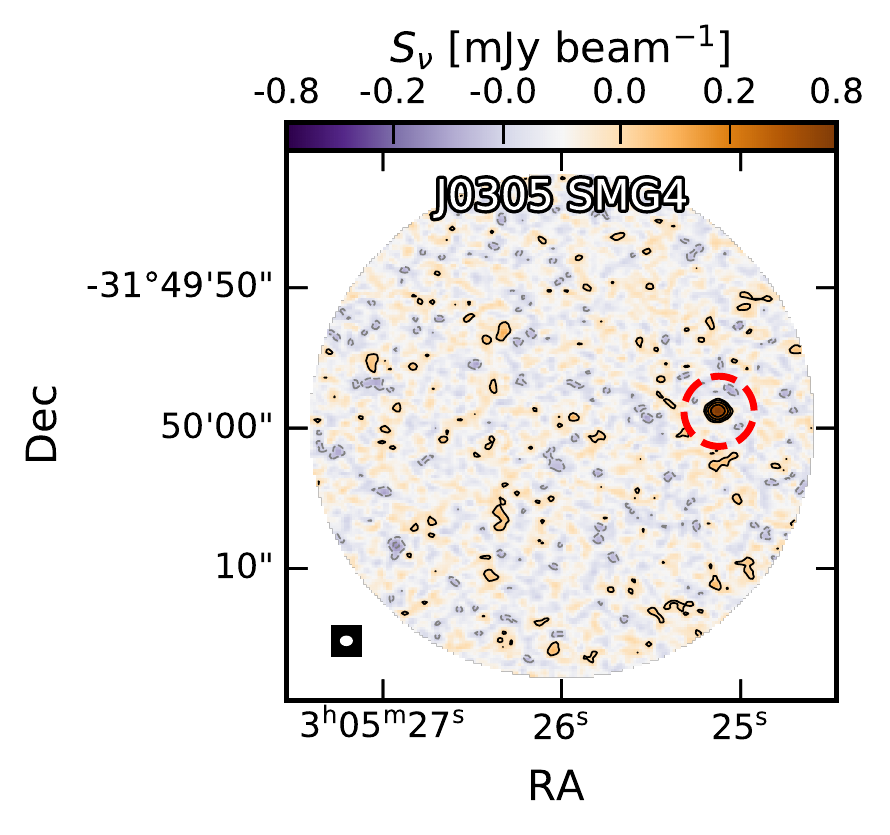}
    \includegraphics[width=0.32\textwidth]{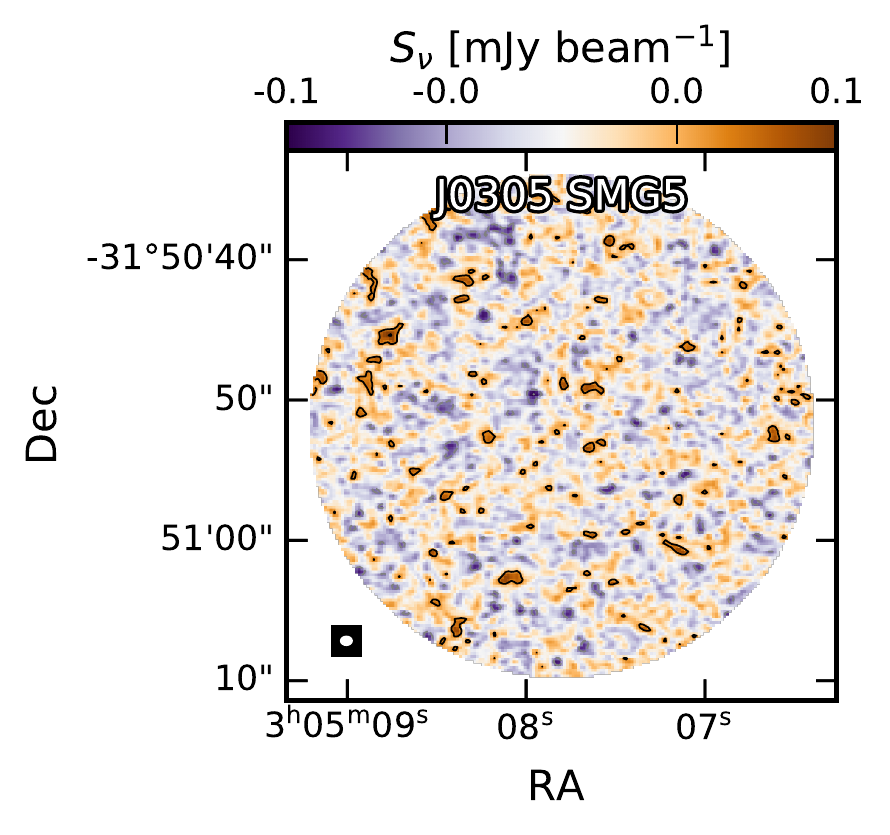} 
    \caption{ALMA continuum imaging of the SCUBA2--identified SMGs in the fields around PJ231--20 and J0305--3150. The color scaling is log--linear for better contrast, and the contours are logarithmic: $(-2,2,4,8,16,32) \sigma$, where $\sigma$ is the rms noise (see Table \ref{tab:quasar_obs_params}). The identified continuum sources ($>5\sigma$) are circled in dashed red, and are labeled "C1" and "C2" if multiple sources are detected in the same pointing.  The beam is plotted in the lower left corner (white against black square) and the sizes are tabulated in Table \ref{tab:quasar_obs_params}. The ALMA detections can be offset from the center of the pointing due to the large beam of the SCUBA2 imaging ($13"$ at $850\ \mu\rm{m}$, e.g. of the order of the ALMA Field of View) from which the target were selected. }
    \label{fig:fields_pj231_j0305}
\end{figure*}

\begin{figure*}
    \centering
    \includegraphics[width=0.32\textwidth]{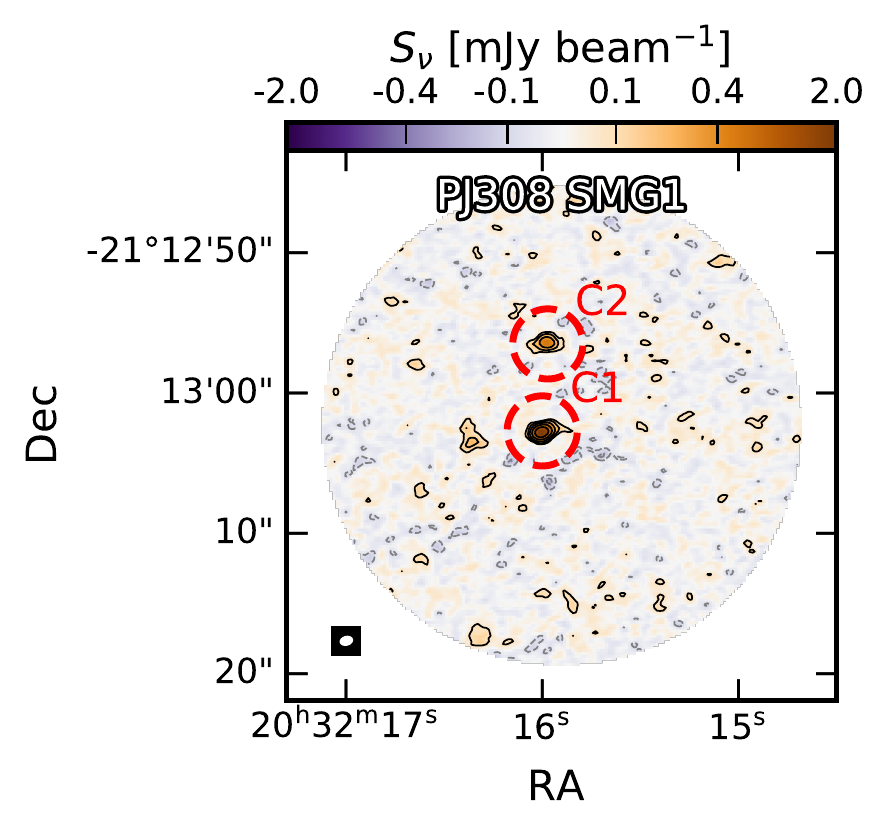}
    \includegraphics[width=0.32\textwidth]{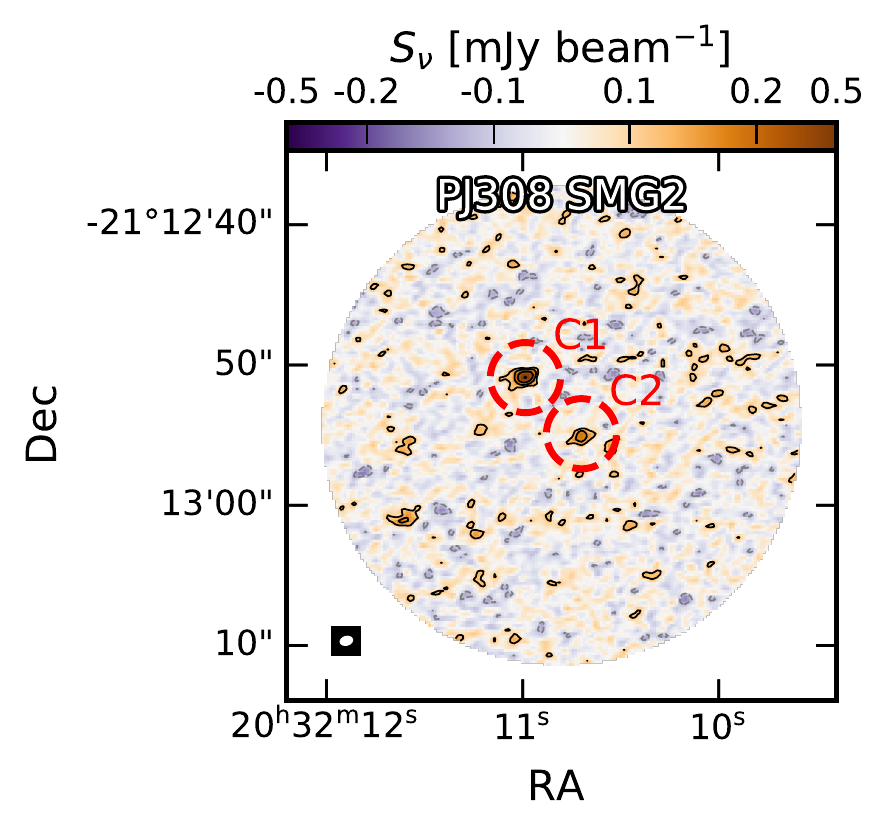}
    \includegraphics[width=0.32\textwidth]{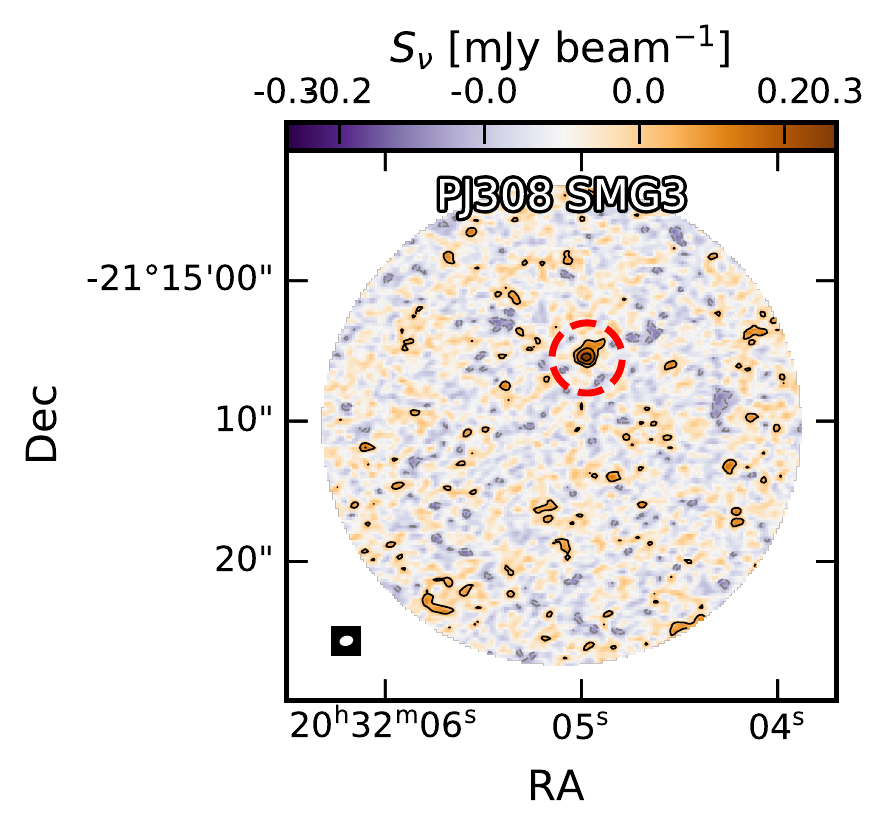}    \includegraphics[width=0.32\textwidth]{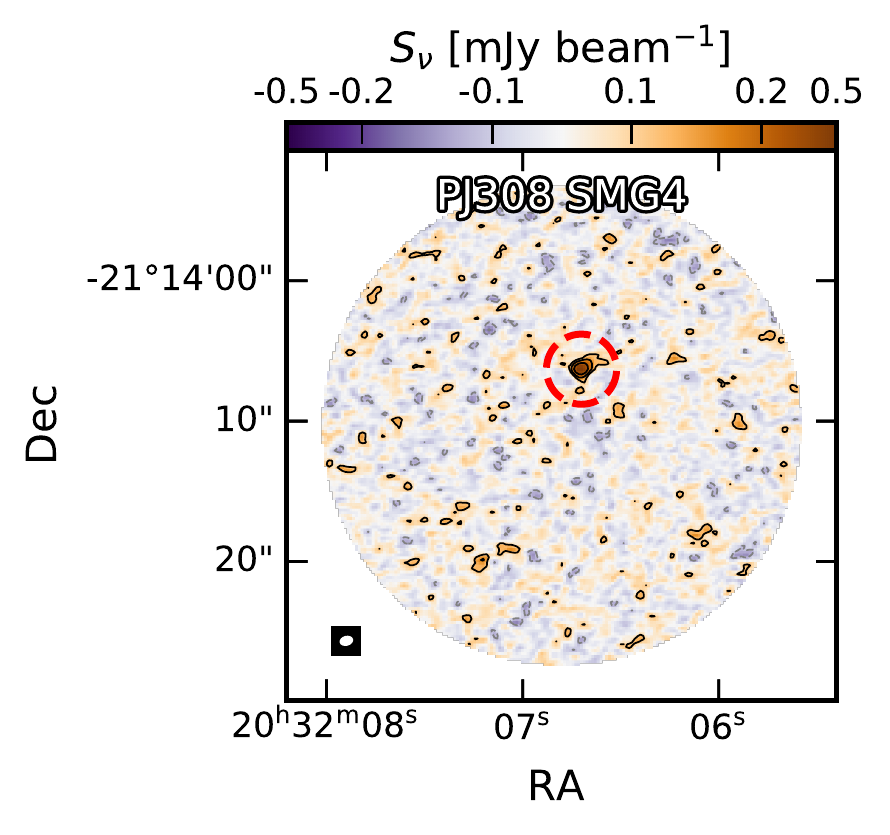} 
    \includegraphics[width=0.32\textwidth]{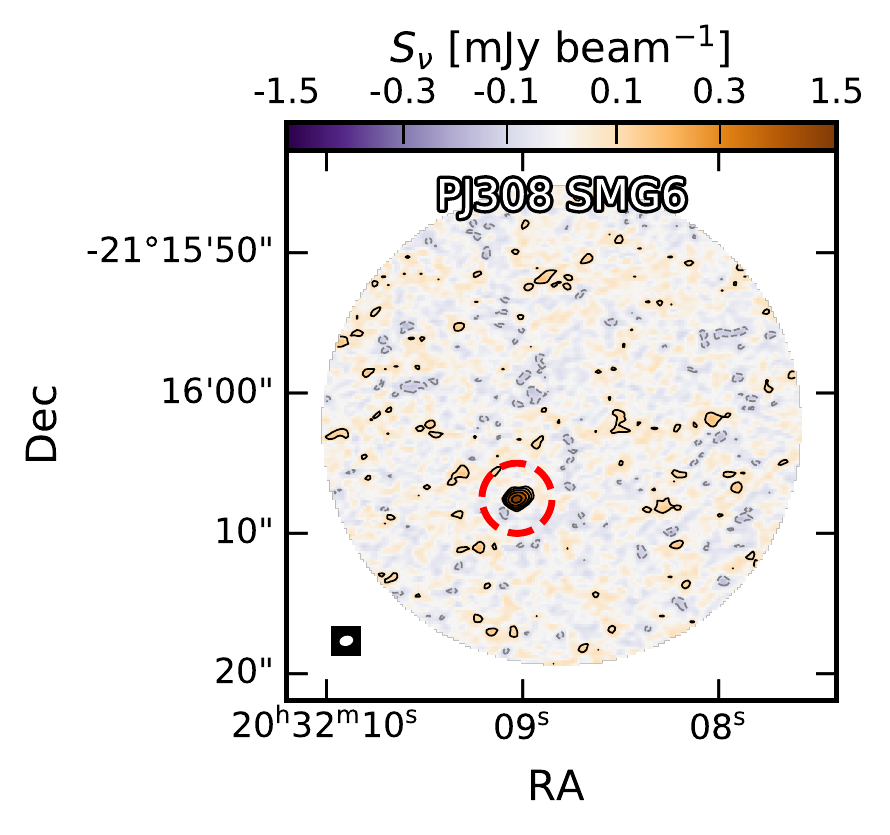}
    \includegraphics[width=0.32\textwidth]{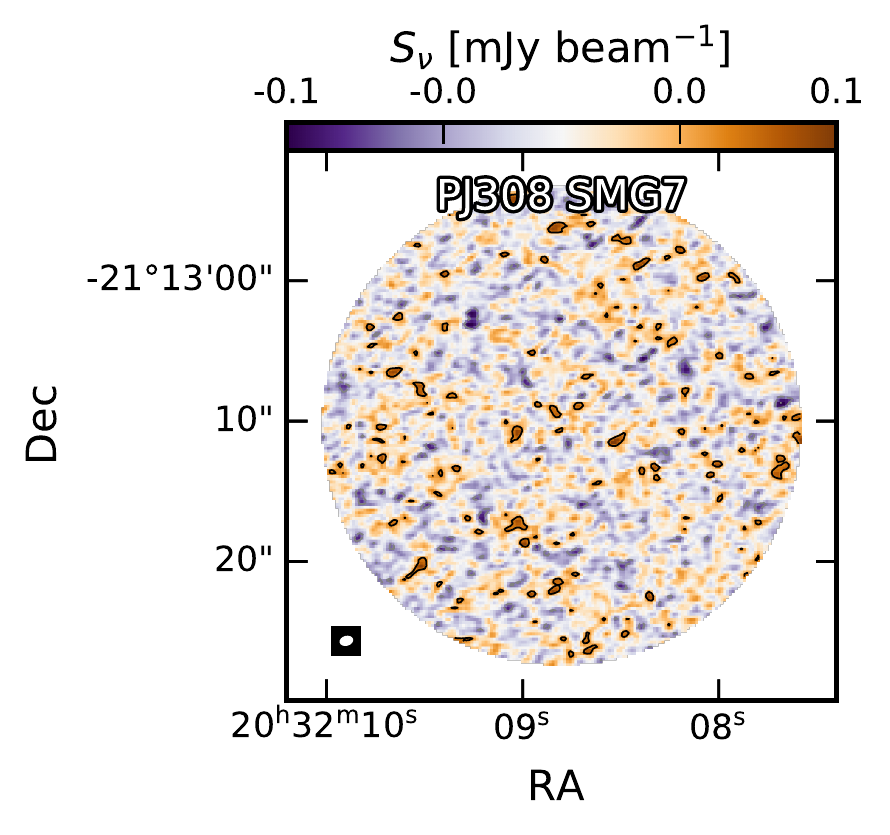} \\
    \includegraphics[width=0.32\textwidth]{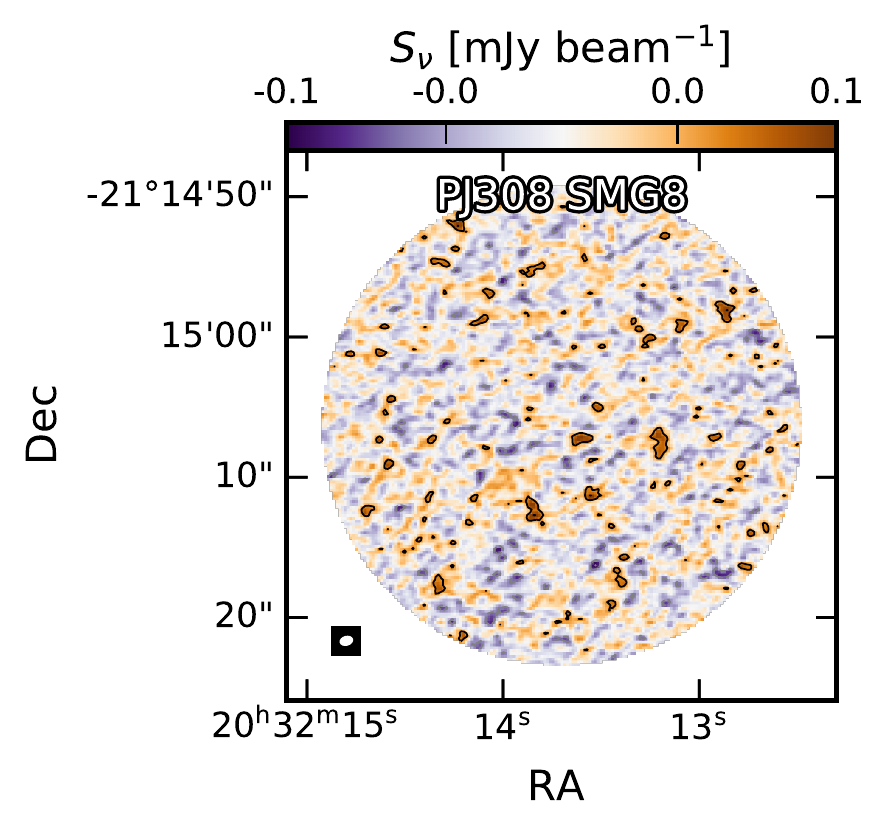}
    \includegraphics[width=0.32\textwidth]{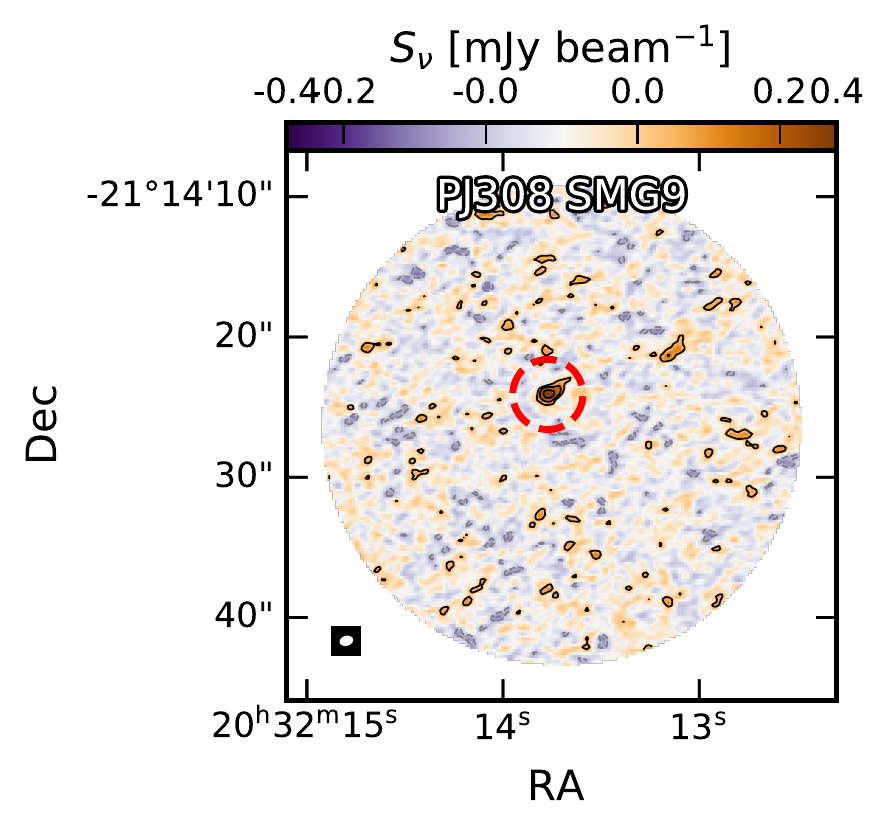}
    \includegraphics[width=0.32\textwidth]{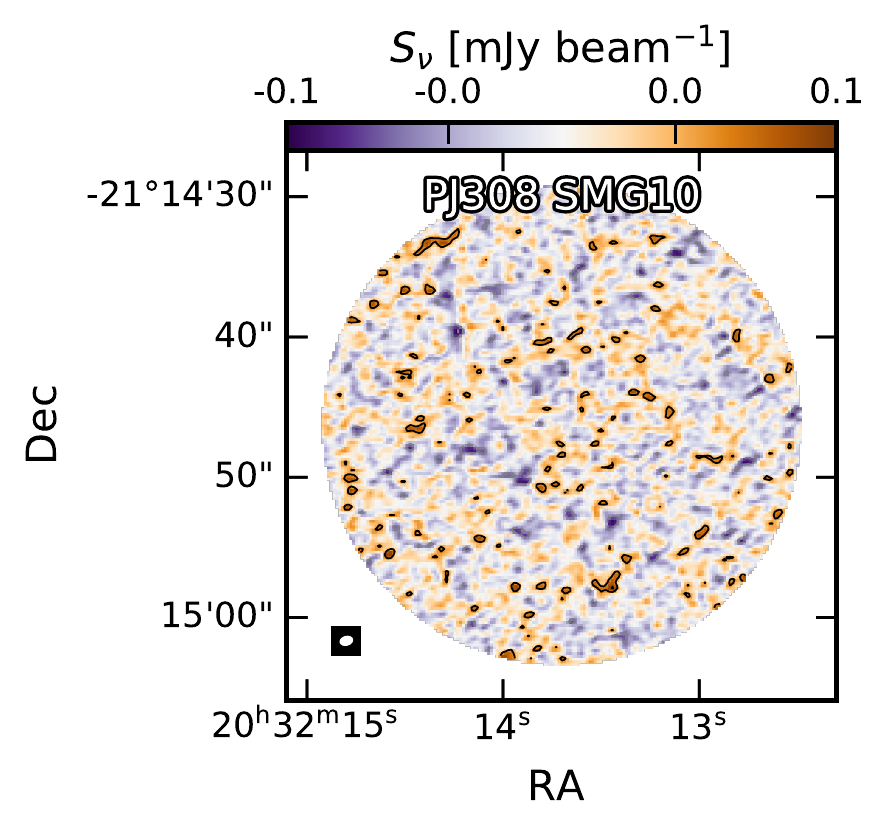}
    \caption{Same as Fig. \ref{fig:fields_pj231_j0305}, but for PJ308--21.} 
    \label{fig:fields_pj308}
\end{figure*}

Imaging and cleaning was performed with CASA, and the final images and datacubes were produced in the following fashion. First, the visibilities were imaged with a natural weighting and cleaned down to $2$ sigma (rms noise) to produce preliminary continuum maps and datacubes. Continuum sources (peak surface brightness $>5\sigma$) were identified in most SMG pointings (see Fig. \ref{fig:fields_pj231_j0305} and \ref{fig:fields_pj308}). Preliminary spectra of the continuum sources were then extracted using a $r=2"$ aperture and fitted with a simple Gaussian profile and a constant continuum to identify prominent lines (if any). Frequencies at $\pm1.25$ times the FWHM of significant lines were subsequently masked to image line--free continuum maps and cubes from the visibilities. The line--free continuum was then subtracted in the UV plane to produce continuum--subtracted cubes with $50\, \kms$ channels (this step was only performed in pointings where continuum sources were detected). The continuum--subtracted cubes were then imaged and cleaned down to $2\sigma$ (rms) with circular masks on the identified continuum sources. We present in Table \ref{tab:quasar_obs_params} the beam size, continuum rms and sensitivity per channel and per beam of the cleaned data products for each quasar field. 

\begin{table*}
    \centering
    \begin{tabular}{l|c|c|c|c|c|c}
        Quasar & $\rm{N}_{SMGs}$ & Continuum rms & Beam &  rms per channel$^a$&   $\nu_{obs}$ \\
        & & [mJy beam$^{-1}$]  & [arcsec$^2$] &  [mJy beam$^{-1}$ ]  & [GHz]
       \\ \hline
        PJ231--20 & 4& $3.0\times 10^{-2}$ & $1.66" \times 1.17"$& 0.55 / 0.61  & $ 234.95-238.62\, / \,248.70-252.35$ \\
        J0305--3150 & 4 & $2.5\times 10^{-2}$ & $0.93"\times0.75"$& 0.44 / 0.49 &
        $234.08-237.74\, /\, 247.77- 251.40 $\\
        PJ308--20 & 9 & $3.2\times 10^{-2}$ & $0.98"\times0.72"$ & 0.57 / 0.61  &
        $246.45-250.18 \,/ \,260.85-264.58$\\
    \end{tabular}
    \caption{Summary of the our ALMA observations of SMGs in the fields of three $z>6.5$ quasars. We report the beam and rms of the continuum images, as well as the sensitivity and frequency coverage of the datacubes. \textit{a)} The rms per channel is given for $50\,\kms$ channels and for both sidebands.}
    \label{tab:quasar_obs_params}
\end{table*}

\section{continuum--detected sources}
\label{sec:cont_sources}
\subsection{ALMA continuum detections}

We present the continuum maps of the $17$ ALMA pointings on SMGs detected in SCUBA2 850$\mu\rm{m}$ images in Figs. \ref{fig:fields_pj231_j0305} and \ref{fig:fields_pj308}. In 12 out of 17 pointings, we detect a continuum source in the ALMA data, and in 3 of those, we even detect two sources. Each source is given a unique identifier consisting of the quasar name, SMG number in the original SCUBA2 catalogue and source name (“C1" or “C2" for continuum sources, where “C1" is the brightest continuum source in the ALMA pointing, and “C2" is the fainter, secondary source in multiple systems).

The continuum fluxes were extracted using a $r=2"$ aperture, which encompasses all of the $2\sigma$ emission for the vast majority of sources (a large fraction of the SCUBA SMGs are resolved in a few beams in the ALMA continuum maps). The fluxes were computed using the residual scaling method \citep[e.g.,][]{Jorsater1995,Walter1999,Walter2008, Novak2019} and are listed alongside the coordinates and detection significance of each continuum source in Table \ref{tab:continuum_sources}.

\begin{table*}[]
    \centering
    \begin{tabular}{l|cccccccc}
ID & RA & DEC & $\rm{SN}_{\rm{Cont}}$ & $S_\nu$  [mJy] & $\rm{SN}_{\rm{Line}}$ & $\nu_{\rm{Line}}$ [GHz] & $S_\nu\Delta v$ [Jy $\kms$] & FWHM [$\kms$]      \\ \hline 
PJ231-SMG1-C1 & 15:26:39.83 & -20:51:12.87 & 29.9 & $1.09\pm0.08$ &  5.2 & $237.27\pm0.20$ & $1.85\pm0.63$  & 1960$^a$  \\
PJ231-SMG1-C2 & 15:26:40.22 & -20:51:14.28 & 19.8 & $0.74\pm0.08$ & - & - & - & - \\ 
PJ231-SMG2-C1 & 15:26:38.19 & -20:50:43.08 & 41.2 & $1.52\pm0.07$  &5.8& $236.01\pm0.03$ & $0.47\pm0.11$   & 291\\ 
PJ231-SMG3-C1 & 15:26:38.93 & -20:51:38.89 & 60.6 & $2.94\pm0.08$ & 9.6& $235.40\pm0.10$ & $2.16\pm0.40$   & 1167$^a$ \\ 
PJ231-SMG5-C1 & 15:26:30.75 & -20:48:17.40 & 45.4 & $1.84\pm0.08$ &8.9&  $236.92\pm0.08$ & $1.84\pm0.31$   & 1270$^a$\\ 
J0305-SMG2-C1 & 03:05:21.14 & -31:49:51.02 & 15.7 & $0.53\pm0.07$ & - & - & - & -\\ 
J0305-SMG4-C1 & 03:05:25.12 & -31:49:58.82 & 31.3 & $0.74\pm0.07$ & - & - & - & -\\ 
PJ308-SMG1-C1 & 20:32:16.00 & -21:13:02.69 & 58.3 & $2.40\pm0.11$  &4.3& $262.54\pm 0.02$   & $0.27\pm0.08$   & 129  \\ 
PJ308-SMG1-C2 & 20:32:15.97 & -21:12:56.49 & 13.9 & $0.82\pm0.22$ & - & - & - & -\\ 
PJ308-SMG2-C1 & 20:32:10.99 & -21:12:50.89 & 16.6 & $0.67\pm0.10$ & - & - & - & -\\ 
PJ308-SMG2-C2 & 20:32:10.70 & -21:12:54.88 & 5.42 & $0.92\pm0.38$ & - & - & - & -\\ 
PJ308-SMG3-C1 & 20:32:04.97 & -21:15:05.49 & 10.7 & $0.44\pm0.09$ & - & - & - & -\\ 
PJ308-SMG4-C1 & 20:32:06.70 & -21:14:06.30 & 15.1 & $0.65\pm0.09$ & - & - & - & -\\ 
PJ308-SMG6-C1 & 20:32:09.03 & -21:16:07.48 & 46.1 & $1.54\pm0.10$ & - & - & - & -\\ 
PJ308-SMG9-C1 & 20:32:13.77 & -21:14:24.09 & 11.4 & $0.43\pm0.09$ & 3.1 & $263.11\pm 0.02$ & $0.081\pm0.067$ & 298

    \end{tabular}
    \caption{Properties of the detected continuum sources. Flux densities are computed by taking the integrated flux in an aperture of $r=2"$ and applying residual scaling \citep[e.g.,][]{Jorsater1995,Walter1999,Walter2008, Novak2019}. The line fluxes are rescaled by $1/0.84$ because the continuum--subtracted (in the uv--plane) data are averaged over $1.2\times \rm{FWHM}$, which contain $84\%$ of the flux for a Gaussian line \citep[see][Appendix A]{Novak2020}. \textit{a)} The quoted width is that of the best--fit single Gaussian profile. In Appendix \ref{app:disc_z24_pj231} we discuss how these can be attributed to CO7--6 and \ci at $z\simeq 2.4$.} 
    \label{tab:continuum_sources}
\end{table*}

\subsection{ALMA/SCUBA2 continuum flux density comparison}
\label{sec:disc_scuba_sources}
Only $70\%$ of the SMGs selected in the SCUBA2 imaging have a continuum detection in the ALMA data. We now investigate the different continuum fluxes at $\lambda =850\ \mu\rm{m}$ (SCUBA2) and $\lambda \simeq 1.3\ \rm{mm}$ (ALMA) to determine whether this is expected for sources at various redshifts.

In order to do so, we model the dust in the optically thin limit using a modified black body spectrum. We include the prescription of \citet[][]{DaCunha2013} to model the effect of Cosmic Microwave Background (CMB) heating and correct for contrast against the CMB. The opacity is assumed to follow the best fit relation and coefficients of \citet{Dunne2003} $\kappa_{\nu_{rest}} = \kappa_{\nu_0}(\nu_{rest}/\nu_0)^\beta$ with $\nu_0=c/(125\mu\rm{m})$, where $\beta$ is the power--law dust emissivity index. We assume a fiducial dust temperature $T=30 \, \rm{K}$ and dust emissivity $\beta=1.5$ following the common values found in $0.1\lesssim z\lesssim 2.8$ \citep[e.g.,][]{Hwang2010}, or $T=47\, \rm{K}$ as commonly used for high--redshift quasars \citep[e.g.,][]{Beelen2006,Venemans2020}.  

We account for flux deboosting by correcting the observed SCUBA2 fluxes by mean flux boosting factor of 1.19 for sources at Signal to Noise Ratio (SNR) of $4$ (Li et al., in prep.), which is similar to the SNR of our sources in the SCUBA2 images. We take into account the impact of multiplicity on the observed ALMA fluxes. This phenomenon is well--known from earlier ALMA follow--up of SMGs \citep[][]{Barger2012,Smolcic2012, Hodge2013}: unresolved bright SMGs in single--dish observations are often resolved in interferometric observations and break into multiple sources. Consequently, the total flux might not be recovered as some resolved sources are below the sensitivity of the ALMA data. The number of SMGs showing multiple counterparts in high--resolution mm observations (e.g. the observed multiplicity) varies from $16\%$ to $45\%$ \citep[][]{Barger2012,Smolcic2012, Hodge2013}. These are only lower limits since secondary or tertiary sources in multiple systems might not be bright enough to be detected in the shallow ALMA observations. In this work, three of the $15$ SMGs have two corresponding ALMA detections, implying an observed multiplicity of the SMGs of $>20\%$. Following \citet[][]{Hodge2013}, we assume an intrinsic multiplicity of $50\%$. We further assume that when a source is resolved in multiple components, the strongest source accounts for $65\%$ of the total flux following what is observed in our multiple detections (see Table \ref{tab:continuum_sources}), which is an upper limit considering some sources will not be detected in the ALMA continuum maps.
 
\begin{figure}
    \centering
    \includegraphics[width=0.47\textwidth]{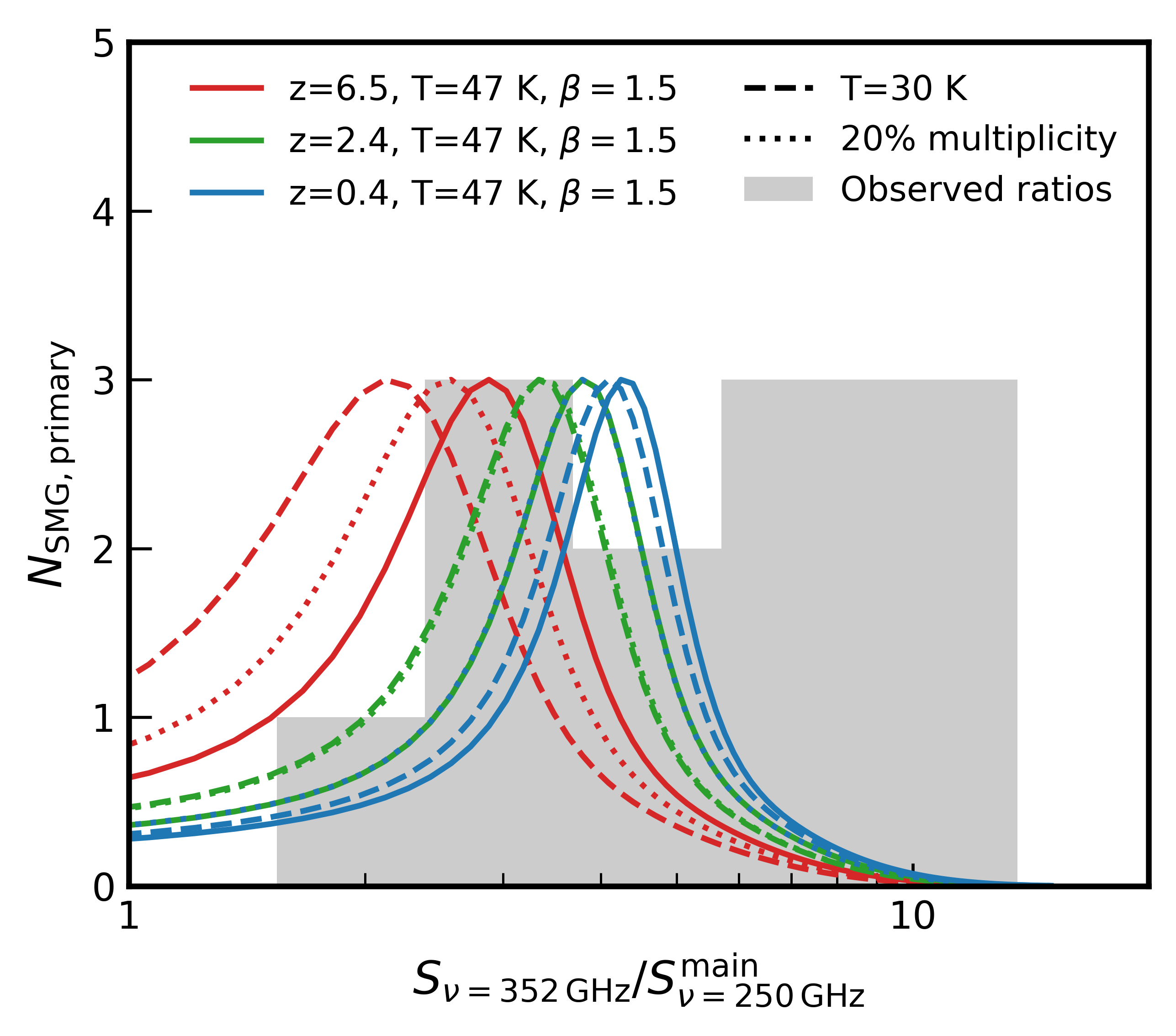}
    \caption{Observed continuum flux ratios (grey) between the SCUBA2 detections ($352$ GHz) and the primary ALMA counterpart ($\sim 250$ GHz). Predicted ratio based on modified black body SED models (see Section \ref{sec:disc_scuba_sources}) are plotted for a dust temperature $T=47 \rm{\, K}$ and dust emissivity index $\beta=1.5$ at $z=0.4,2.4,6.5$ in blue, green and red, respectively. Dashed and dotted lines indicate similar predictions with a lower temperature ($T=30 K$) or source multiplicity ($20\%$). The excess of $352$ GHz to $250$ GHz flux is expected if our assumptions for the source multiplicity and fraction of flux in the brightest sources are conservative (see Sec. \ref{sec:disc_scuba_sources}).}
    \label{fig:cont_ratios}
\end{figure}

We can now predict the expected continuum densities at $\sim 250$ GHz from the observed SCUBA2 continuum flux densities. For simplicity, we consider only the brightest source detected with ALMA. We compare in Fig. \ref{fig:cont_ratios} the observed flux density ratios against the ones extrapolated from modified black body SEDs for various dust temperatures, source multiplicity and redshift. 
We find that we cannot constrain the redshift of the SMGs using the available ALMA and SCUBA2 continuum fluxes, as the redshift has a minimal impact on the flux ratios compared to multiplicity, which is poorly constrained. The excess of faint ALMA counterparts to bright SCUBA2 detections suggests that a significant fraction of the $850\ \mu\rm{m}$ flux density comes from sources undetected in the higher resolution ALMA observations. This hypothesis is in agreement with the conservative assumptions of multiplicity and fraction of flux in the brightest source made above. In conclusion, the ALMA continuum detections are broadly in line with measured flux densities in the SCUBA2 imaging, and we attribute any discrepancies to faint sources that are resolved and undetected in the ALMA observations.

\subsection{SCUBA2/ALMA SMGs redshifts from emission lines}

\begin{figure}
    \raggedright
    \includegraphics[height=0.154\textheight,trim={0.3cm 0 0.25cm 0},clip]{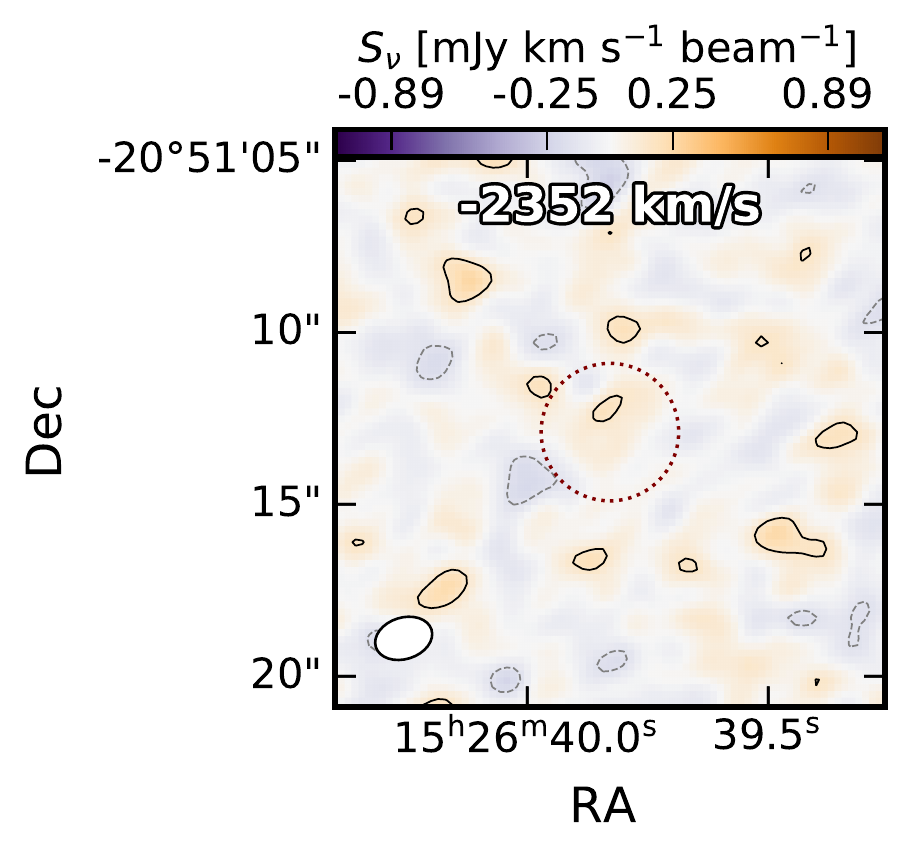}
    \includegraphics[height=0.154\textheight,trim={3.25cm 0 0.25cm 0},clip]{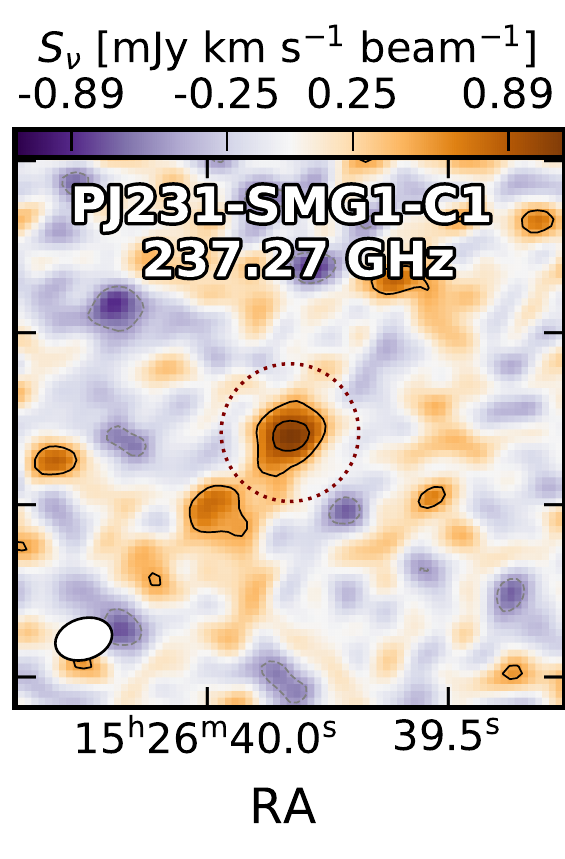} \\
    \includegraphics[height=0.154\textheight,trim={0.3cm 0 0.25cm 0},clip]{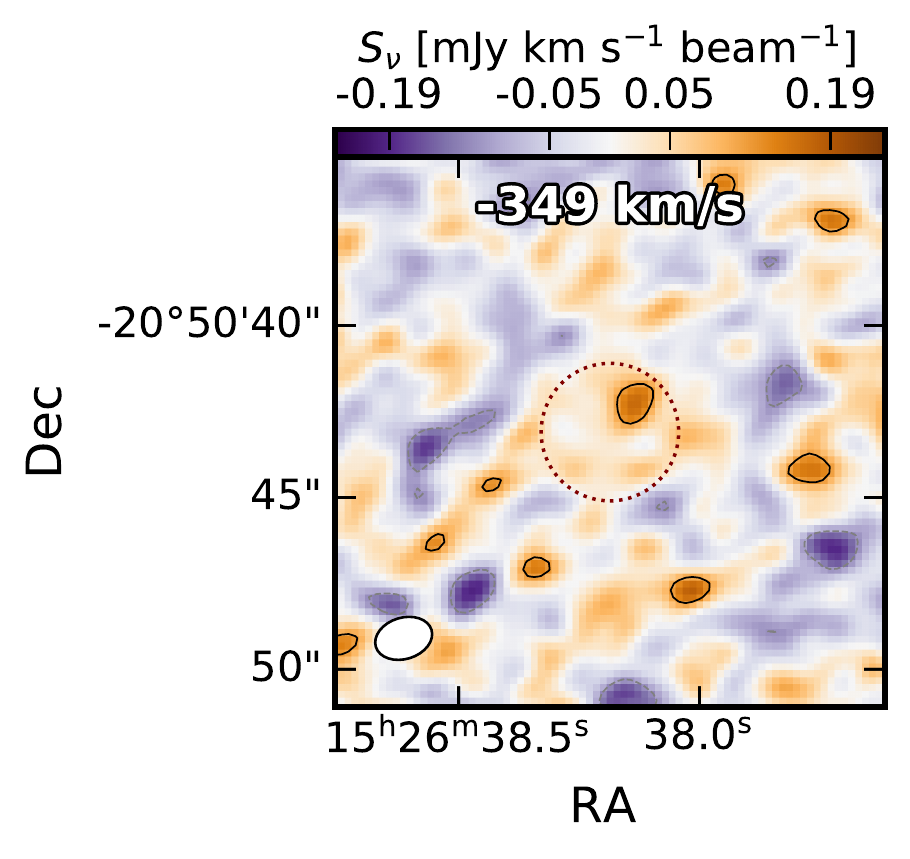}
    \includegraphics[height=0.154\textheight,trim={3.25cm 0 0.25cm 0},clip]{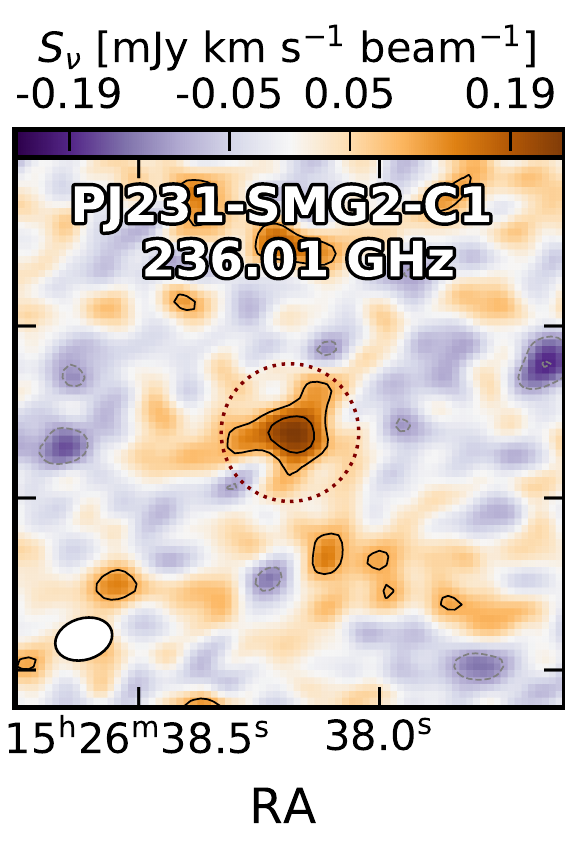}
    \includegraphics[height=0.154\textheight,trim={3.25cm 0 0.25cm 0},clip]{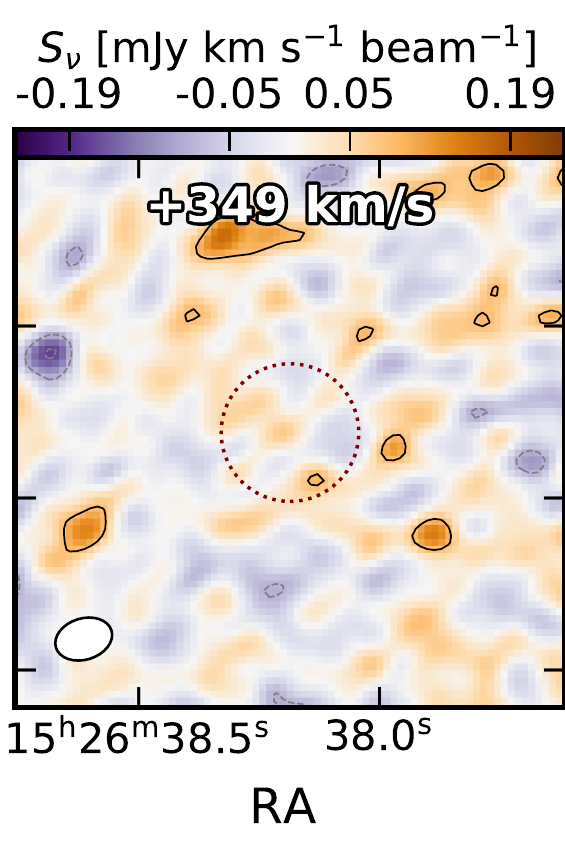} \\
    \hspace{2.35cm}
    \includegraphics[height=0.154\textheight,trim={0.3cm 0 0.25cm 0},clip]{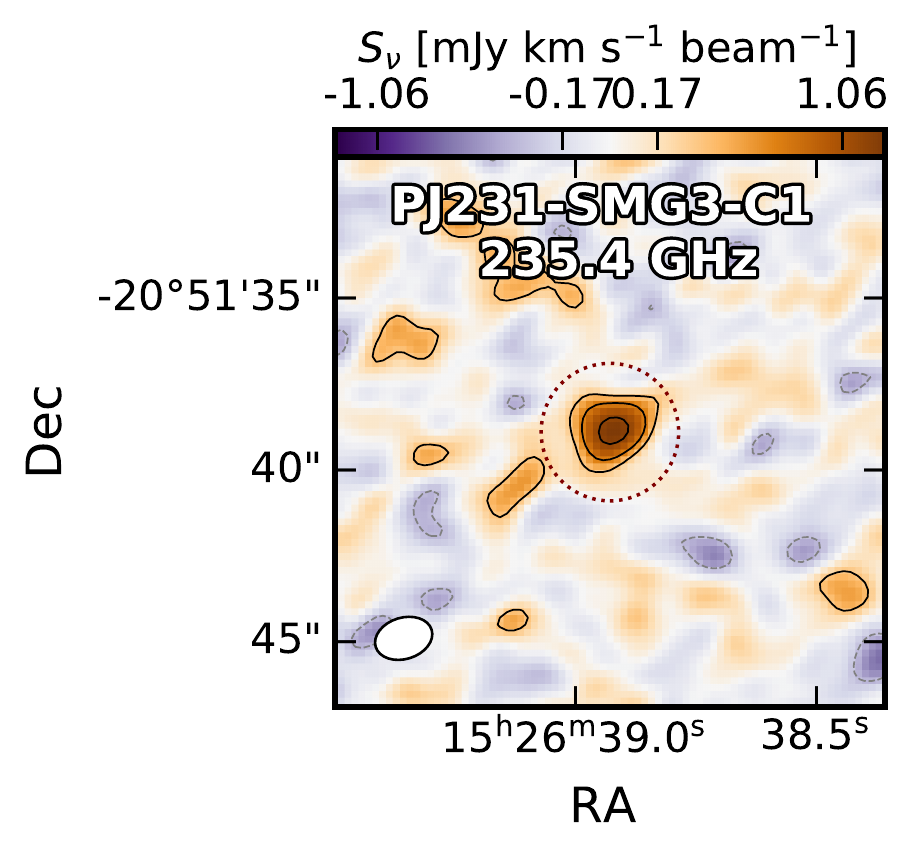}
    \includegraphics[height=0.154\textheight,trim={3.25cm 0 0.25cm 0},clip]{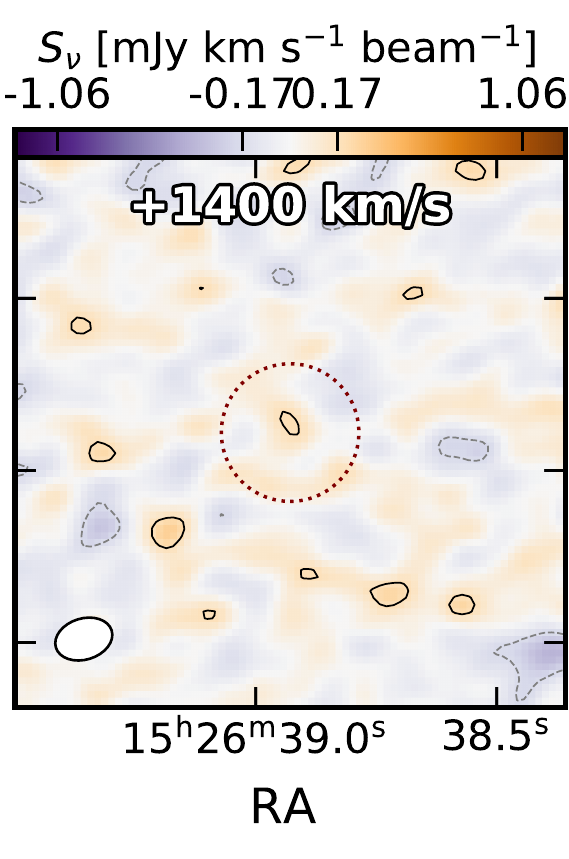} \\
    \includegraphics[height=0.154\textheight,trim={0.3cm 0 0.25cm 0},clip]{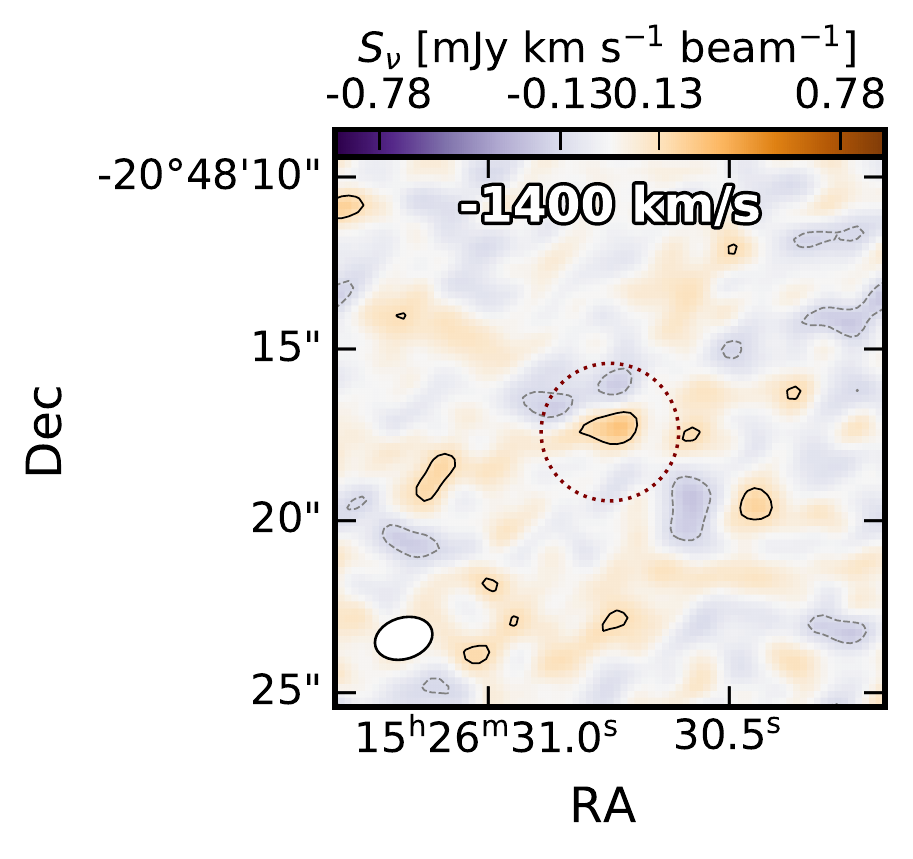}
    \includegraphics[height=0.154\textheight,trim={3.25cm 0 0.25cm 0},clip]{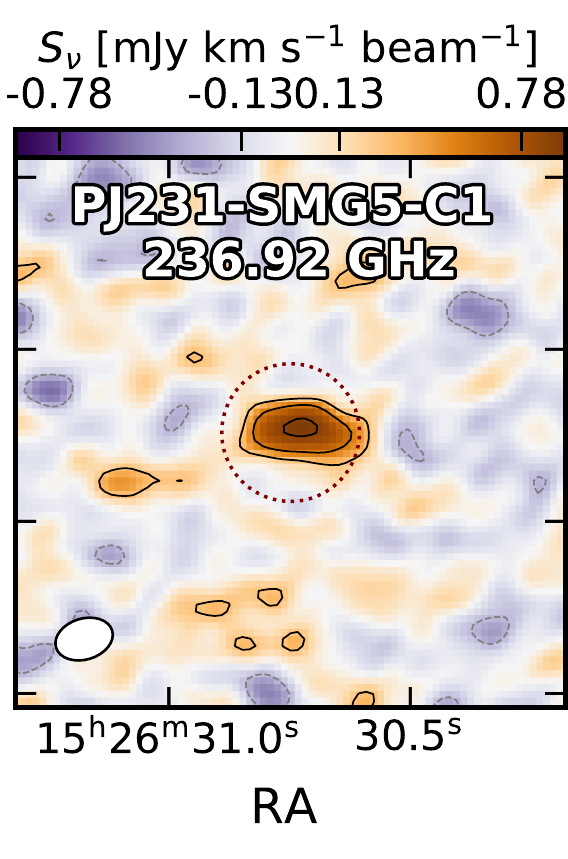}
    \includegraphics[height=0.154\textheight,trim={3.25cm 0 0.25cm 0},clip]{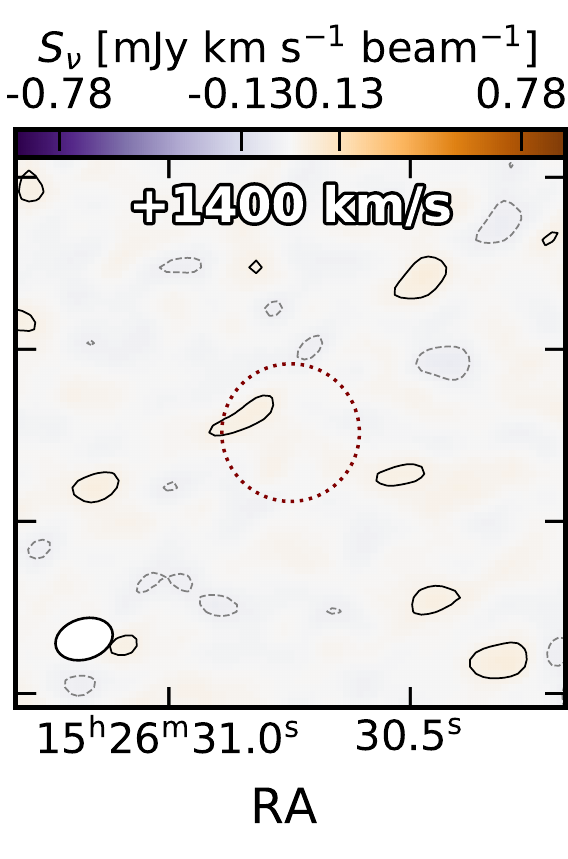} \\
    \includegraphics[height=0.153\textheight,trim={0.3cm 0 0.35cm 0},clip]{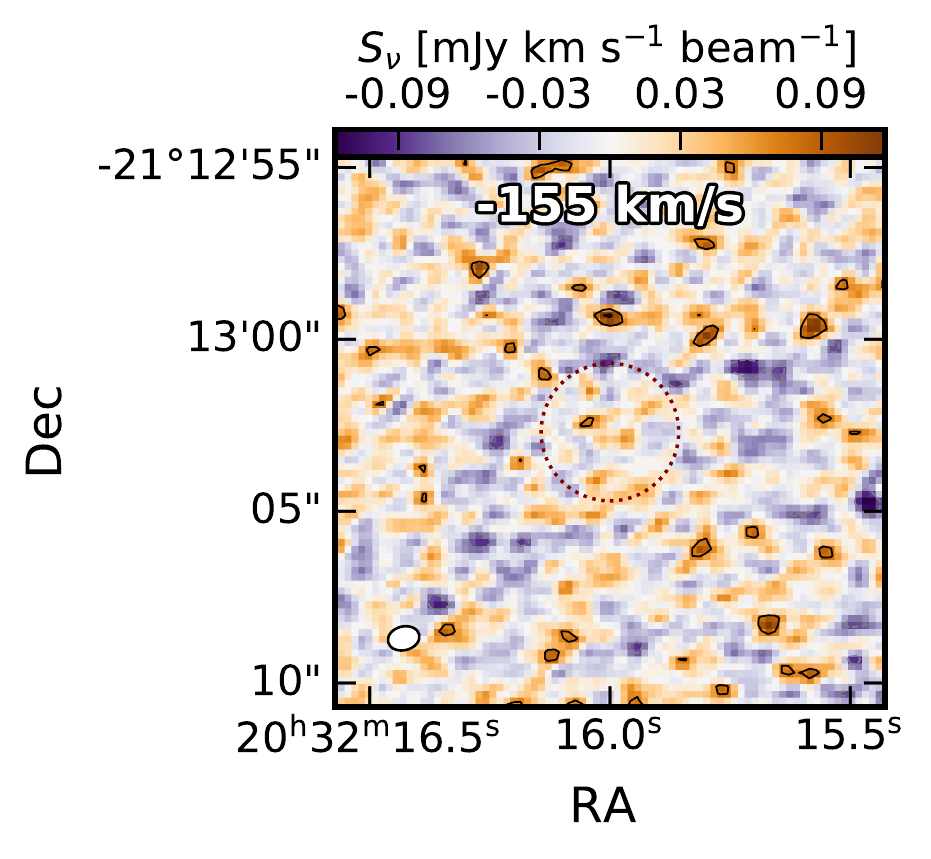}
    \includegraphics[height=0.153\textheight,trim={3.25cm 0 0.35cm 0},clip]{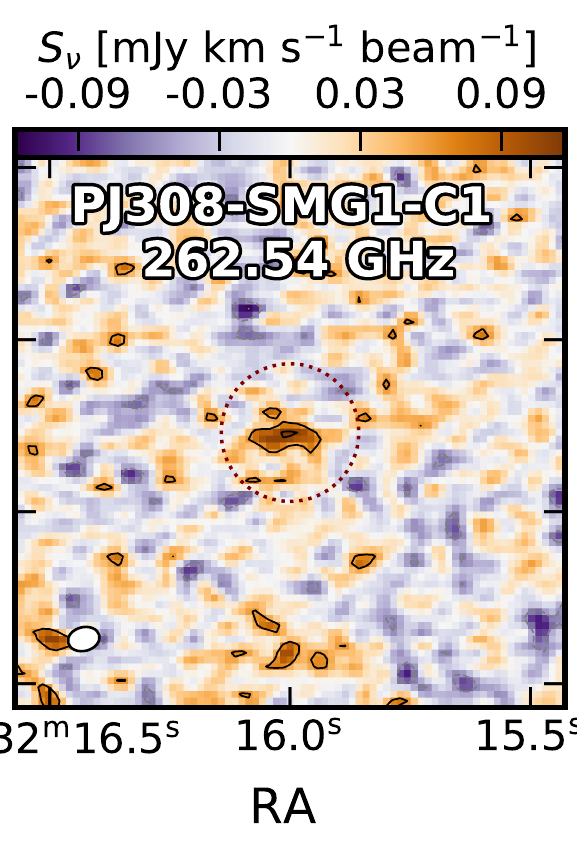}
    \includegraphics[height=0.153\textheight,trim={3.25cm 0 0.35cm 0},clip]{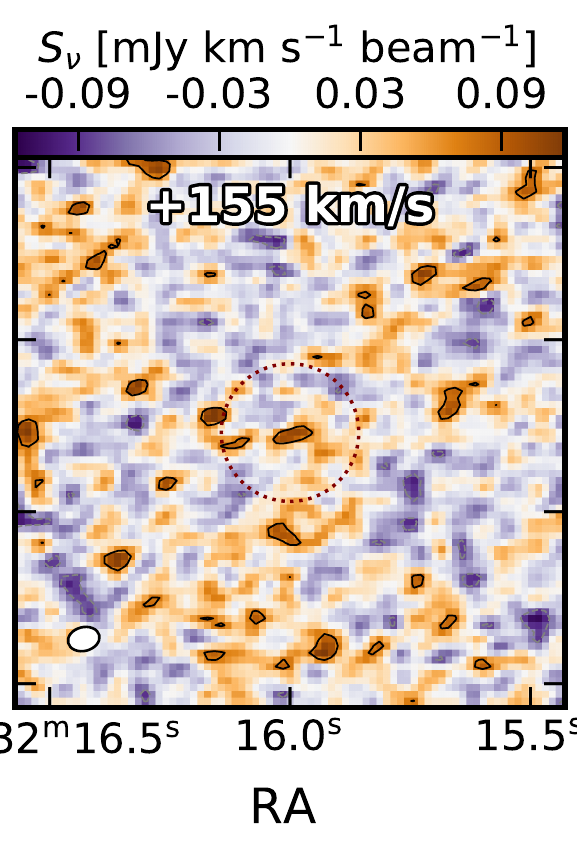} \\
    \includegraphics[height=0.154\textheight,trim={0.3cm 0 0.25cm 0},clip]{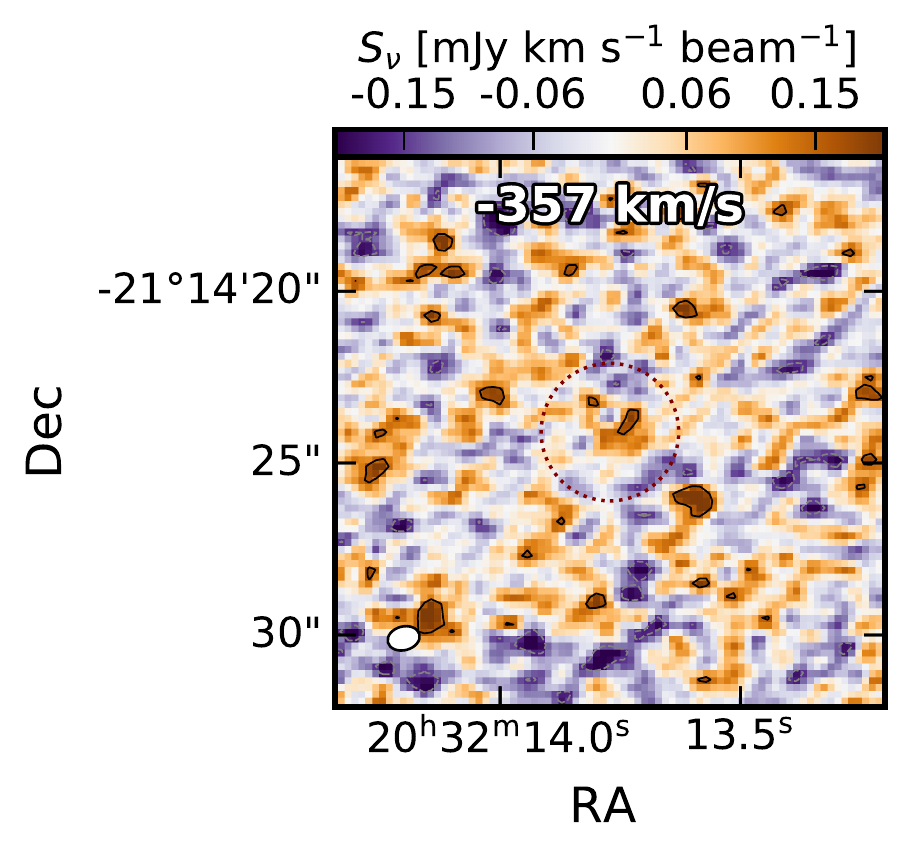}
    \includegraphics[height=0.154\textheight,trim={3.25cm 0 0.25cm 0},clip]{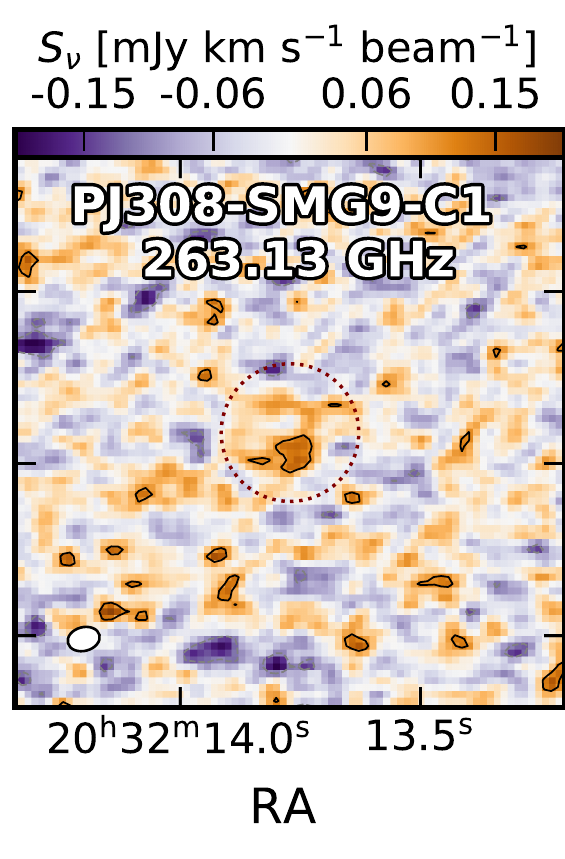}
    \includegraphics[height=0.154\textheight,trim={3.25cm 0 0.25cm 0},clip]{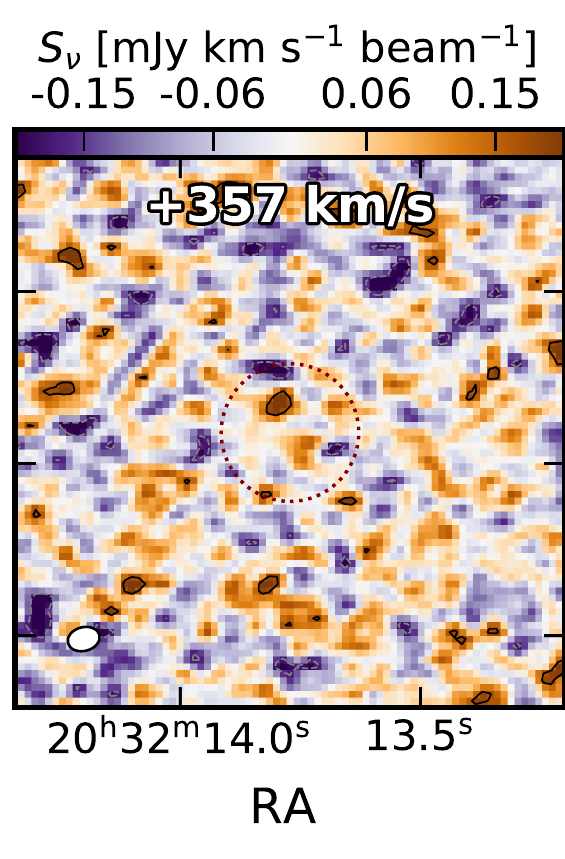}
    
    \caption{Emission line map, velocity--integrated over $1.2\times$FWHM of the fitted Gaussian profile to the extracted spectrum. Each row features a significant emission line found in one of the ALMA continuum sources. For each line, we also provide two additional maps (left/right) integrated over the same velocity range but offset by $\pm 1.2$FWHM. In two cases, either of these control maps is missing as the emission is detected close to the edge of the band. The contours are logarithmic (-4,-2,2,4,8)$\sigma$ (rms).  }
    \label{fig:line_dets_continuum_sources}
\end{figure}

The main aim of our observations is to confirm whether the SCUBA2 SMGs are at the redshift of the $z>6$ quasar in the field by detecting their redshifted \cii\ line. Therefore, spectra for each ALMA continuum source were extracted from the continuum--subtracted datacubes using a $r=2"$ aperture and applying residual scaling. A single Gaussian was fitted to each spectra to locate any significant emission feature. For each emission line detected, a velocity--integrated emission line map was produced by integrating channels within $\pm$1.2 $\times$FWHM of the line. Additionally, control maps with the same velocity range, but containing the velocity channels adjacent to the lines were produced to assess visually the significance of the line. These emission line maps are presented in Fig. \ref{fig:line_dets_continuum_sources}. 

We detect emission lines in six of the continuum--detected sources: four of the continuum sources in the field of PJ231--20 show $>5\sigma$ lines, and two of the sources around PJ308--21 have $3-4\sigma$ lines. Only the weak lines in PJ308-SMG1-C1 and PJ308-SMG9-C9 are approximately at the same redshift than that of the quasar, which we consider as marginal. Indeed, the emission lines of the PJ231 SMGs are detected in the lower sideband of the ALMA setup and can be ascribed to CO7--6 and \ci\ at $z=2.4$ coincident with the redshift of a Mg{~\small II} absorber in the spectrum of the quasar (see Appendix \ref{app:disc_z24_pj231} for a more detailed discussion of this result).  All the detected line significances, FWHM and frequencies are presented with the continuum sources information in Table \ref{tab:continuum_sources}. 

\begin{figure}
    \centering
    \includegraphics[width=0.47\textwidth]{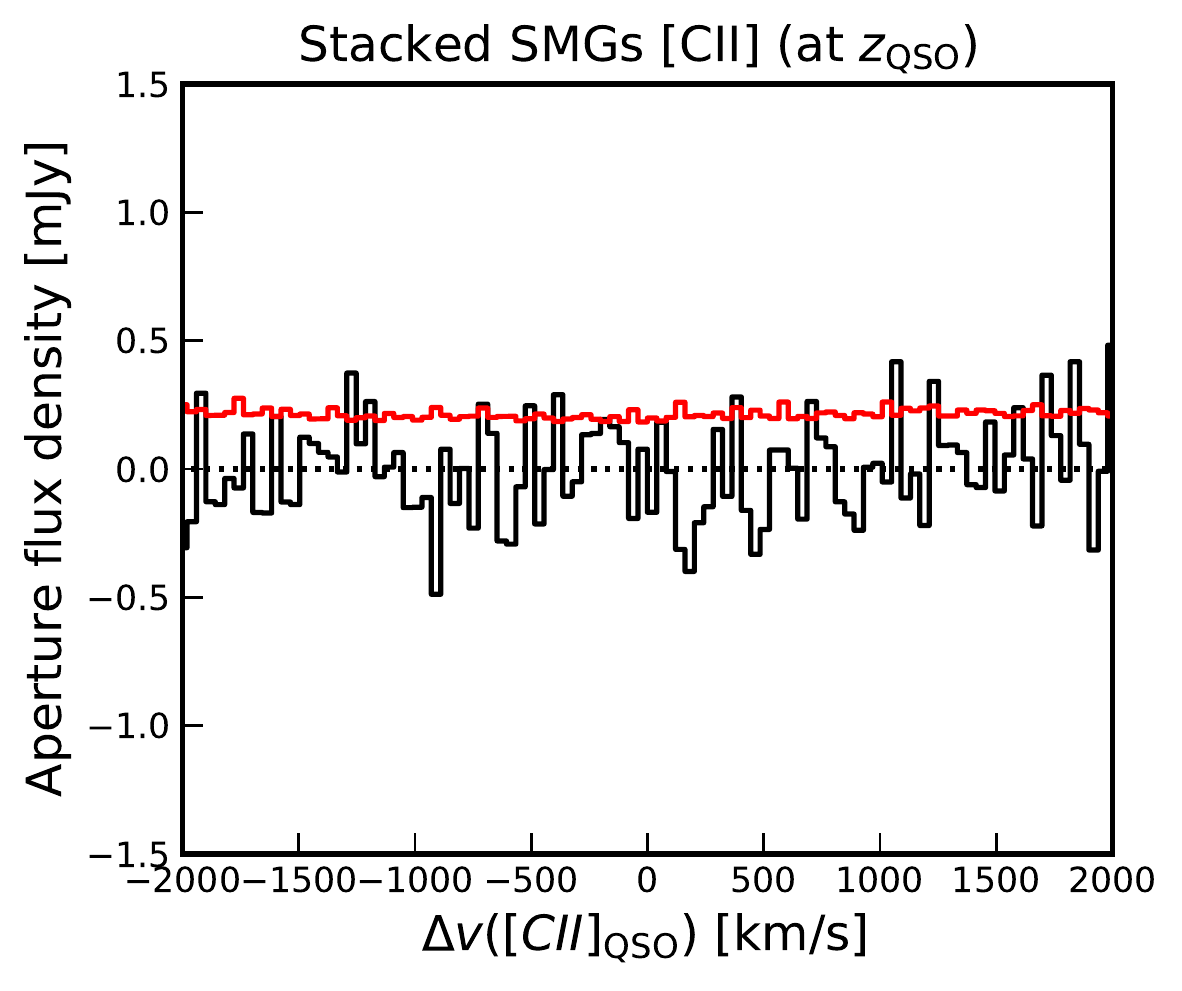}
    \caption{Stacked aperture--integrated ($r=2"$) spectrum (black, error in red) of all ALMA--continuum sources (with the exception of the $z=2.4$ sources in the foreground of PJ231), centered at the redshifted frequency of the quasars \cii\ emission line. The error array (red) is measured for each spectrum using the rms per beam rescaled to the number of beams in the $r=2"$ aperture and subsequently stacked. We find no evidence of \cii\ emission in the stacked spectrum. }
    \label{fig:stacked_CII_continuum_sources}
\end{figure}

For the continuum sources without lines, we have stacked the spectrum at the rest--frame frequency of the \cii\ emission of the background quasar. The result is shown in Fig. \ref{fig:stacked_CII_continuum_sources}. We do not find evidence for \cii\ emission at the redshift of the quasar in the stacked spectrum.

The absence of lines close to the frequency of \cii\ at the redshift of the quasar cannot plausibly be attributed to a “weak" \cii\ line emission. At the mean continuum flux density ($1.2\, \rm{mJy}$) of the ALMA detections, the FIR luminosity (modelled as described above using a modified black body assuming standard dust parameters and $z=6.5$) is $\sim 3.4 \times 10^{12} \ \rm{L}_\odot$. The \cii--to--FIR luminosity ratio in low--redshift ULIRGs and high--redshift quasar varies from $10^{-2}$ to $10^{-4}$ \citep[][]{Diaz-Santos2017,Novak2019,Pensabene2021,Li2020b}. We would thus expect \cii\ lines in our sources with a luminosity $\sim 3.4 \times 10^{8}-10^{10}\ \rm{L}_\odot$. Assuming a linewidth of $300\ \kms$, this is equivalent to a \cii\ line flux of $1-100\, \rm{mJy}$ over the full linewidth. With a sensitivity of $\sim0.5\ \rm{mJy}$ per $50\ \kms$ channel (see Table \ref{tab:quasar_obs_params}), the \cii\ lines should have been detected at least at the $4.4\sigma$ level in the sources with the strongest \cii\ deficit. In conclusion, the non-detection of \cii\ emission lines strongly suggests that the SMGs are not associated with the quasar and probably foreground sources.

\subsection{SCUBA2/ALMA SMGs photometric redshifts}
\label{sec:photoz}
In this section, we complement our analysis of the absence of \cii\ line detections in the SMGs targeted by studying their photometric redshifts. The $3$ quasar fields of interest have ancillary \textit{HST}/WFC3 F140W, \textit{Spitzer}/IRAC [$3.6\mu\rm{m}$] and [4.5$\mu\rm{m}$] imaging (see Table \ref{tab:ancillary_photometry}) which we can add to the SCUBA2 850, 450 $\mu\rm{m}$ imaging (Li et al., in prep.) and ALMA Band 6 observations presented in this work. Now that the positions of the SMGs is determined from the ALMA data, we can combine these datasets to constrain their Spectral Energy Distribution (SED) and obtain reliable photometric redshifts using \emph{MAGPHYS} \citep[][]{DaCunha2015}, a UV--to--FIR SED modelling code.

\begin{figure*}
    \centering
    \includegraphics[width=\textwidth]{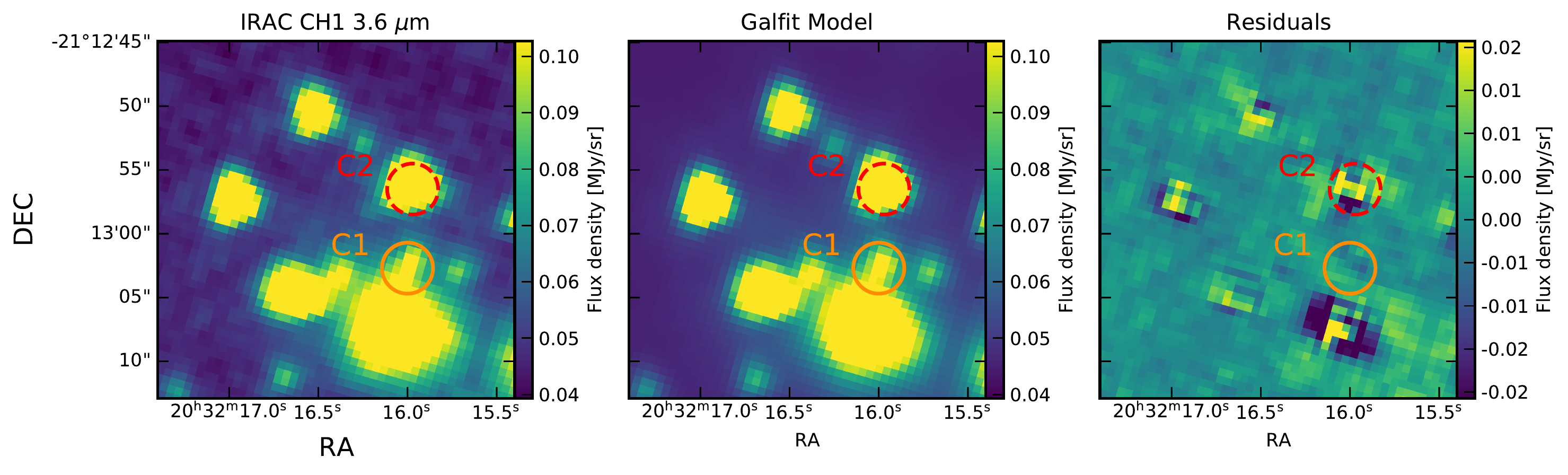}
    \includegraphics[width=\textwidth]{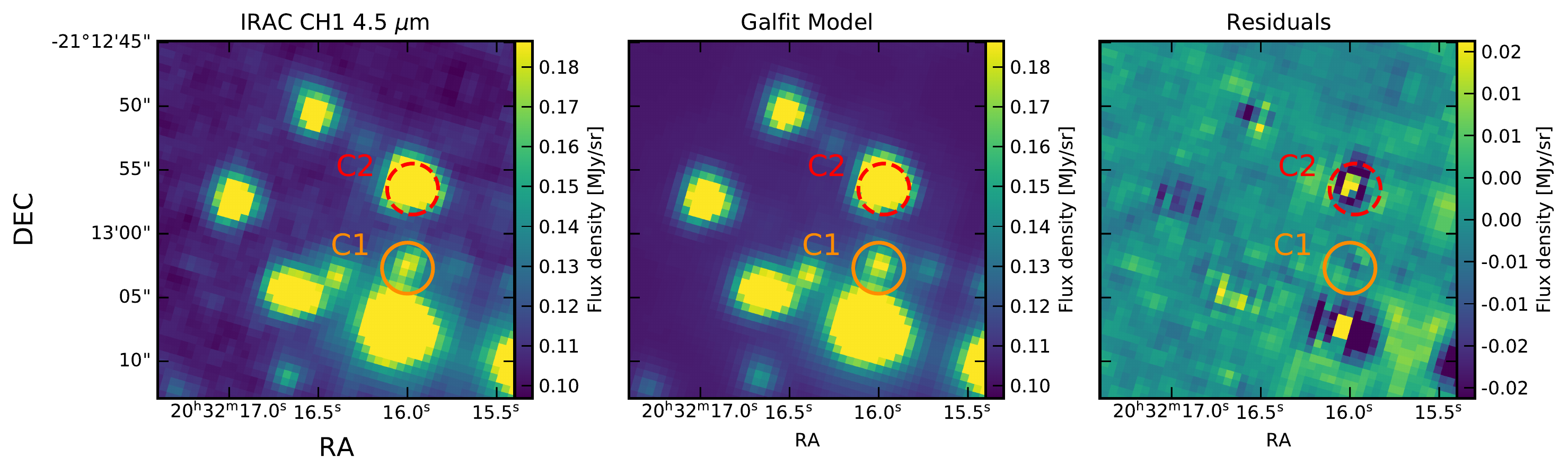}
    \caption{IRAC imaging, \emph{GALFIT} model and residuals for the sources PJ308-SMG1-C1/C2 (indicated with dashed red and solid orange circles, respectively). The model photometry for PJ308-SMG1-C2 (dashed red circle, not blended) is consistent with that derived by \emph{Sextractor}. Note the significant difference in flux density range between the image and the final residual map.}
    \label{fig:galfit_final}
\end{figure*}

\begin{table*}
    \centering
    \begin{tabular}{l|c|c|c|c}
         Quasar & Program ID/PI & Telescope/Instrument & Filters/$\Delta \lambda$ & Exp. Time\\ \hline\hline
         J0305--3150 & 11030 / R. Decarli & \textit{Spitzer}/IRAC & 3.6$\mu\rm{m}$, 4.5$\mu\rm{m}$ & 1000s \\
         & 094.B--0893(A) / B. Venemans & VLT/MUSE & 0.465--0.93$\rm{mu}$ & 2h30m \\
         PJ231--20 & 14876 / E. Bañados & \textit{HST} / WFC3 & F140W & 2612s \\
         & 099.A--0682(A) / E. Farina & VLT/MUSE & 0.465--0.93$\rm{mu}$ & 3h20m \\
        & 13066 / C. Mazzucchelli & \textit{Spitzer}/IRAC & 3.6$\mu\rm{m}$, 4.5$\mu\rm{m}$ & 7200s  \\
         PJ308--21 & 14876 / E. Bañados & \textit{HST} / WFC3 & F140W & 2612s \\
         & 13066 / C. Mazzucchelli & \textit{Spitzer}/IRAC & 3.6$\mu\rm{m}$, 4.5$\mu\rm{m}$ & 7200s  \\
        & 099.A--0682(A) / E. Farina & VLT/MUSE & 0.465--0.93$\rm{mu}$ & 5h \\
    \end{tabular}
    \caption{Ancillary infrared imaging and IFU spectroscopic data available for the three quasars fields.}
    \label{tab:my_label}
\end{table*}

\begin{table*}[]
    \centering
    \begin{tabular}{c|c|c|c|c|c}
    ID  &  F140W  & [$3.6\mu\rm{m}$] &  [$4.5\mu\rm{m}$] & $S_{850\mu\rm{m}}$ [mJy] &  $S_{450\mu\rm{m}}$ [mJy]\\
  PJ231-SMG1-C1  &  --$^{a}$             & $22.08\pm0.08$  & $21.58 \pm 0.13$   &  $<7.36$ &  $<46.8$ \\
  PJ231-SMG1-C2  &  --                   & $22.25\pm0.09$  & $21.18 \pm 0.09$ &  $<7.36$ &  $<46.8$ \\
  PJ231-SMG2-C1  & $25.080\pm0.075^{b}$  & $22.79\pm 0.12$ & $21.55 \pm 0.11$ & $<6.66$ &  $<46.8$ \\
  PJ231-SMG3-C1  &  --                   & $20.05\pm 0.02$ & $19.34 \pm 0.02$ &  $<6.96$ &  $<46.8$ \\
  PJ231-SMG5-C1  &  -- & -- & -- & $<4.8$ &  $<127.69$ \\
  J0305-SMG2-C1  &  --                   & $22.08\pm0.10$ & $20.66\pm0.10$ & $<7.8$ &  $<46.8$ \\
  J0305-SMG4-C1  &  --                   & $20.80\pm0.05$ & $19.97\pm0.06$ & $<7.38$ & $<46.8$ \\
  PJ308-SMG1-C1  &  --                   & $21.58\pm0.55$ & $20.77\pm0.44$  & $<11.29$  & $<46.8$ \\
  PJ308-SMG1-C2  &  --                   & $19.23\pm0.01$ & $18.52\pm0.01$ & $<11.29$ & $< 46.8$ \\
  PJ308-SMG2-C1  &  --                   & $20.42\pm0.03$ & $19.70\pm0.02$ & $<7.35$ & $<78.64$\\
  PJ308-SMG2-C2  &  --                   & $21.33\pm0.06$ & $21.00\pm0.06$ & $<7.35$ & $<78.64$ \\
  PJ308-SMG3-C1  &  --                   & $20.58\pm0.03$ & $19.85\pm0.02$ & $<6.22$ & $<46.8$ \\
  PJ308-SMG4-C1  & $24.170\pm0.050$      & $22.06\pm0.08$ & $21.22\pm0.06$ & $<4.41$  & $<46.8$ \\
  PJ308-SMG6-C1  & --                    & $21.98\pm0.07$ & $21.01\pm0.05$ & $<5.02$  &  $<82.40$ \\
  PJ308-SMG9-C1  & $<$$23.559\pm0.022^{c}$  & $20.56\pm0.02$ & $19.81\pm0.02$ & $<5.35$ & $<46.8$\\
    \end{tabular}
    \caption{Ancillary photometry for the continuum--detected ALMA sources. The \textit{HST} and IRAC photometry are given in AB magnitudes. We use the detections in SCUBA2 850--450$\mu\rm{m}$ maps as upper limits since the sources are unresolved (see discussion on the multiplicity in Section \ref{sec:disc_scuba_sources}). $^{a}$ Sources without \textit{HST} or \textit{Spitzer} photometry are not covered by the imaging available and thus no upper limits are available. $^{b}$ Resolves to multiple sources in the \textit{HST} images (see Appendix \ref{app:magphys_SEDs}) $^{c}$ The source is on the edge of the F140W imaging\, with $\gtrsim50\%$ of the flux lost,} and the F140W magnitude is therefore not considered for the SED fitting. 
    \label{tab:ancillary_photometry}
\end{table*}

The \textit{HST} images were reduced using the standard pipeline \citep[see][for more details]{Mazzucchelli2019, Decarli2019}. The photometry was extracted for all sources in the fields using \emph{Sextractor} \citep[][]{Bertin1996} using standard parameters \footnote{DETECT\_MINAREA=3, DETECT\_THRES=3, DEBLEND\_NTHRES=64, DEBLEND\_MINCONT=0.005, BACK\_SIZE=64, BACK\_FILTERSIZE=4, GAIN=6530, MAG\_ZEROPOINT=26.46, where the effective gain is computed from the exposure time and the instrument gain, and the F140W zeropoint AB magnitude (26.46) is taken from the WFC3 Handbook \url{http://www.stsci.edu/hst/wfc3/analysis/ir phot zpt}} and $MAG\_AUTO$ magnitudes. We adopt the pipeline--reduced \textit{Spitzer} images from \citet[][]{Mazzucchelli2019} with a refined astrometric solution based on the Gaia DR1 catalogue. Due to the spatial resolution of the data ($\sim 1.2-1.8"$) and the limited sampling ($0.6"\times0.6"$), one of the sources (PJ308-SMG1-C1) is blended with foreground objects and the photometry cannot be retrieved accurately using \emph{Sextractor}. Therefore we fit the blended SMG and the foreground objects simultaneously using the latest version of \emph{Galfit} \citep[][]{Peng2002,Peng2010}. We create a PSF image by rescaling, interpolating and upsampling images of $4-10$ (depending on the channel) stars in the field. Most sources are modelled with a single point source profile, except the most extended ones with a Sersic profile (which have the half--light radius and Sersic index profiles $n$ as additional parameters to the integrated magnitude and position). We show in Fig. \ref{fig:galfit_final} the \textit{Spitzer} imaging, the best--fit \emph{Galfit} model and the residuals. We check that the magnitude of the nearby unblended source (PJ308-SMG1-C2) measured with \emph{Galfit} ($m_{\rm{CH1}}=19.28\pm0.03$, $m_{\rm{CH2}}=18.49\pm0.03$) is consistent with that measured by \emph{Sextractor} ($m_{\rm{CH1}}=19.23\pm0.01$, $m_{\rm{CH2}}=18.49\pm0.17$). Table \ref{tab:ancillary_photometry} lists the ancillary photometry and fluxes for all our continuum sources.

We fit the photometry of the ALMA continuum sources using \emph{MAGPHYS-photoz} \citep[][]{DaCunha2015, Battisti2019} and show the resulting posterior redshift distributions in Fig. \ref{fig:magphys_photoz}. The best--fit SEDs are presented in Appendix \ref{app:magphys_SEDs}. None of the continuum--detected sources have a photometric redshift consistent with that of the background quasar ($z>6$). We conclude that the weak emission lines in PJ308-SMG1-C1 and PJ308-SMG9-C1 are either noise or low--redshift CO lines, and that none of the SCUBA2--detected SMGs are likely to be at the background quasar redshift. The detection of $z\simeq2.4$ CO7--6 and \ci\ lines in the SMGs in the field of PJ231--20 support this conclusion (see Appendix \ref{app:disc_z24_pj231}) although the peak of the photometric redshift posterior is at $\sim 2$ rather than $z\simeq2.4$, a difference that could be explained by the low number of datapoints available to constrain the SEDs. 
 
\begin{figure}
    \includegraphics[width=0.47\textwidth]{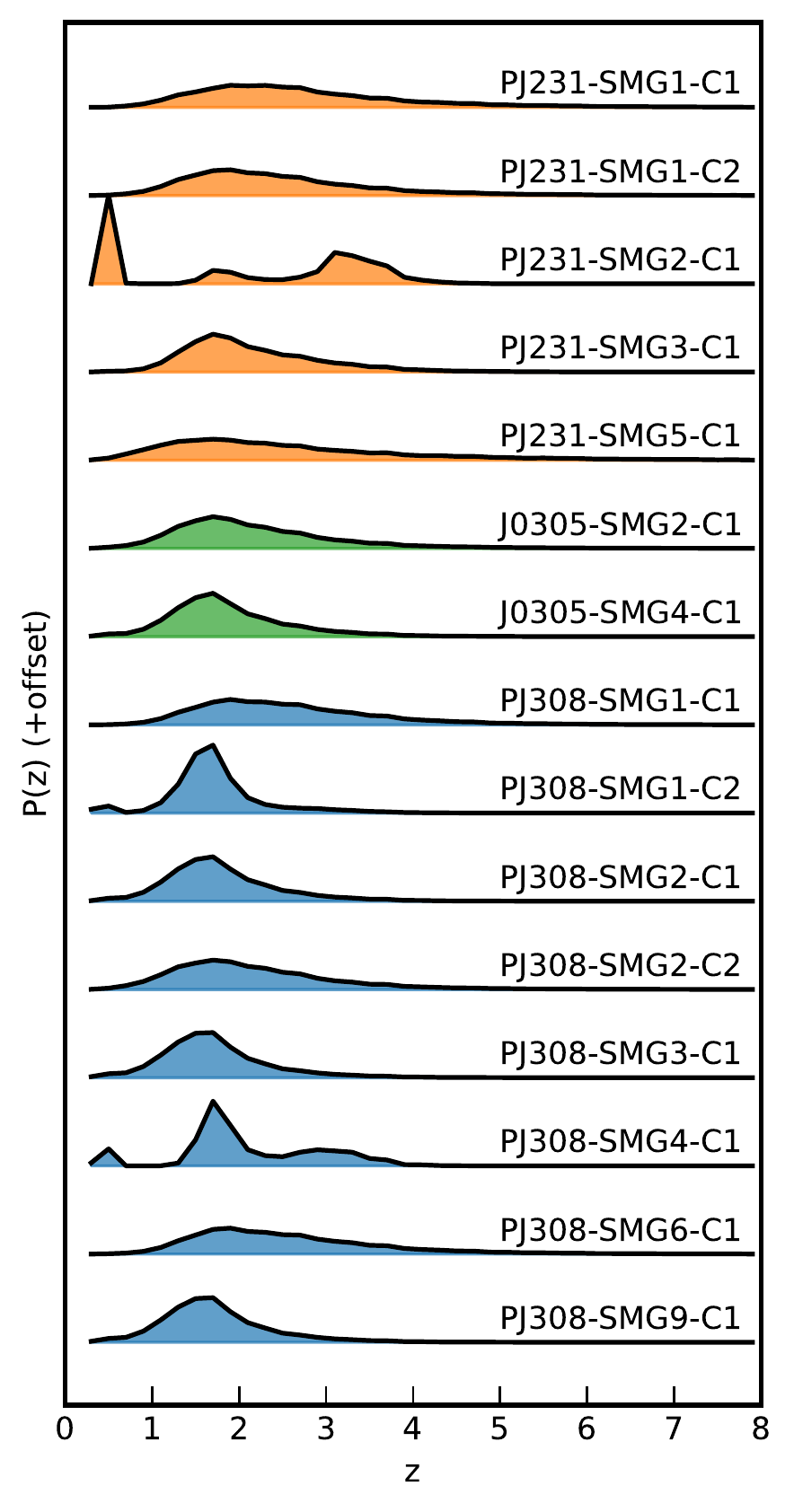}
    \caption{Posterior distributions for the photometric redshifts of the ALMA--detected SMGs obtained from \emph{MAGPHYS-photoz} \citep[][]{Battisti2019} using \textit{HST} F140W, IRAC, ALMA Band 6 and SCUBA2 imaging. None of the sources have a photo--z consistent with those of the background quasars ($z>6$). }
    \label{fig:magphys_photoz}
\end{figure}

\section{Searching for serendipitous \cii\ Line Emitters}
\label{sec:line_sources}
We have demonstrated in the previous section that the targeted SMGs are most likely foreground sources not associated with the background quasars. Nonetheless, \cii\ emitting sources without a detection in the ALMA continuum maps could also be associated with the background quasar. We thus set out to search for line emitters (e.g., without continuum emission) to constrain the large--scale structure around the three quasars studied here. In order to do so, we make use of two line--finding algorithms developed for interferometric data. 

The first algorithm is \emph{FindClumps} \citep[][]{Walter2016,Decarli2019}. \emph{FindClumps} convolves the imaged and cleaned datacube with boxcar filters of a given width in the spectra direction (e.g., it effectively slices and averages a number of continuous channels) to produce velocity--integrated line map which is then passed through \emph{Sextractor} \citep[][]{Bertin1996} to find significant sources. The operation is repeated for different boxcar widths (in this case $3$ to $19$ channels of $50 \kms$), after which the sources are grouped together (sky separation $<2"$) and the minimum offset in frequency between two emission features (here $0.1$ GHz). The procedure is repeated for the negative emission in the cleaned datacube. The fidelity criterion \citep[e.g.,][]{Walter2016}
\begin{equation}
    \rm{fidelity}(SNR,\sigma_{\rm{kernel}}) = 1 - \frac{N_{\rm{neg}}(SNR,\sigma_{\rm{kernel}})}{N_{\rm{pos}}(SNR,\sigma_{\rm{kernel}})}
\end{equation}
is used to determine the fraction of credible (e.g., not due to noise) detections as a function of SNR. Note that by definition the fidelity score is an empirical estimate of so--called “true positive" fraction in the final sample. In this work, we use the latest version of \emph{FindClumps} implemented in \emph{interferopy} \citep*[][]{interferopy}. 

\emph{LineSeeker} \citep[][]{Gonzalez-Lopez2017} takes a slightly different approach than \emph{FindClumps}, which we summarise here briefly. It starts by creating velocity--integrated maps using a Gaussian kernel rather than a boxcar. Sources are then searched using the DBSCAN algorithm on $\rm{SNR}>5$ pixels \citep[][]{Ester1996}. Finally, \emph{LineSeeker} performs a contamination analysis by injecting and retrieving mock sources in the created maps/cubes. The threshold to select significant emitters is then adjusted for each datacube by choosing the maximum acceptable false--positive rate.

Both methods were compared on the ASPECS--HUDF 3mm data in \citet{Gonzalez-Lopez2019}, who find that both methods agree relatively well, although \emph{FindClumps} tends to slightly overestimate (by $\sim7\%$) the SNR of faint sources (SNR$\sim 4$) compared to \emph{LineSeeker}, which could be explained by the different convolution kernels.

We run \emph{LineSeeker} and \emph{Findclumps} on the continuum--subtracted cleaned datacubes of the targeted 17 SMGs to generate a first list of candidate line--emitters. We select a fidelity/true--positive threshold of $90\%$ for \emph{Findclumps}, and, accordingly, a false--positive probability threshold of $10\%$ for \emph{LineSeeker} (both criteria are roughly equivalent to a nominal SNR$\sim5$ cut). The candidates close ($<2"$) to the edge of the map or near any continuum sources ($<4"$ from the center of any continuum source) are both discarded to remove artifacts due to the edge of the map or poor continuum subtraction, respectively.  This results in $10$ \emph{LineSeeker} detections and in $11$ \emph{Findclumps}, for a total of unique $11$ candidates emitters in all the fields (e.g. \emph{FindClumps} only finds one additional candidate at the SNR threshold chosen \footnote{This line emitter (J0305-SMG2-spwA-4636), is nominally detected by LineSeeker at SNR$=5.3$ and the false positive probability based on simulated cubes is $5\%$. However, the \emph{LineSeeker} Poisson statistic false probability is $31\%$, which removed this emitter from the final \emph{LineSeeker} selection.}). As for continuum sources (see Sec. \ref{sec:cont_sources}), a spectrum is then extracted with a $r=2"$ aperture at the position of each candidate and we fit the emission line with a Gaussian profile to create a velocity--integrated line map (see Appendix \ref{app:cii_line_search_candidates}). The secure line emitters, tentatively identified as \cii\ emitters at $z>6$, and their properties are listed in Table \ref{tab:line_emitters}. 

The line emitters found with \emph{LineSeeker} and \emph{FindClumps} are not necessary \cii\ at the quasar redshift. Indeed, a large fraction of them are found in the second spectral tuning of ALMA, e.g., $\sim 16\ \rm{GHz}$ from the high--redshift quasar \cii\ emission. Even candidate \cii\ lines in the correct spectral tuning could be low--redshift CO interlopers or simply spurious noise. Indeed, the fact that all detections have a SNR only slightly above the threshold chosen to select line emitters suggests a large fraction are not real sources. We do not perform a photometric redshift analysis similar to that done for the continuum sources in Section \ref{sec:photoz} as all of our line emitters (except one, PJ231-SMG2-spwAmm.03, $m_{F140W}=25.89\pm0.08$) are undetected in the SCUBA2, \textit{Spitzer}/[3.6]--[4.5] or \textit{HST}/F140W imaging.

\begin{figure}
    \centering
    \includegraphics[width=0.45\textwidth]{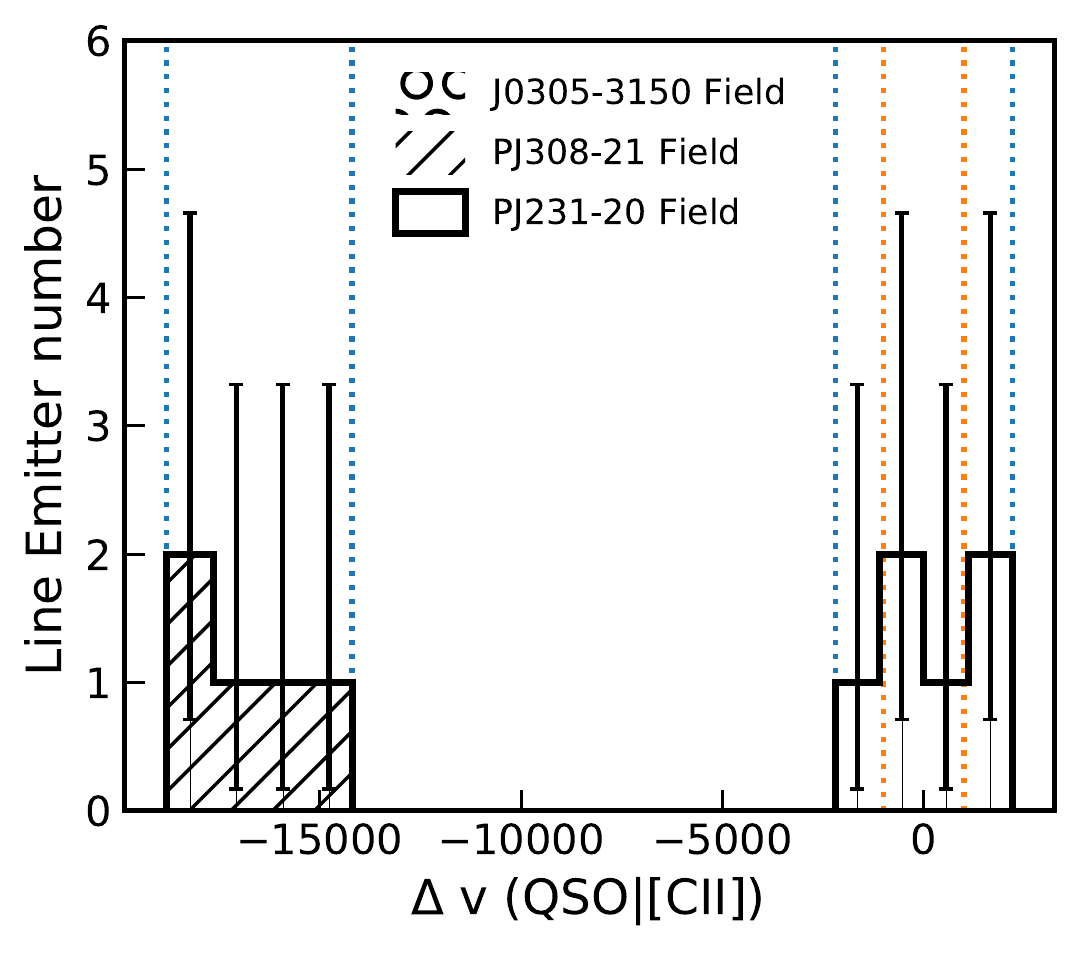}
    \caption{Velocity offset distribution of line (continuum--free) emitters detected in the SMG fields, assuming the line is \cii. The vertical blue bars show the velocity range of the two spectral windows of the ALMA tunings, and the vertical orange bars the interval $(-1000,1000)\ \kms$ around the quasar redshift where an overdensity would be expected. The errorbars are Poisson uncertainties on the number of detected sources \citep[][]{Gehrels1986}. The absence of clustering around the quasar in velocity spacecould suggest that most emitters are foreground CO sources.}
    \label{fig:emitters_dv_distrib}
\end{figure}

Nonetheless, if some or all of these emitters are indeed \cii\ close to the quasar redshift, they should cluster in velocity space around the quasar redshift. We present the distribution in velocity space of the line emitters, assuming the line \cii, in Fig. \ref{fig:emitters_dv_distrib}. We find no evidence for an increased number of line emitters in the vicinity of the quasar redshifts, suggesting that most or all of these detections are not associated with the quasars. This is in agreement with the findings of \citet{Decarli2020} who find that in a blank field observed with ALMA Band 6, \cii\ emission at $z>6$ accounts for $<1\%$ of the flux seen in the lines which are overwhelmingly lower redshift CO emitters.

\begin{table*}[]
    \centering
    \begin{tabular}{c|cccccc}
Name & RA & DEC & Freq [GHz] & SNR & $r_\perp$ [cMpc] & $\Delta v$ [$\kms$] \\ \hline 
PJ231-SMG1-spwAmm.01 & 15:26:40.30 & -20:51:03.70 & 250.924 & 5.5 & 2.95 & 490\\
PJ231-SMG2-spwAmm.01 & 15:26:39.33 & -20:50:38.90 & 249.909 & 5.2 & 1.79 & -730\\
PJ231-SMG2-spwAmm.03$^{a}$ & 15:26:37.89 & -20:50:54.90 & 248.698 & 4.9 & 2.23 & -2180\\
J0305-SMG2-spwAmm.01 & 03:05:23.38 & -31:49:44.40 & 250.565 & 5.7 & 4.49 & 1160\\
J0305-SMG2-spwAmm.02 & 03:05:21.78 & -31:49:50.40 & 249.53 & 5.7 & 3.71 & -80\\
J0305-SMG2-spwA-4636$^{b}$ & 03:05:22.30 & -31:50:03.52 & 250.838 & 5.7 & 3.55 & 1490\\
PJ308-SMG4-spwBmm.01 & 20:32:07.33 & -21:14:03.30 & 249.164 & 5.4 & 1.60 & -15470\\
PJ308-SMG6-spwBmm.01 & 20:32:09.62 & -21:16:00.90 & 246.781 & 5.6 & 4.80 & -18200\\
PJ308-SMG6-spwBmm.02 & 20:32:08.69 & -21:16:12.70 & 249.418 & 5.4 & 5.34 & -15180\\
PJ308-SMG6-spwBmm.03 & 20:32:08.41 & -21:16:14.50 & 247.504 & 5.3 & 5.43 & -17370\\
PJ308-SMG9-spwBmm.02 & 20:32:13.08 & -21:14:31.70 & 246.937 & 5.4 & 2.03 & -18020\\

    \end{tabular}
    \caption{Line emitters recovered with \emph{LineSeeker} and \emph{Findclumps} in the ALMA pointings studied in this work. The last two columns give the proper transverse distance (assuming the line is \cii\ close to the redshift of the quasar) and velocity offset from the quasar \cii\ redshift. $^a$ Detected in the \textit{HST} imaging with $m_{F140W}=25.89\pm0.08$$^b$ Only selected with \emph{FindClumps}.}
    \label{tab:line_emitters}
\end{table*}

\section{MUSE archival data and LAEs}
\label{sec:LAEs}
The three quasars studied with ALMA and SCUBA2 in this paper all have archival MUSE observations (see Table \ref{tab:ancillary_photometry}) published and discussed in various papers \citep[e.g.][]{Farina2017, Venemans2019,Drake2019,Farina2019}. As discussed in the introduction, the presence of a LAE in the field of J0305--3150 was reported by \citet[][]{Farina2017}. However, systematic searches for MUSE LAEs at the redshift of PJ231--20 and PJ308--21 have not been reported. \citet{Meyer2020} searched several quasars fields, including that of PJ231--20 and PJ308--21, to find LAEs in the redshift range probed by the Lyman--$\alpha$ forest of the quasars, and only published those detections. The search for LAEs also covered the redshift range around the quasars, and we now report the result of this search.

The MUSE archival observations of the three quasar fields were reduced and searched for LAEs in \citet{Meyer2020}, to which we refer for further details. Briefly, the datacubes were reduced using the standard ESO pipelines recipes and sky emission contamination was removed using the Zurich Atmospheric Purge code \citep[ZAP][]{Soto2016}. The reduced datacubes were then searched for Lyman--$\alpha$ emitters using two different software packages: \emph{MUSELET} \citep[][]{Bacon2016} and \emph{LSDCat} \citep[][]{Herenz2017}. On the one hand, \emph{MUSELET} creates NB slices from the IFU data, then uses \textit{Sextractor} to identify significant sources in the sub--images and finally groups detections at close separation in wavelength--adjacent images. On the other hand, \emph{LSDCat} runs a 3D Gaussian--matched filter on the median--filtered cube. \citet{Meyer2020} concluded that the use of the two algorithms is beneficial since they are complementary and do not perform similarly for faint emitters or emitters close to bright continuum sources. 

We searched for LAEs in the MUSE data in the field of PJ231--20, PJ308--21 and J0305--3150 using \emph{LSDCat} and \emph{MUSELET} with parameters described in \citet[][]{Meyer2020}. Briefly, we use the standard $6.25\rm{\AA}$--width for the NBs created by \emph{MUSELET} with $DETECT\_MINAREA=4,DETECT\_THRESH=2.0, ANALYSIS\_THRESH 2.0$ \emph{Sextractor} parameters for the NB search. We use the standard continuum width (4 times wider than the NBs) for continuum subtraction. For \emph{LSDCat} we use a gaussian convolution kernel with the default polynomial coefficients for the PSF FWHM dependence on wavelength, make use of the optional median filtering, and impose a SNR threshold $>8$ to select pixels with significant flux. All candidates were subsequently visually inspected. Typical contaminants are artifacts due to poor continuum subtraction, nearby extended sources with strong emission lines, low-redshift [OII] doublets and cosmic rays (see further Appendix B of \citet[][]{Meyer2020}). Low-redshift [OII]$ 3727,3729 \AA$ doublets (with a peak separation of $\sim 220\ \kms$) are easily resolved in the MUSE data with a resolution of $75\ \kms$ at 9300 $\AA$ where $z=6.6$ Lyman-$\alpha$ is searched for. Whilst double-peaked emission could potentially be a high-redshift double-peaked Lyman-$\alpha$ profile \citep[e.g.][]{Hu2016,Songaila2018, Matthee2018, Bosman2020, Meyer2021}, such profiles are expected to be exceedingly rare at $z>5.5$ \citep[e.g.][]{Gronke2020,Garel2021}. Additionally, the [OII] doublet presents a relatively constant ratio \citep[e.g.][]{Paulino-Afonso2018}, making it easy to discard such contaminants. Other low-redshift interlopers (e.g H$\beta$+[OIII], H$\alpha$) are identified and removed due to the presence of other multiple lines in the MUSE wavelength range. We have verified that LAEs are not detected in the continuum image produced from the MUSE cube.

\begin{figure*}
    \centering
    \includegraphics[width=0.42\textwidth]{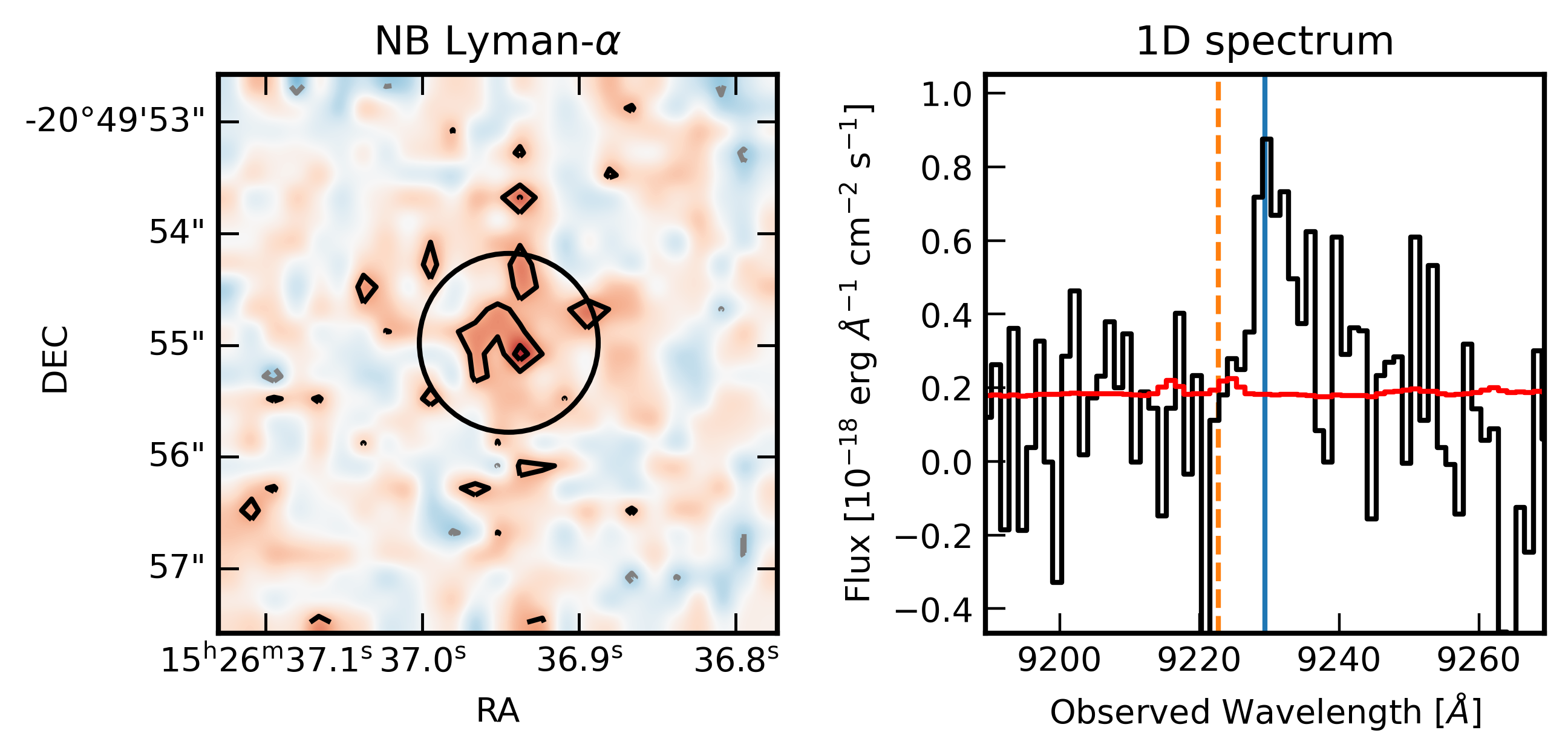}
    \includegraphics[width=0.57\textwidth]{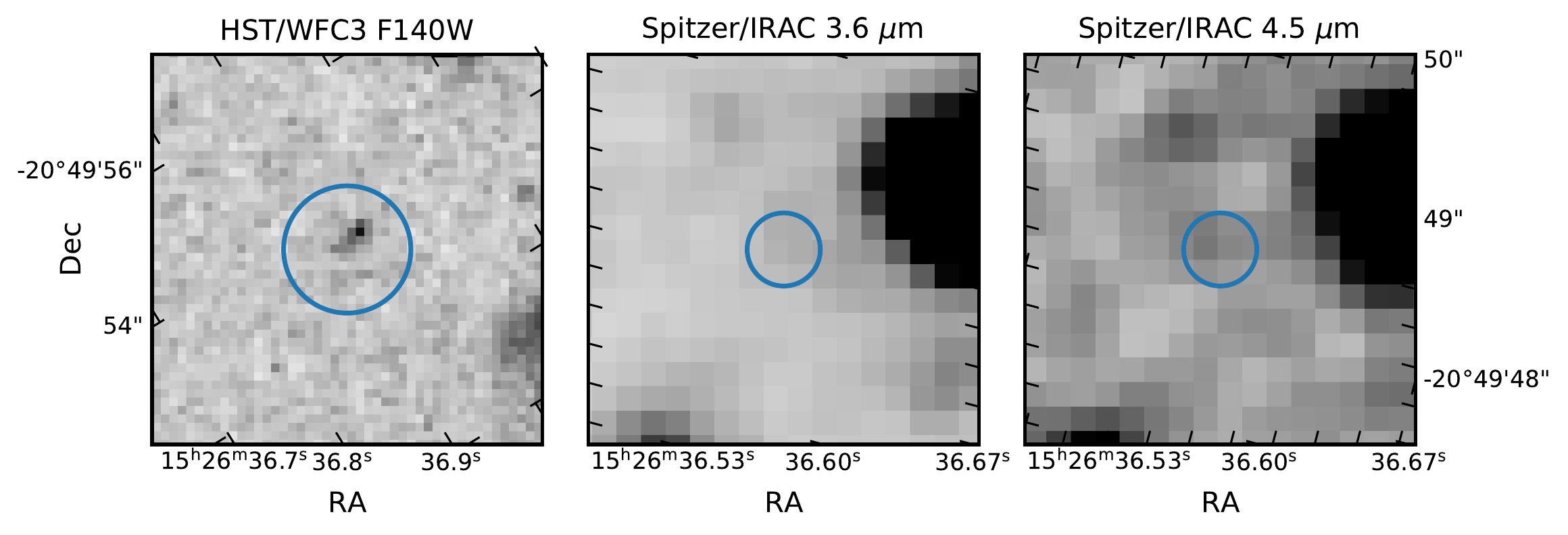} \\
    \includegraphics[width=0.42\textwidth]{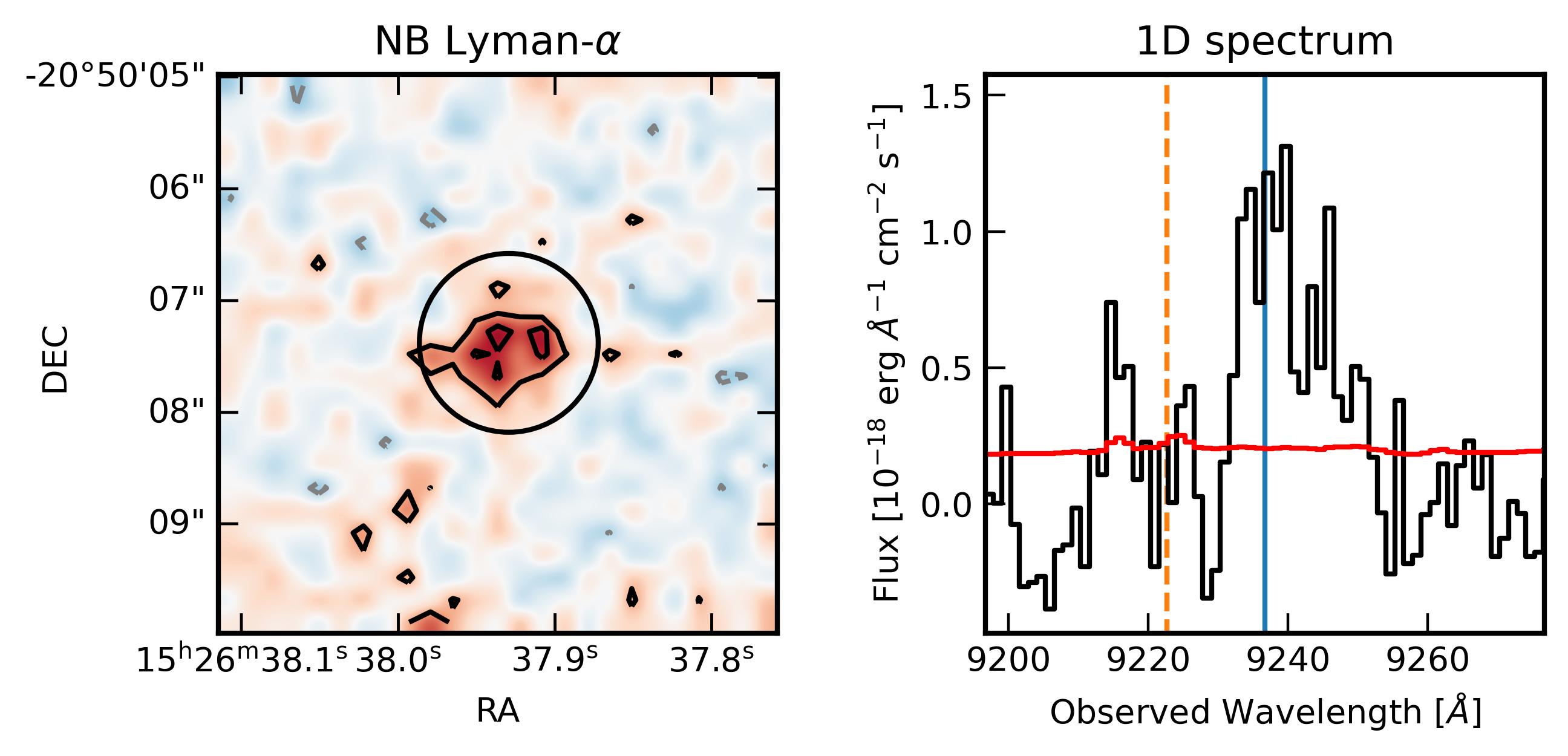}
    \includegraphics[width=0.57\textwidth]{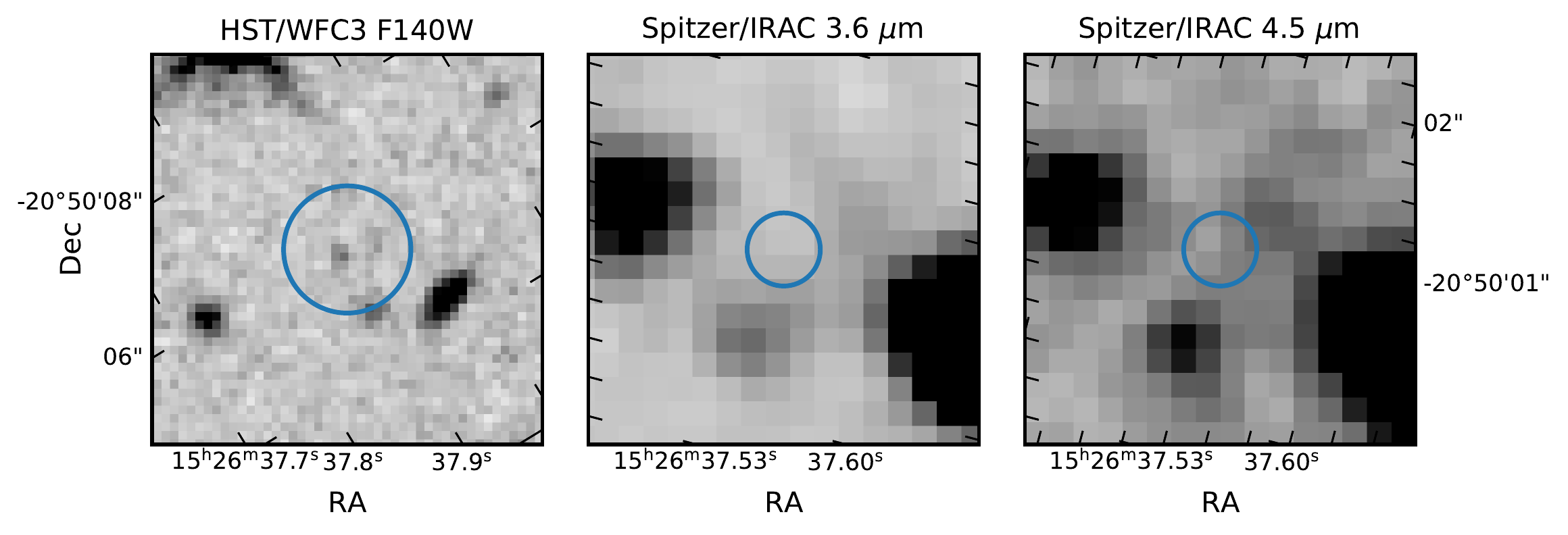} \\
    \caption{LAEs found at the redshift of PJ231--20 in the archival MUSE data. The leftmost panel shows the pseudo--narrowband image centered on the peak of the emission. 
    The contours (negative in dashed grey, positive in black) mark the (-4,-2,2,4) surface brightness rms levels. The second leftmost panel shows the extracted Lyman--$\alpha$ spectrum (black) and error array (red), extracted in a $r=0.8"$ aperture shown in the leftmost panel. The blue vertical line indicates the redshift of the Lyman-$\alpha$ emitter determined by \emph{LSDCat} from the Gaussian-filtered cube. The wavelength of the redshifted Lyman--$\alpha$ line of the quasar is indicated by a vertical orange dashed line. The last panels show the F140W and IRAC imaging, with the position of the LAE marked with a $r=1"$ aperture circle (blue). }
    \label{fig:LAEs}
\end{figure*}

From this search, we selected galaxies with velocity separation $\pm 1000\ \kms$ from the central quasar  \cii\ redshift \footnote{Note that the Lyman--$\alpha$ redshift of the emitters is corrected towards the systemic redshift using the apparent FWHM--method of \citet{Verhamme2018}}. Only two such LAEs were found, both in the field of PJ231--20, with no candidates at the redshift of PJ308--21 and J0305--3150. PJ231--20 is thus the only quasar with two relatively bright LAEs (within $\pm 1000 \kms$) at $z\gtrsim5.5$. The LAEs around PJ231--20 were found at a distance of $r_\perp=0.562\, \rm{cMpc}$ and $r_\perp=0.287\, \rm{cMpc}$ from PJ231--20 and have a Lyman--$\alpha$ luminosity $L_{\rm{Ly}\alpha} > 3\times 10^{42} \rm{erg\ s}^{-1}$ (see Table \ref{tab:LAE_PJ231}). Both LAEs were detected with \emph{LSDCat} but not recovered by \emph{MUSELET} using standard parameters \footnote{\emph{MUSELET} runs \emph{Sextractor} on $6.25\AA$-wide NB images which would only contain a fraction of the flux of a high--redshift Lyman-$\alpha$ line, therefore detecting only the strongest lines. We have verified that reducing the detection significance and mininum area thresholds in the \emph{Sextractor} parameters lead to the recovery of the two LAEs with \emph{MUSELET}.}. One of the LAEs is detected in the F140W image, and both are undetected in the IRAC data (see Fig. \ref{fig:LAEs}). At these Lyman--$\alpha$ luminosities, the number density of LAEs in a blank MUSE field is $\sim 1-5 \times 10^{-4}\ \rm{cMpc}^{-3}$ \citep[][]{Drake2017,DeLaVieuville2019a, Konno2018}, which for the volume probed ($50"\times50"\times 2000\ \kms \sim 80\ \rm{cMpc}^3$ at $z=6.6$) implies $0.009-0.04$ LAEs per field, making the environment of PJ231--20 extremely overdense.

The search did however not recover the LAE reported by \citet[][]{Farina2017} near the redshift of the quasar J0305-3150. The reason for this is that the LAE is extremely close to the quasar (2.3", e.g. $\sim9$ pixels offset). As the quasar and the LAE emits at the same wavelength, they are connected in the $>8\sigma$ thresholded SNR cube, and considered as one unique source by \emph{LSDCat}. Similarly, \emph{MUSELET} did not recover the emitter as it is very faint and not extended. \citet[][]{Farina2017} recovered the LAE by using wider pseudo--NB images ($10\rm{\AA}$ vs $6.25\rm{\AA}$ in MUSELET) and by using a lower threshold for the NB search ($1.5\sigma$ vs $2.0\sigma$). In either case, the parameters of \emph{LSDCat} and \emph{MUSELET} can be adapted to recover this specific emitter at the cost of losing others or increasing the number of spurious sources over the full datacube. While this would suggest that combining the different LAE search methods is the way forward, this would results in a complex selection function which might be a hindrance to future statistical analyses. We therefore do not attempt to perform a  search similar to that of \citet[][]{Farina2017} in the field of PJ308--21 and PJ231--20.

\begin{table*}[]
    \centering
    \begin{tabular}{c|c|c|c|c|c|c|c|c}
         RA & DEC & $z_{\rm{Ly}\alpha}$ & FWHM [$\kms$]  & $z_{corr}^{a}$ & $L_{\rm{Ly}\alpha}^{b}$ [$10^{42}\ \rm{erg\ s}^{-1}$] & $m_{\rm{F140W}}$ & EW$_{\rm{rest}}(\rm{Ly}\alpha)$ [$\rm{\AA}$] & $r_\perp$ [cMpc]  \\
         15:26:36.91 & -20:49:53.47 & 6.592 & 121.9 & 6.590 & $3.3\pm0.5$ & $25.9\pm0.1$ & $23.5\pm 3.3$ & 0.562 \\
         15:26:37.90 & -20:50:05.87 & 6.598 & 162.4 & 6.596 & $5.4\pm0.5$ & $<27.1$ & $>118$ & 0.287 \\
    \end{tabular}
    \caption{Properties of new Lyman--$\alpha$ emitters discovered in the field of PJ231--20. Limits are given at the $3\sigma$ level. The EW width is computed using the continuum UV flux at $1500\rm{\AA}$ (rest-frame) derived from the $m_{\rm{F140W}}$ magnitude assuming a flat $f_\nu$ spectrum. a) The corrected redshift is derived using the observed FWHM following the empirical correction calibrated on low--redshift LAE by \citet[][]{Verhamme2018}. b) The luminosities are derived using the flux extracted in a $(+300,-300)\kms$ window centered on the peak of the emission.}
    \label{tab:LAE_PJ231}
\end{table*}

\section{The varied environments of high--redshift quasars}
We now characterize the cross-correlation of $z>6$ quasars and galaxies with our updated constraints. A common approximation of the galaxy--quasar cross-correlation is a simple power--law relation \citep[see e.g.,][]{Hennawi2006, Kayo2012,Farina2017, Eftekharzadeh2017,Eftekharzadeh2019, Garcia-Vergara2017,Garcia-Vergara2019a, Farina2019}
\begin{equation}
    \xi^{QG}(r) = \left(\frac{r}{R_{0}^{QG}}\right) ^{-\gamma}
\end{equation}
where $r$ is the 3D comoving distance between the quasar and the galaxy, $\gamma=1.8-2.0$ is the slope of the clustering strength and $R_{0}^{QG} = \sqrt{R_{0}^{GG}R_{0}^{QQ}}$ is the quasar-galaxy cross-correlation length, which can be inferred from the quasar and galaxy auto-correlation length. For the purpose of this discussion, we use the LAE auto-correlation has been measured by \citet[$r_0^{GG}=10.3^{+4.7}_{-8.6}$ cMpc][]{Ouchi2010} and assume the \citet[][]{Shen2007} quasar correlation length ($r_0^{QQ}=17.4^{+2.5}_{-2.8}$ cMpc) derived from $z>2.9$ SDSS quasars. 

\begin{figure*}
    \centering
    \includegraphics[width=0.8\textwidth]{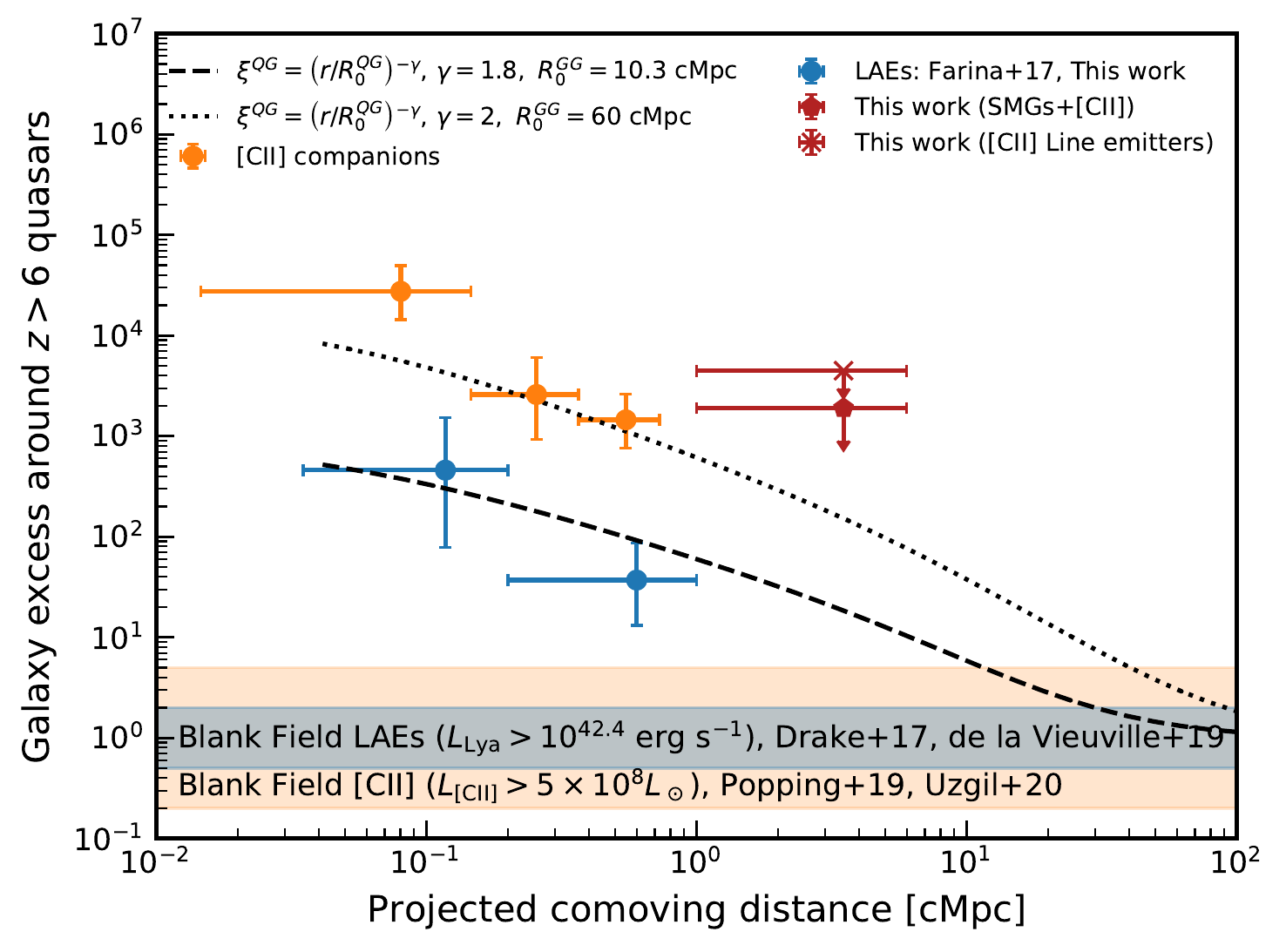}
    \caption{Excess number density of galaxies around $z>6$ quasars searched for \cii\ or LAE companions. The observed number densities of \cii\ emitters and LAEs is divided by the blank field number densities of \citet{Popping2019, Uzgil2021} and \citet[][]{Drake2017,DeLaVieuville2019a}, respectively, integrated down to the limits of $L_{\rm{[CII]}} = 5\times 10^{8} L_\odot$ and $L_{\rm{Lya}} = 10^{42.4} \rm{erg s}^{-1}$, respectively. The upper limits from this work are derived using the $2\sigma$ single--sided upper limit Poisson errors \citep[][]{Gehrels1986} on the number of continuum+\cii\ (0, e.g. $<3.783 (2\sigma)$) and \cii--only emitters (3, e.g. $<8.9(2\sigma)$) divided by the volume surveyed and the blank field \cii\ number density. The dashed and dotted lines give the expectation for an approximated quasar-galaxy cross-correlation (see text for details). }
    \label{fig:overdensity_constraints}
\end{figure*}

We summarize the updated constraints on the overdensity of galaxies in the fields of high--redshift quasars in Fig. \ref{fig:overdensity_constraints}. On the one hand, our constraints at large--scale ($\sim 1.8-5.8$ cMpc) on $17$ pointings show that the overdensity of \cii\ emitters declines with distance, in agreement with the simple model described above. In fact, the detection of a single SMG in \cii\ at the redshift of the quasar would have implied a number density of \cii\ emitters close to that found on smaller scales, and thus an extremely high correlation strength or length. On the other hand, our LAE search results support earlier findings that LAEs are overdense in quasar fields \citep[][]{Farina2017, Mignoli2020}. However, their overdensity is an order of magnitude below that of \cii\ companions found in single ALMA pointings centered on the quasars. The fiducial model of the quasar-LAE cross-correlation  ($\gamma=1.8, R_0^{GG}=10.3\ \rm{cMpc}$) matches very well the observed number densities, whereas the quasar-\cii\ clustering is only well reproduced with a much larger clustering length ($R_0^{GG}=60\ \rm{cMpc}$) and $\gamma=2$. 

This difference in clustering strength (one order of magnitude at $\sim 1\ \rm{cMpc}$) could either signal a strong bias towards dusty sources around quasars, or a large host halo mass for \cii\ emitters. Unfortunately, the $z\sim 6$ \cii\ auto-correlation and host halo mass is not well constrained and we cannot definitely conclude. However, \citet[][]{Garcia-Vergara2017,Garcia-Vergara2019a,Garcia-Vergara2021} find similar results at $z\sim 4$ where CO emitters and LBGs are more strongly clustered around quasars than LAEs. They suggest that a bias towards dusty galaxies in the environment of massive quasars could suppress the number of galaxies detected with the Lyman-$\alpha$ line which strongly absorbed by dust. However, in their measurement the number of LAEs around quasars is inferior to that inferred from the cross-correlation model detailed above, whereas in our case the fiducial model matches the data well (although the uncertainties and cosmic variance are still large). One possibility is that the quasar radiation carves out an ionised region boosting the Lyman-$\alpha$ transmission in nearby galaxies \citep[e.g.][]{Bosman2020a}, offsetting some of the bias towards more dusty galaxies which should decrease the number of LAE detections.

In the context of this hypothesis, it is interesting to note that the closest LAE to the PJ231--20 was not detected in previous ALMA \cii\ observations of the quasar \citep{Decarli2017,Venemans2020}, despite being within the ALMA field of view. This sets a lower limit for the \cii\ flux density of this source at $S_\nu \Delta v \lesssim 0.1\ \rm{Jy km s}^{-1}$ at the 5$\sigma$ level (assuming a FWHM of $300\ \kms$), which can be translated to a line luminosity $L_{\rm{[CII]}} \lesssim 10^{8} L_\odot$ and $SFR_{IR}\lesssim 10-30\ M_\odot\ \rm{yr}^{-1}$ using standard scaling relations \citep[e.g.][]{Herrera-Camus2018a}. Conversely, the  \cii\ companions reported in \citet[][]{Decarli2017, Neeleman2019a, Venemans2020} around the three quasars are not seen in Lyman--$\alpha$ in the MUSE data, and therefore have $L_{\rm{Ly}\alpha} \lesssim 2 \times 10^{42}\ \rm{erg\ s}^{-1}$. Following e.g., \citet[][]{Sobral2018}, assuming case B recombination, $10\%$ escape fraction of Lyman continuum photons \citep[][]{Meyer2020}  and a $20\%$ escape fraction of Lyman-$\alpha$ photons, we can put an approximate upper limit on the SFR$_{Ly\alpha}\ \lesssim 10\ M_\odot \ \rm{yr}^{-1}$. 
On the one hand, the [CII]-SFR relationship used above assumes that most of the UV emission is absorbed by dust and re-emitted in the infrared, and thus becomes inefficient for dust- and metal-poor galaxies. On the other hand, the Ly$\alpha$-SFR relation assumes an average escape fraction of Lyman-$\alpha$ photons and would thus underestimate the SFR if the LAEs were dust-rich (thus absorbing more  Lyman-$\alpha$ photons than expected). The fact that \cii--detected sources are not detected in Lyman-$\alpha$, at the same nominal SFR limits, suggests that we are looking at a dichotomy of metal-poor and metal-rich, dusty galaxies detected at different wavelengths. This dichotomy might be only apparent as we sample the extremes of the obscured-to-unobscured SFR distribution, and deeper observations might change this picture. What is interesting however is that high--redshift quasars cluster more strongly with \cii--emitters (dust-rich) than LAEs, in agreement with results at $z\sim 4$ \citep[][]{Garcia-Vergara2021}. This suggests that more evolved and dusty galaxies are found around quasars, which would be expected if they trace the first large-scale structure in the Universe. Finally, the clustering strength discrepancy between the LAEs and \cii--emitters might be even larger, as the number of LAEs could preferentially be increased by the presence of the quasar ionisation zone facilitating the escape of Lyman-$\alpha$ photons \citep[e.g.][]{Bosman2020a}. 
 
\section{Conclusion} 
\label{sec:conclusion}

We searched the fields of three high--redshift quasars (J0305--3150, PJ2310--20, PJ308--21) for galaxy overdensities using three approaches: 1) confirming the redshift of 17 bright SCUBA2--selected SMGs with ALMA, 2) a blind search for \cii\ emitters in these 17 ALMA pointings, and 3) a search for LAEs using archival MUSE observations. We report the following findings:

\begin{itemize}
    \item With ALMA (Band 6) we detect the continuum of 12 out of $17$ SCUBA2 SMGs targeted, and find that $3$ have two detected counterparts in the ALMA continuum maps. The confirmation rate with ALMA and the multiplicity fraction are in good agreement with that of earlier SMG follow--up studies with ALMA \citep[e.g.,][]{Hodge2013}.
    \item We find no \cii\ lines in the SMGs at a similar redshift of the $z>6$ quasars in the field. The absence of \cii\, is unlikely if these sources are indeed at the quasar redshift, considering their continuum flux densities and the usual \cii/FIR luminosity ratio in low-- and high--redshift galaxies. Moreover, the photometric redshifts derived using \textit{HST}, \textit{Spitzer}, SCUBA2 and ALMA Band 6 imaging disfavour any high--redshift ($z>3$) solutions. We thus conclude that the all SMGs are foreground sources.
    \item We report the detection of emission line in four SMGs in the field of PJ231--20 consistent with CO7--6 and [CI]$_{809\mu\rm{m}}$ at $z\simeq2.4$. This overdensity of sources is located at the redshift of a Mg{~\small II} absorber in the quasar spectrum. This supports the previous finding that none of the SMGs are high--redshift ($z>6$) sources.
    \item Our blind search for \cii\ emitters at the redshift of the quasar in the SMG ALMA pointings finds no excess of sources around the quasar redshift. We conclude that most detections are low--redshift CO interlopers.
    \item We report the discovery of two previously unpublished LAEs at the redshift of PJ231--20, indicating an overdensity in this field. We found however no LAEs around PJ308--21 and J0305--3150. We did not recover the LAE found by \citet[][]{Farina2017}, which can be explained by the different search methods. Overall, we find an overdensity of LAEs in our quasars fields, supporting earlier results on quasar-LAE overdensities \citep[][]{Farina2017,Mignoli2020}.
\end{itemize}

Although our \cii\ non--detections could be due to cosmic variance, our results suggest that targeting bright SMGs is not the most promising way forward to characterise the overdensities of galaxies around high--redshift quasars. Additionally, the combination of MUSE and ALMA data on the same field suggests a dichotomy of dust-rich/dust-poor sources which cannot yet be satisfyingly explained. Blind searches taking advantage of wide--field and frequency coverage, such as provided by large mosaics with ALMA, MUSE and \textit{JWST} might thus be the only way to map and understand the large--scale environment of quasars in the first billion years.

\acknowledgments

The authors thank the referee for useful suggestions which improved the manuscript. RAM, FW acknowledge support from the ERC Advanced Grant 740246 (Cosmic\_Gas). 

This paper makes use of the following ALMA data: ADS/JAO.ALMA\#2019.1.01003.S. ALMA is a partnership of ESO (representing its member states), NSF (USA) and NINS (Japan), together with NRC (Canada), MOST and ASIAA (Taiwan), and KASI (Republic of Korea), in cooperation with the Republic of Chile. The Joint ALMA Observatory is operated by ESO, AUI/NRAO and NAOJ. 
Based on observations made with ESO Telescopes at the La Silla Paranal Observatory under programme ID 094.B-0893(A), 099.A-0682(A), 099.A-0682(A).

These HST observations are associated with program 14876. Support for this work was provided by NASA through grant No. 10747 from the Space Telescope Science Institute, which is operated by AURA, Inc., under NASA contract NAS 5-26555.
This work is based [in part] on observations made with the Spitzer Space Telescope (program IDs 11030, 13066), which was operated by the Jet Propulsion Laboratory, California Institute of Technology under a contract with NASA.

The James Clerk Maxwell Telescope is operated by the East Asian Observatory on behalf of The National Astronomical Observatory of Japan; Academia Sinica Institute of Astronomy and Astrophysics; the Korea Astronomy and Space Science Institute; the National Astronomical Research Institute of Thailand; Center for Astronomical Mega-Science (as well as the National Key R\&D Program of China with No. 2017YFA0402700). Additional funding support is provided by the Science and Technology Facilities Council of the United Kingdom and participating universities and organizations in the United Kingdom and Canada. Additional funds for the construction of SCUBA-2 were provided by the Canada Foundation for Innovation. 

The authors wish to recognize and acknowledge the very significant cultural role and reverence that the summit of Maunakea has always had within the indigenous Hawaiian community. We are most fortunate to have the opportunity to conduct observations from this mountain.

\facilities{ALMA, VLT (MUSE), HST (WFC3), Spitzer (IRAC), JCMT (SCUBA2) }

\software{astropy \citep{TheAstropyCollaboration2018}, scipy \citep[][]{Virtanen2020},
          numpy \citep[][]{Numpy2020} , matplotlib \citep[][]{Hunter2007},
          interferopy \citep*[][]{interferopy}
          }

\clearpage
\appendix

\section{SCUBA2 quasar fields images}
\begin{figure}
    \centering
    \includegraphics[width=0.49\textwidth]{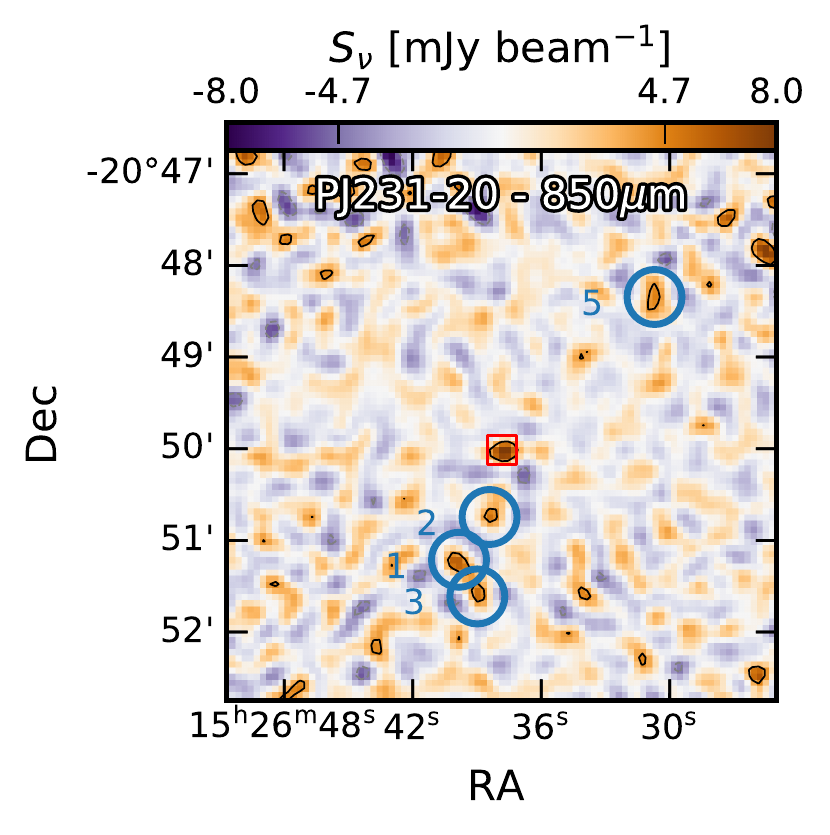}
    \includegraphics[width=0.49\textwidth]{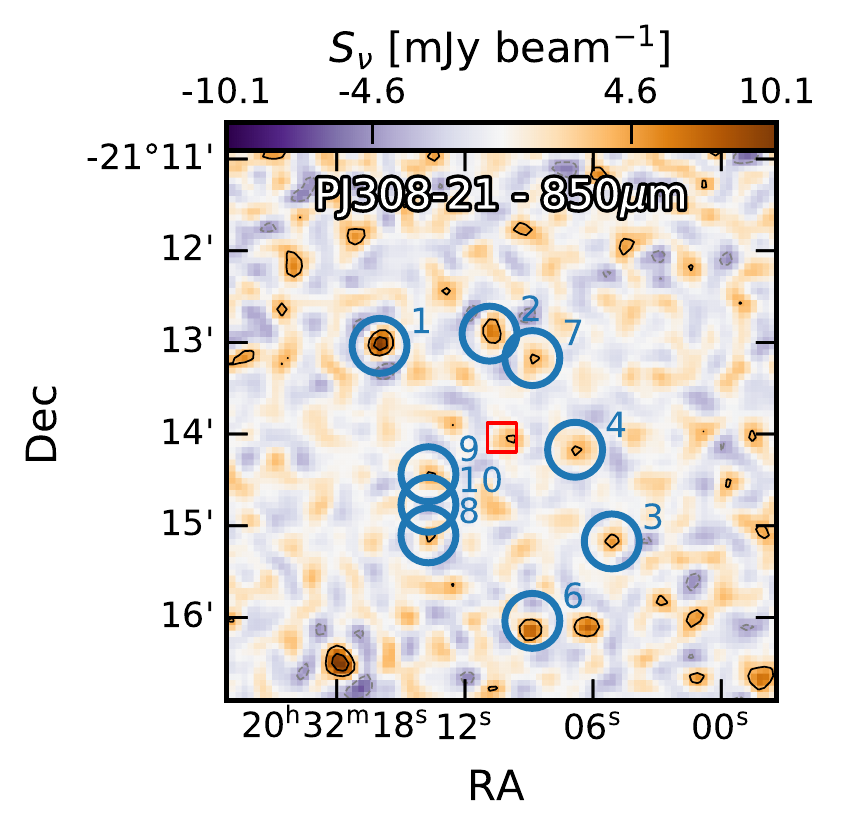} \\
    \includegraphics[width=0.49\textwidth]{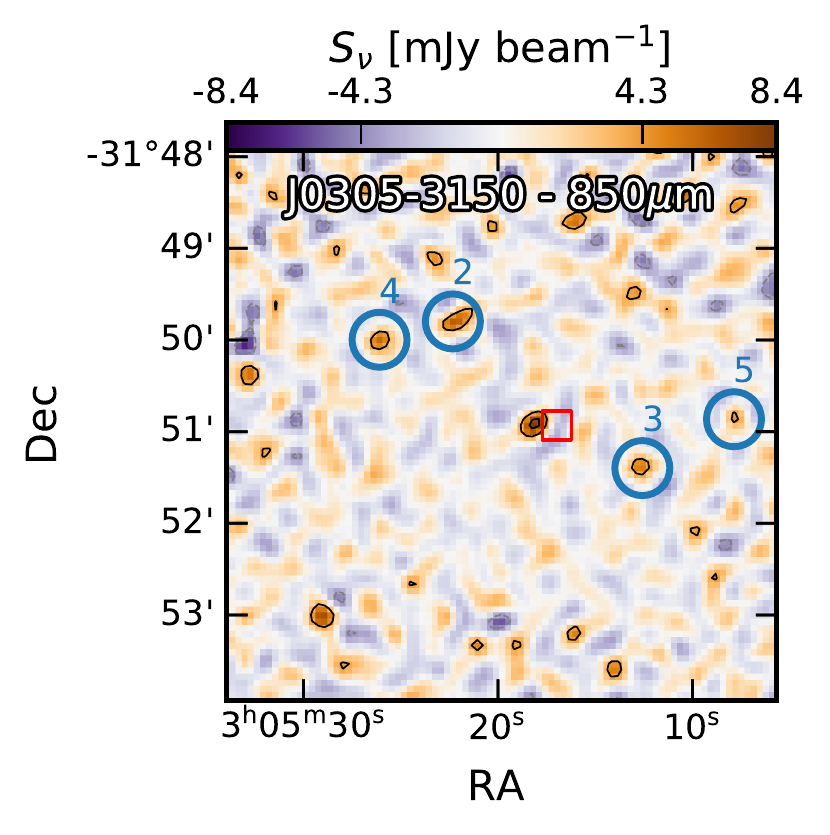}
    \caption{SCUBA2 850 $\mu$m imaging of the quasar fields of PJ231--20, PJ308--21, J0305--3150 from (Li et al., in prep.), reproduced here for completeness. The \cii\ position of the quasar is highlighted in a red box, and the ALMA pointings reported in this work are denoted by blue circle and numbers. }
    \label{fig:scuba_maps_and_pointings}
\end{figure}

In this appendix we reproduce for completeness a zoomed--in version of the SCUBA2 $850\rm{\mu}m$ maps (Fig. \ref{fig:scuba_maps_and_pointings}) with the position of the quasar and the targeted SMGs highlighted. A full description of the SCUBA2 observations and data reductions will be presented in Li et al., (in prep.).
\label{app:scuba_images}

\section{A $z=2.42$ overdensity in the field of PJ231--20}
\label{app:disc_z24_pj231}
\begin{figure*}
    \centering
    \includegraphics[width=0.45\textwidth]{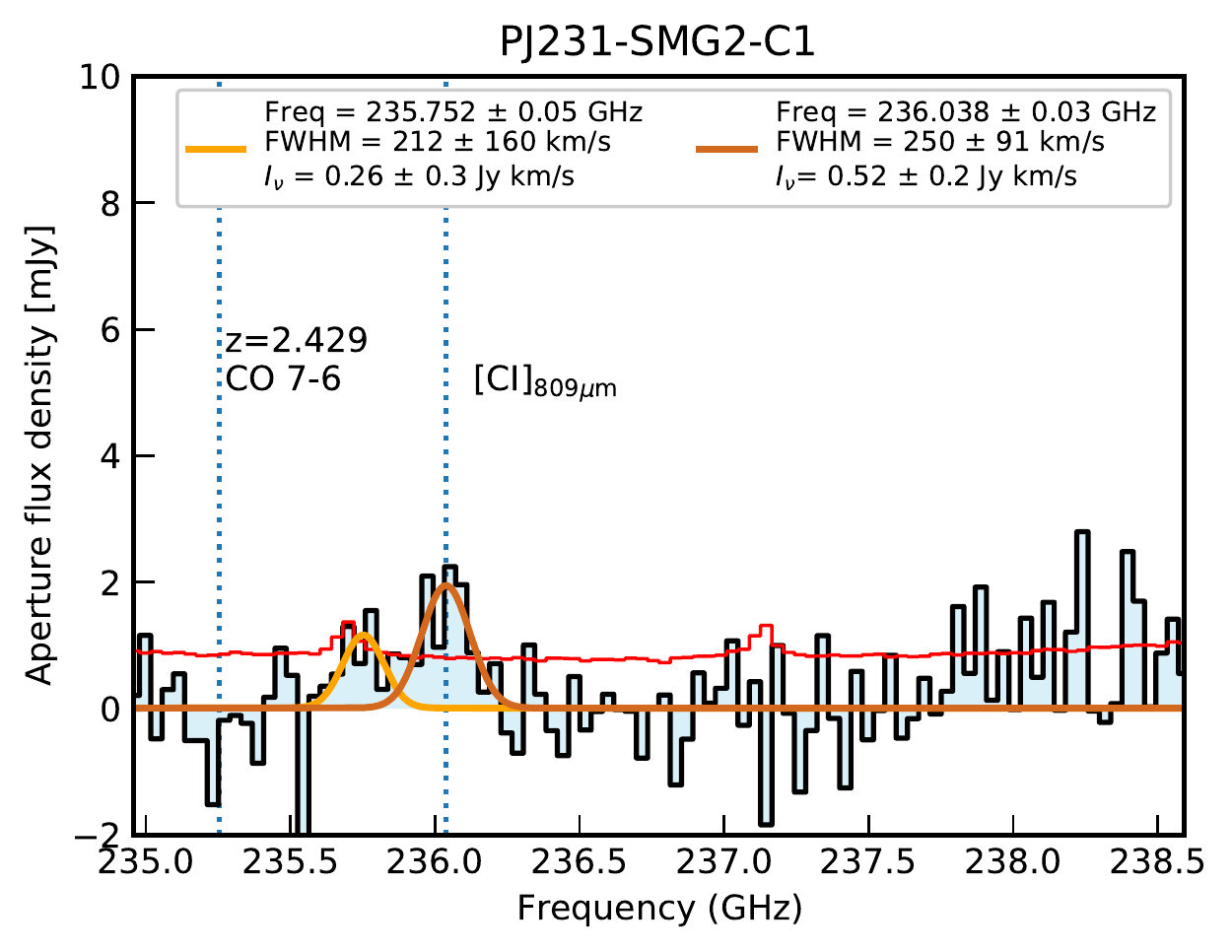}
    \includegraphics[width=0.45\textwidth]{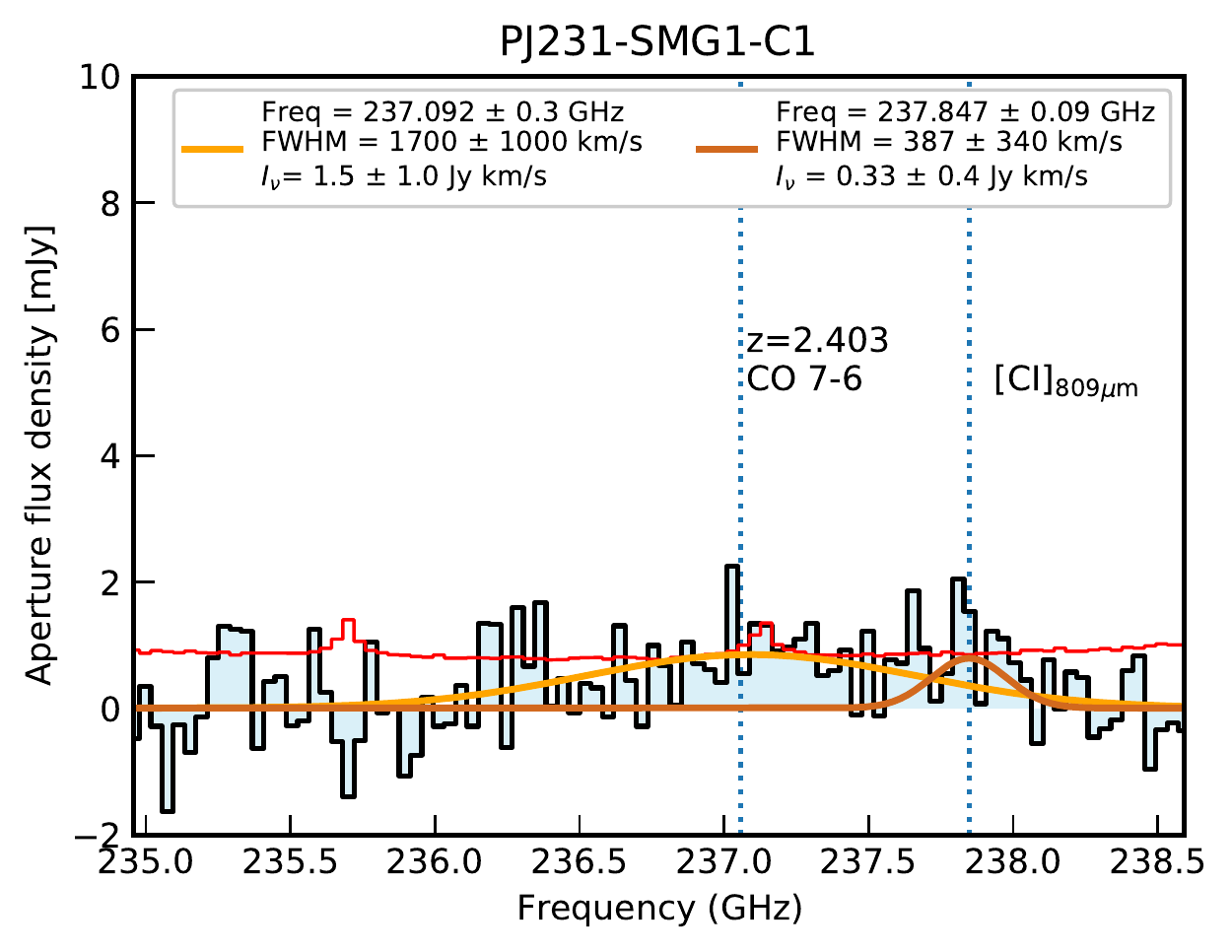}
    \includegraphics[width=0.45\textwidth]{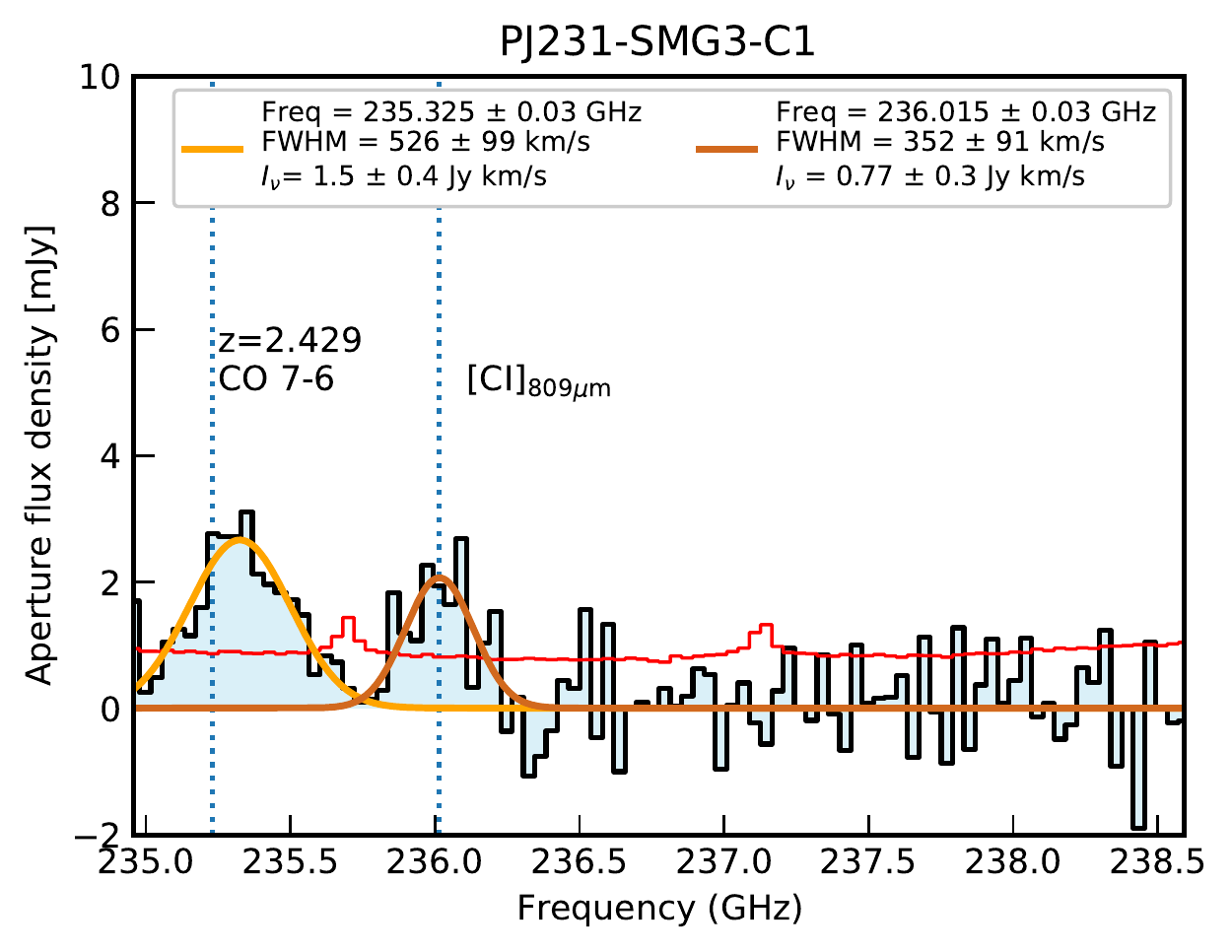}
    \includegraphics[width=0.45\textwidth]{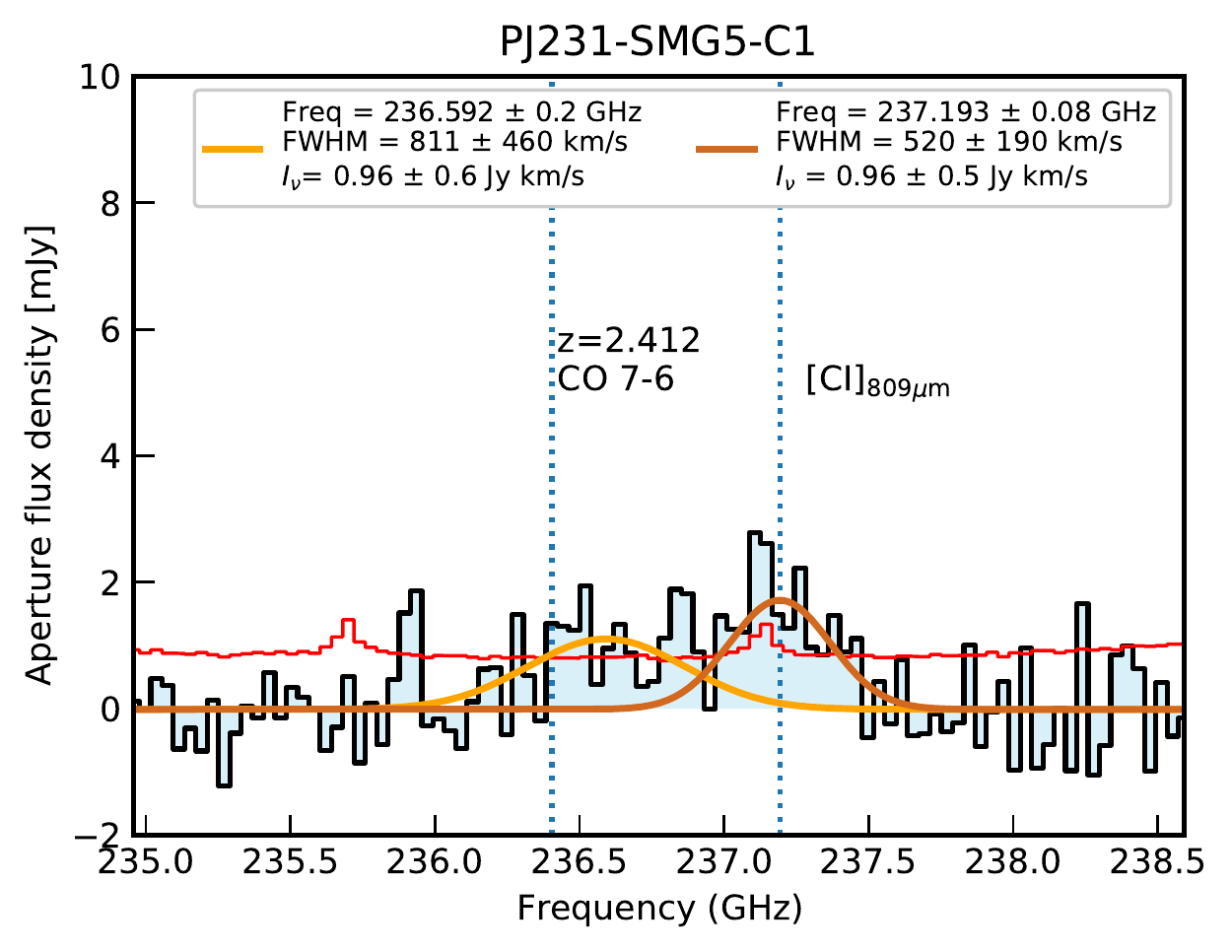}
    \caption{Aperture--integrated (r=2"), continuum--subtracted spectra (black, rms error in red) of the four of the five continuum sources detected in the field of PJ231--20. The error array (red) is measured for each spectrum using the rms per beam rescaled to the number beam in the $r=2"$ aperture. Double Gaussians profiles are fitted to the spectra (orange, dark orange) and in three objects are consistent with [CI]$_{809\mu\rm{m}}$ and CO7--6 ($806\mu\rm{m}$) emission at z=2.41--2.42 (blue dotted vertical lines). }
    \label{fig:pj231_overdens_spectra}
\end{figure*}

The case of the SMGs in the field of PJ231--20 stands out from the analysis presented in Section \ref{sec:cont_sources}. We show the aperture--integrated spectra ($r=2"$) of the PJ231--20 sources in Fig. \ref{fig:pj231_overdens_spectra}. Three of these four targets show evidence for two lines which are consistent with \ci\ and CO7--6 ($806\mu\rm{m}$) at $z=2.412,2.403$ and $z=2.429$. There exists no other pair of strong atomic fine structure lines and CO lines with similar velocity offset, and the $z\sim 2$ interpretation is supported by the photometric redshifts analysis (see Fig. \ref{fig:magphys_photoz}). PJ231-SMG2-C1 (Fig. \ref{fig:pj231_overdens_spectra}, upper left) does not show two clear emission lines, but one of them is exactly at the frequency of \ci\ in PJ231-SGM3-C1 (\ref{fig:pj231_overdens_spectra}, lower left), the source with the most convincing \ci\ -- CO7--6 spectrum, suggesting it could be at the same redshift but with a fainter CO luminosity. Additionally, it should be noted that PJ231-SMG2-C1 resolves in two sources in the \textit{HST} images, and that its photometric redshift might be uncertain (see Fig. \ref{fig:p231_smg2_c1_hst_spitzer}). We provide FIR, CO and \ci luminosities in Table \ref{tab:z24_SMG}. The line luminosity ratios of CO and \ci\ are close to unity in all sources where both are detected, and is $\sim 0.5$ for PJ231-SMG2-C1, in agreement with the values found for local (U)LIRGs \citep[e.g.][]{Lu2017a}.

We hence conclude that an overdensity of $z\simeq2.42$ SMGs lies in the field of PJ231--20. The four objects lie at projected sky distances $r_\perp \simeq 1-4 \rm{cMpc}$ from the quasar. Furthermore, the redshift of the four SMGs matches ($\Delta v \simeq (-820,680,1465)\  \kms$) that of a Mg{~\small II} $\lambda\lambda 2798,2803 \rm{\AA}\ $ absorption system in the spectrum of PJ231--20 \citep[][system 262 in their catalog]{Chen2017}. Given that all the SMGs are at projected distances $r_\perp > 1$ comoving Mpc from the quasar sightline, none of these are expected to be the host of the absorption. Rather, they trace the overdense environment where Mg{~\small II} absorbers are more likely to be found \citep[e.g.,][]{Lee2021a}. We thus conclude that a $z=2.4$ galaxy overdensity is serendipitously aligned with the high--redshift quasar PJ231--20.

\begin{table}
    \hspace{-2.5cm}
    \begin{tabular}{c|c|c|c|c|c|c|c}
ID & $\nu_{\rm{CO}}$ [GHz] & $\rm{FWHM}_{\rm{CO}}$ [$\kms$]  & $L_{\rm{CO}}$ [$10^{8} L_\odot$] & $\nu_{\rm{[CI]}}$ [GHz] & $\rm{FWHM}_{\rm{[CI]}}$ [$\kms$] &  $L_{\rm{[CI]}}$ [$10^{8} L_\odot$] & z  \\ \hline
PJ231-SMG1-C1 & $237.09\pm0.26 $ & $1700\pm1000 $ & $1.4\pm1.0 $ & $237.85\pm0.08 $ & $380\pm340 $ & $0.3\pm0.3$ & 2.403\\ 
PJ231-SMG2-C1 & - & - & - & $236.04\pm0.03 $ & $250\pm90 $ & $0.5\pm0.2$ & 2.429? \\ 
PJ231-SMG3-C1 & $235.33\pm0.03 $ & $530\pm100 $ & $1.42\pm0.34 $ & $236.02\pm0.03 $ & $350\pm90 $ & $0.7\pm0.3$ & 2.429 \\ 
PJ231-SMG5-C1 & $236.60\pm0.20 $ & $810\pm470 $ & $0.90\pm0.58 $ & $237.19\pm0.08 $ & $520\pm200 $ & $0.9\pm0.4$ & 2.412 \\ 

    \end{tabular}
    \caption{Best--fit CO7--6 and \ci emission line properties of the $z\sim 2.4$ SMGs in the field of PJ231--20.}
    \label{tab:z24_SMG}
\end{table}

\section{MAGPHYS SED best--fits}
\label{app:magphys_SEDs}

\begin{figure*}
    \centering
    \includegraphics[width=0.45\textwidth]{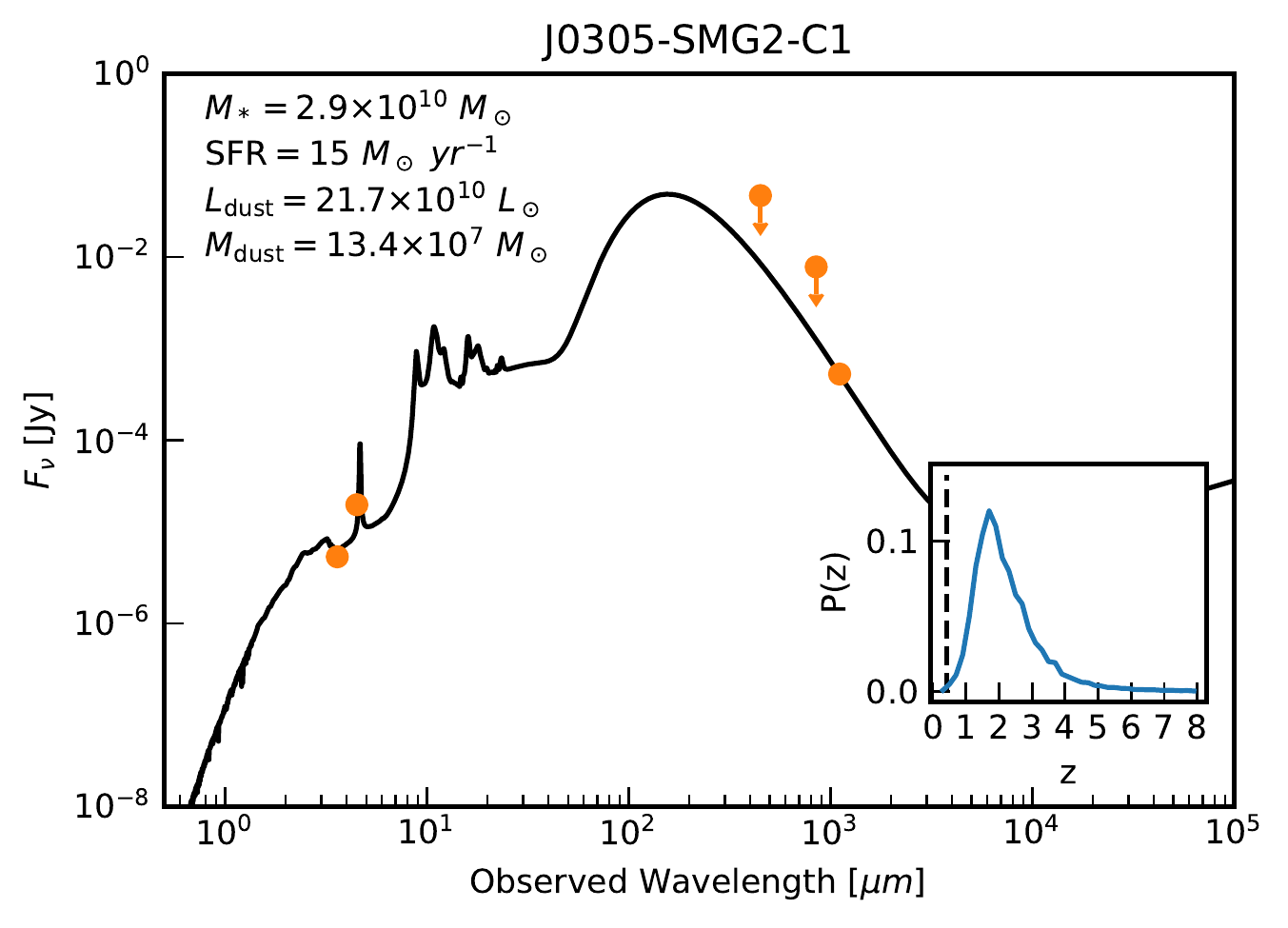}
    \includegraphics[width=0.45\textwidth]{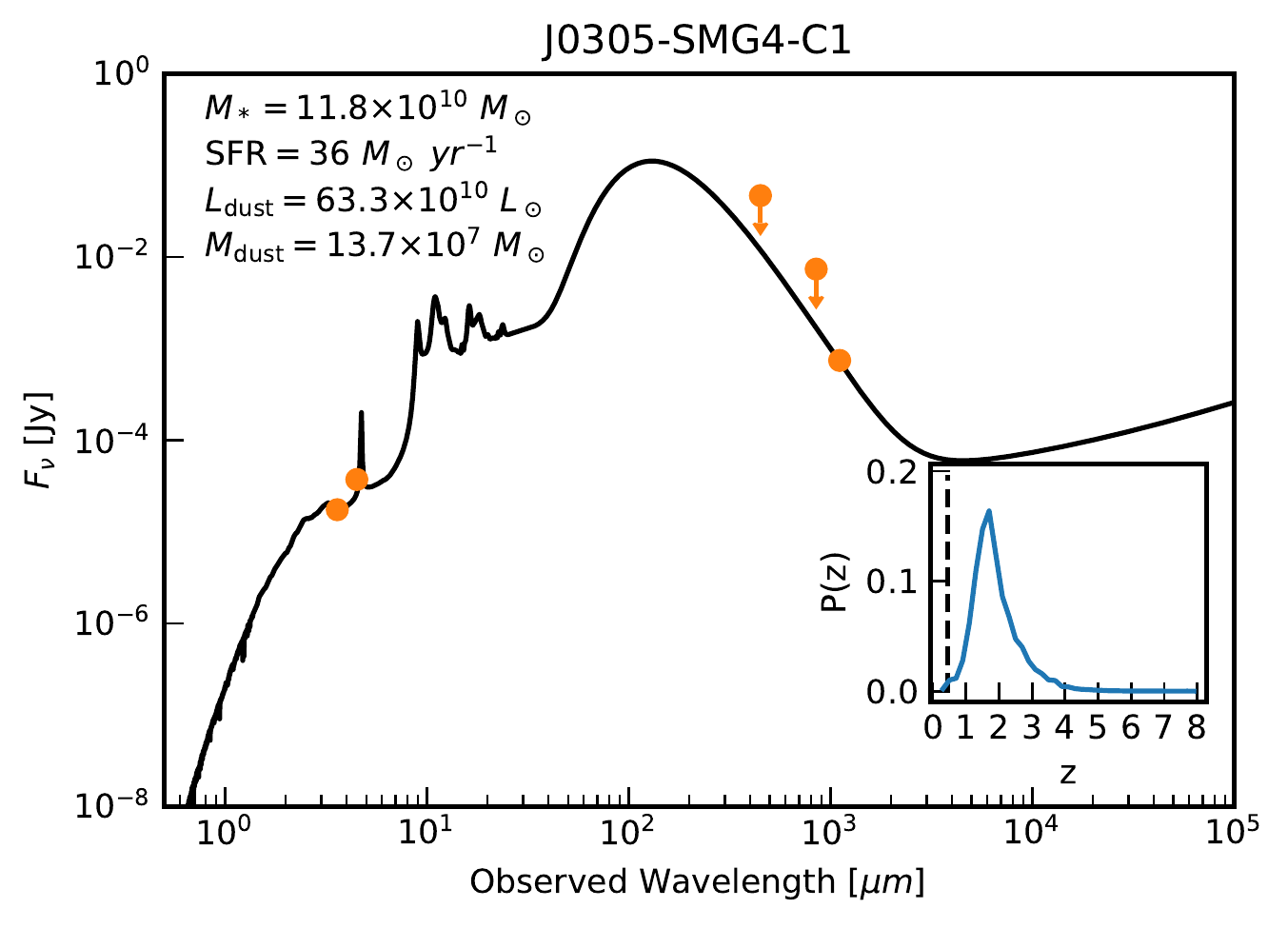} \\
    \includegraphics[width=0.45\textwidth]{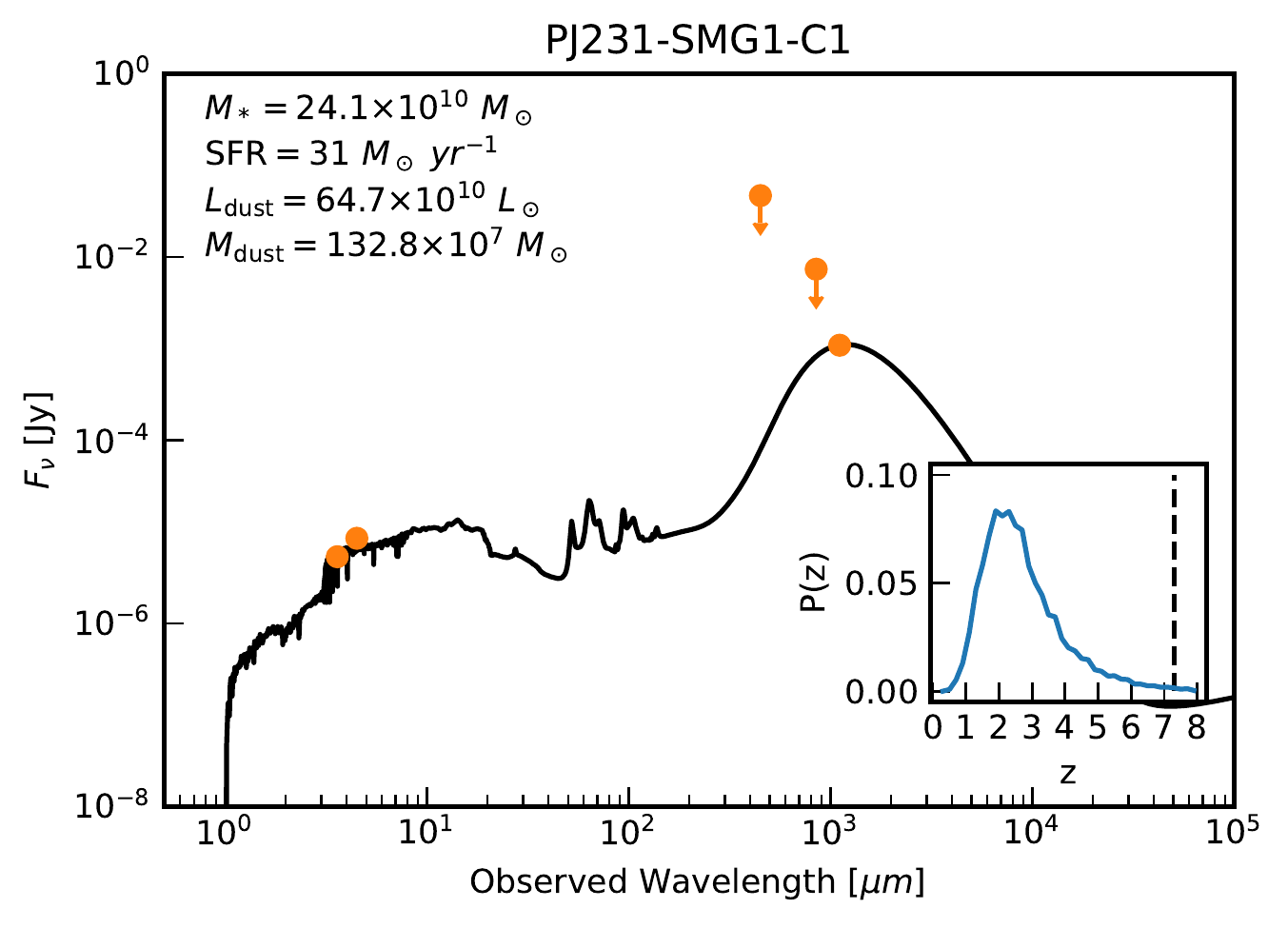}
    \includegraphics[width=0.45\textwidth]{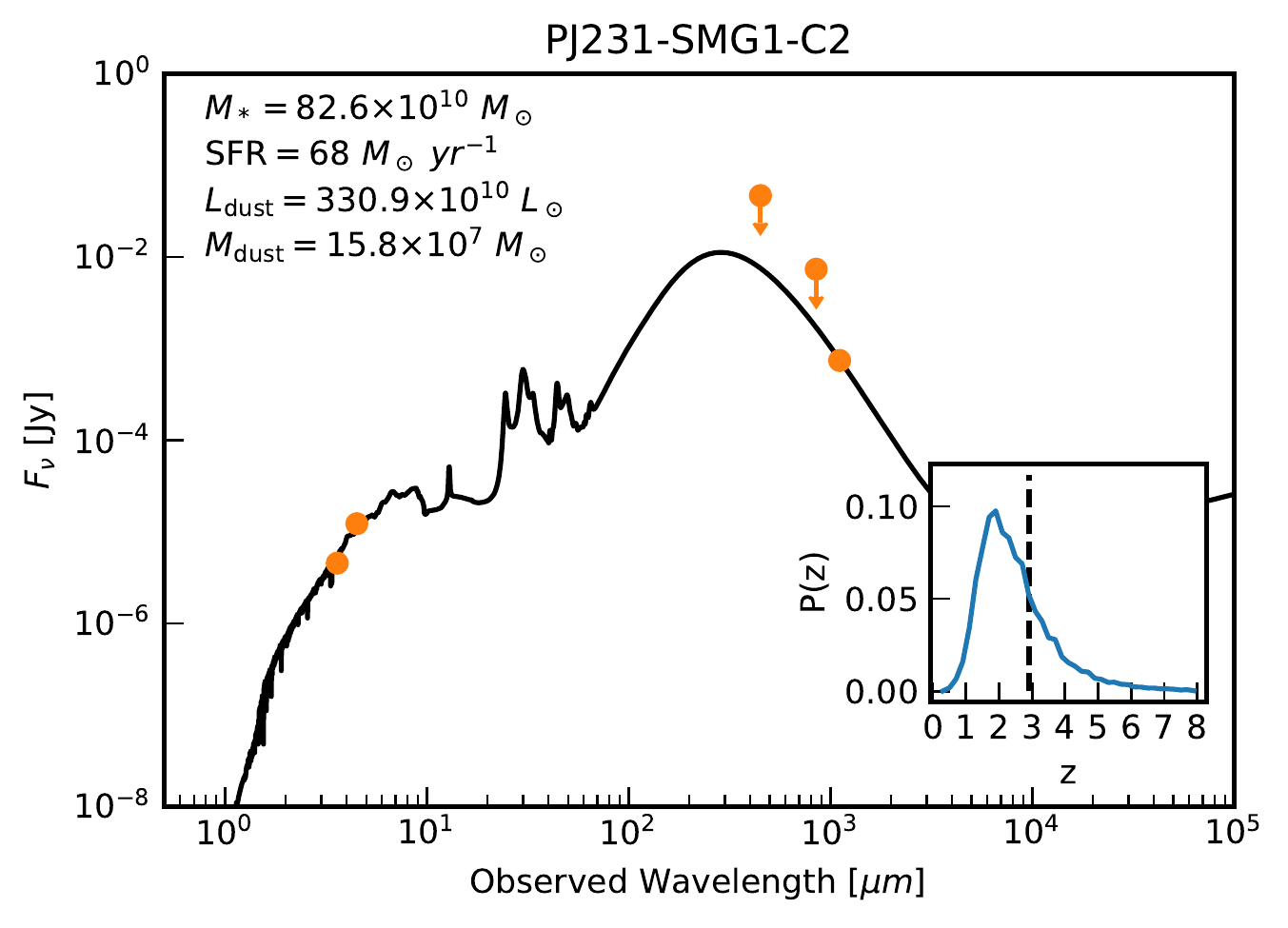} \\
    \includegraphics[width=0.45\textwidth]{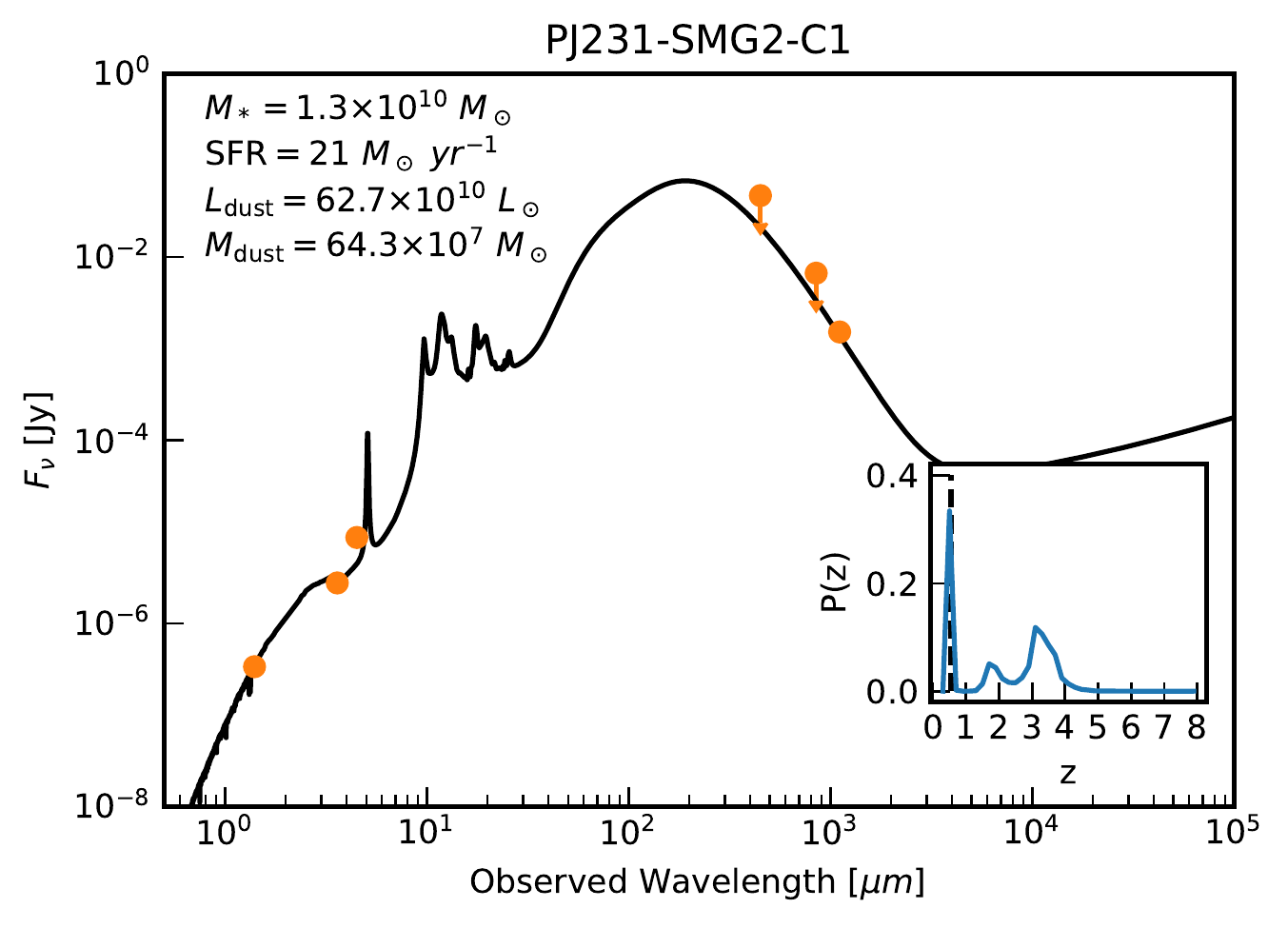}
    \includegraphics[width=0.45\textwidth]{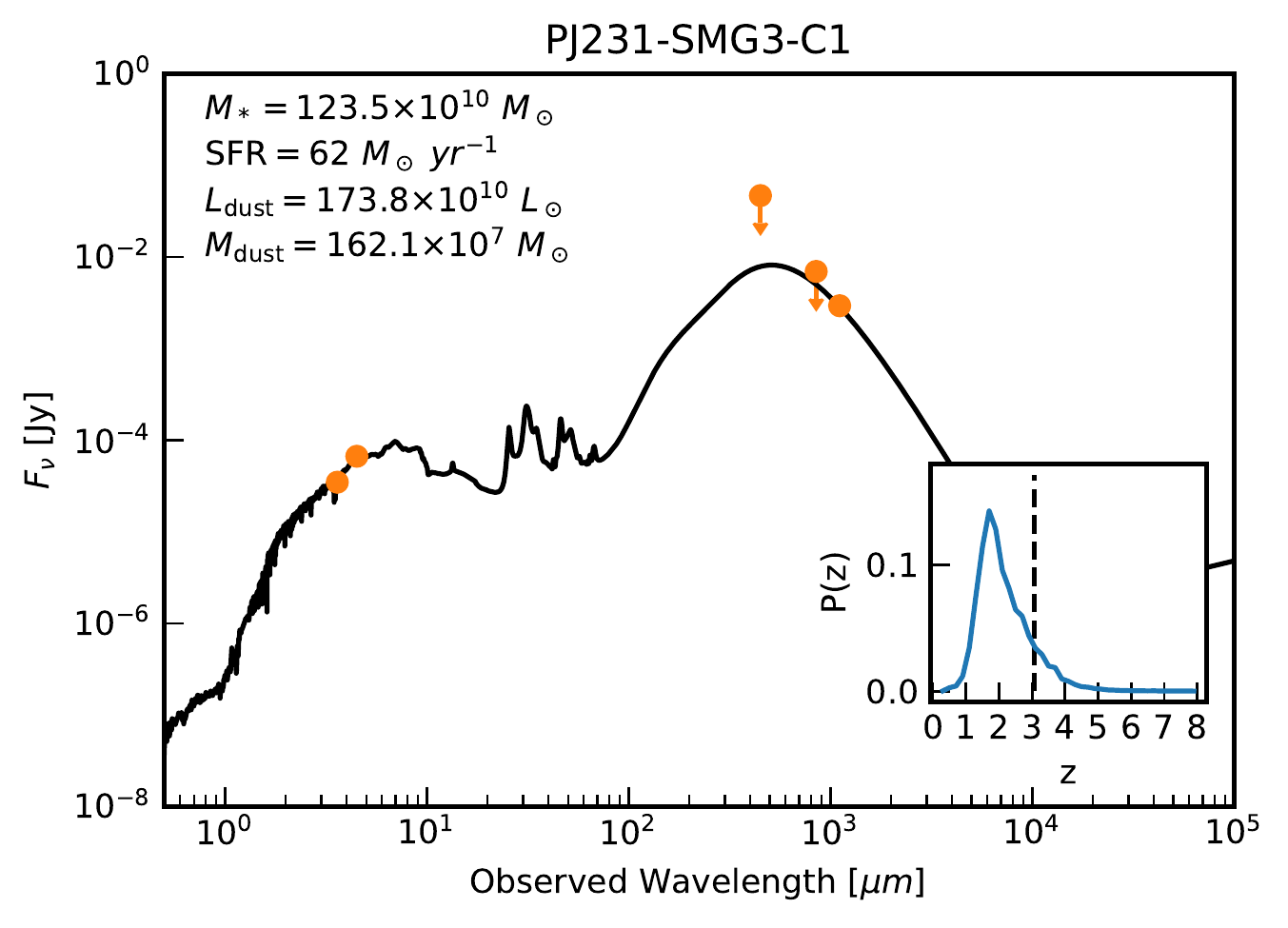} \\
    \includegraphics[width=0.45\textwidth]{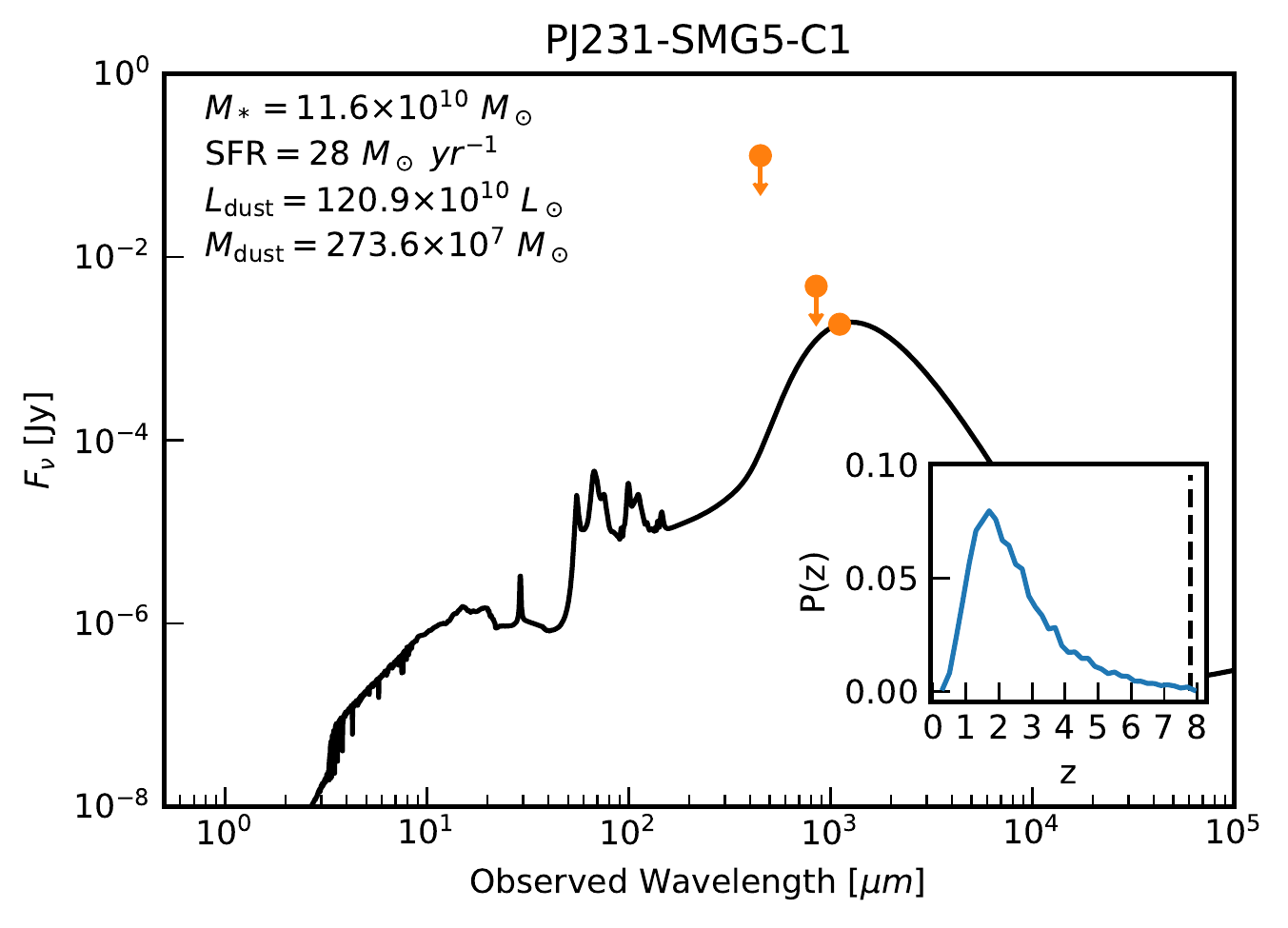}
    \caption{MAGPHYS bestfit SEDs, flux density measurements (orange) and photometric redshift posterior (inset, blue) for the continuum sources detected in our ALMA pointings (see Table \ref{tab:ancillary_photometry} for the ancillary photometry). The redshift of the best-fit SED is indicated as vertical dashed line in the inset and its main physical parameters are indicated in the upper left corner of the main plot. }
    \label{fig:magphys_SED_1}
\end{figure*}

\begin{figure*}
    \centering
    \includegraphics[width=0.45\textwidth]{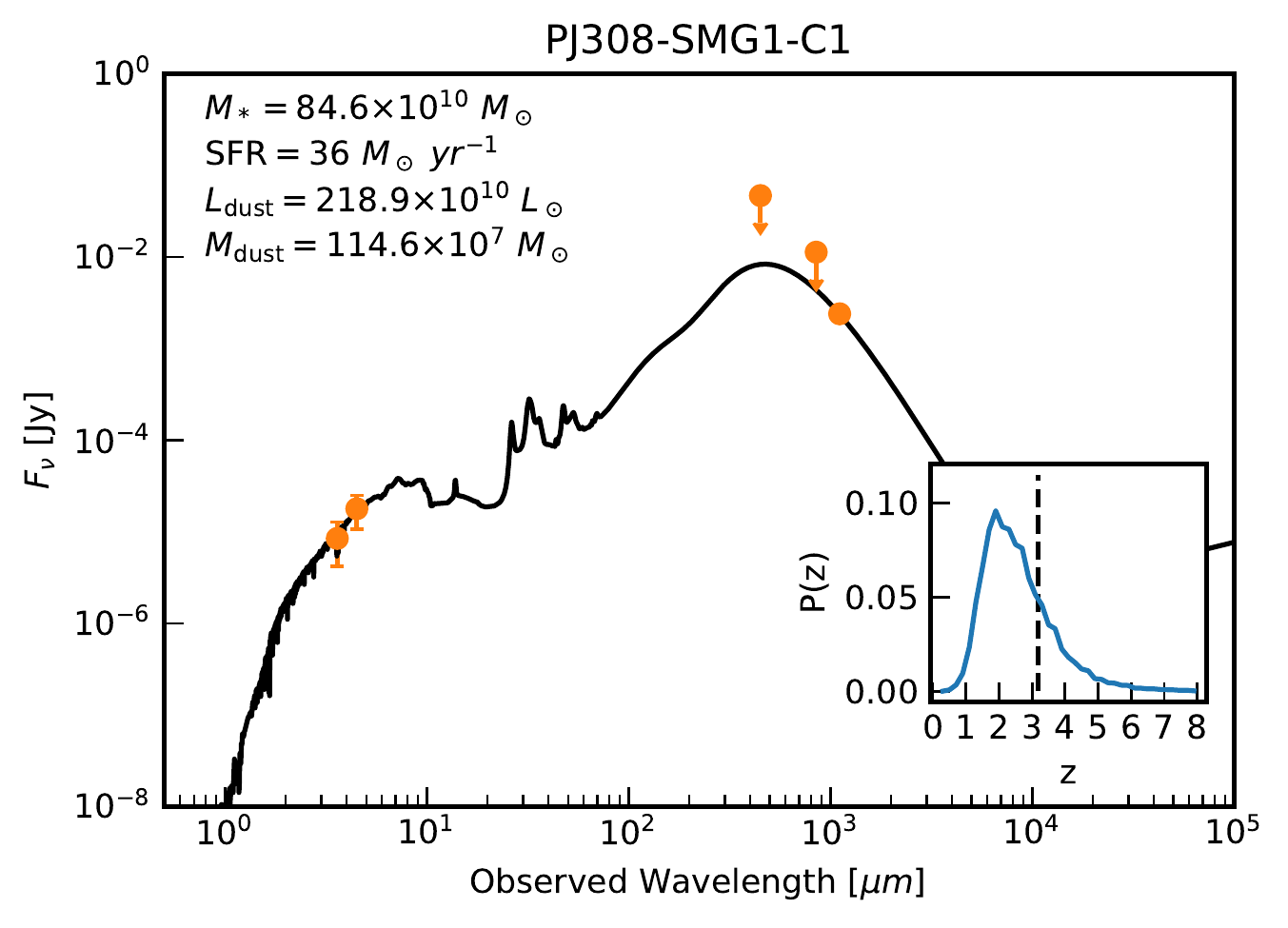}
    \includegraphics[width=0.45\textwidth]{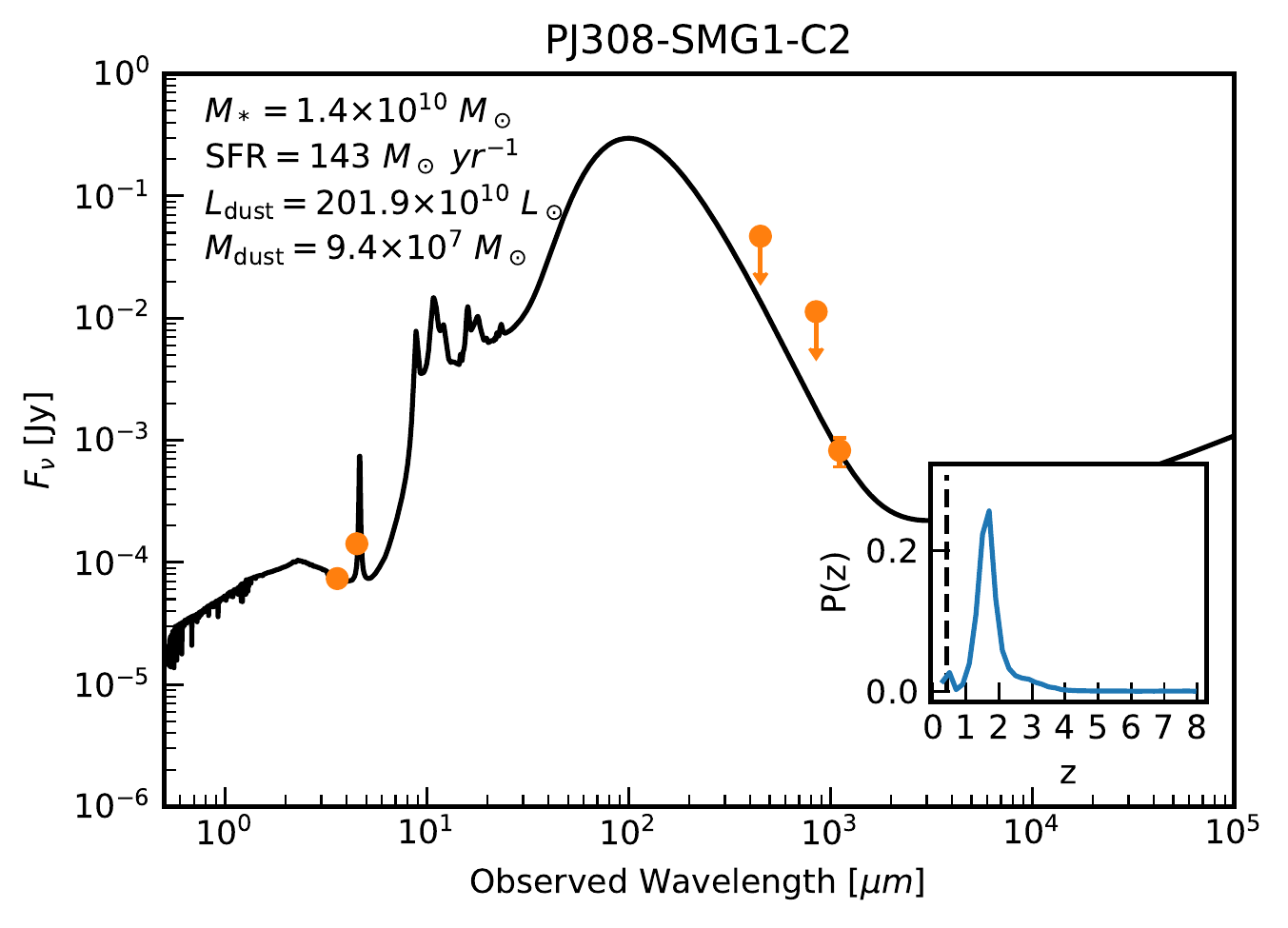} \\
    \includegraphics[width=0.45\textwidth]{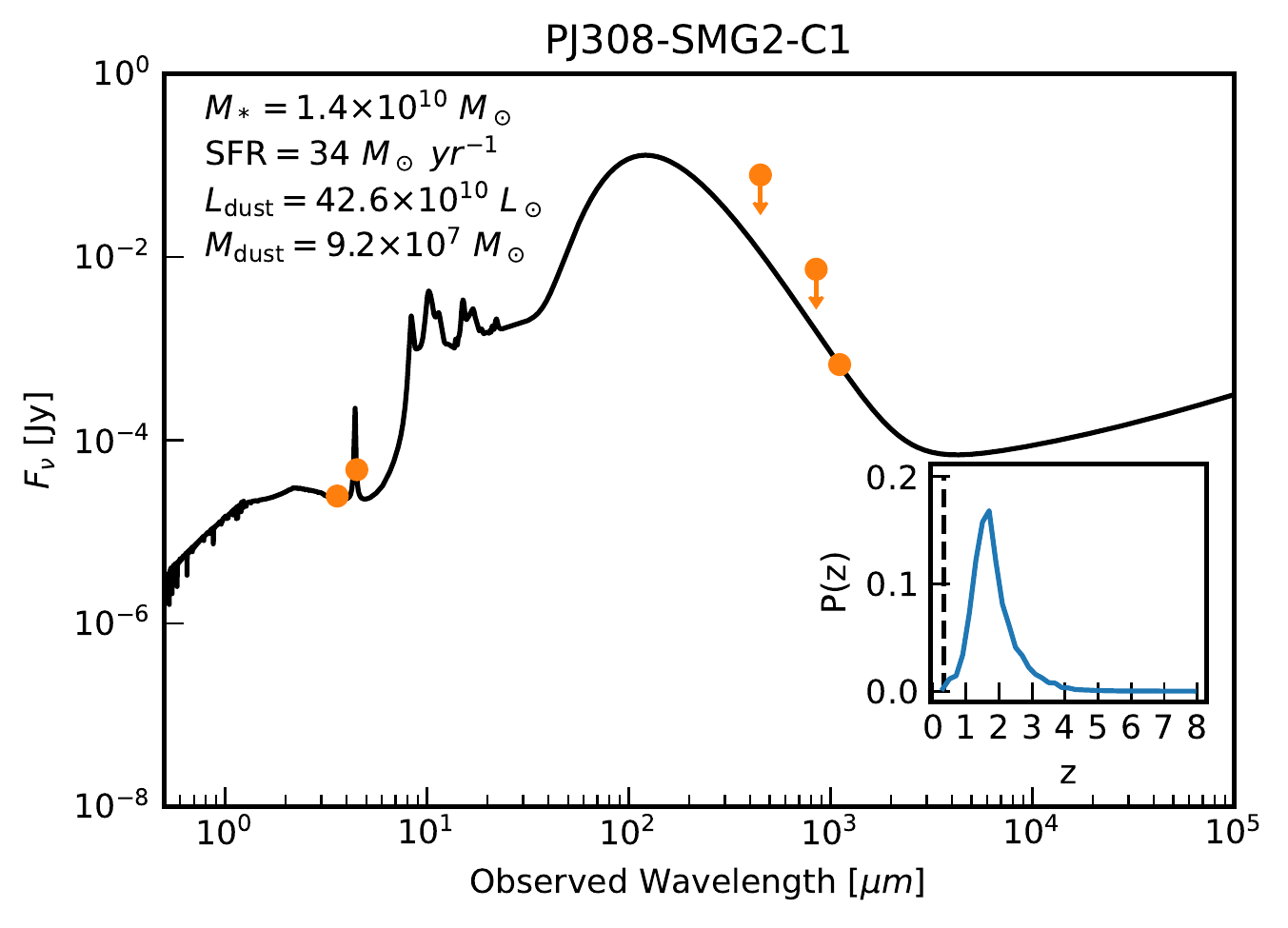}
    \includegraphics[width=0.45\textwidth]{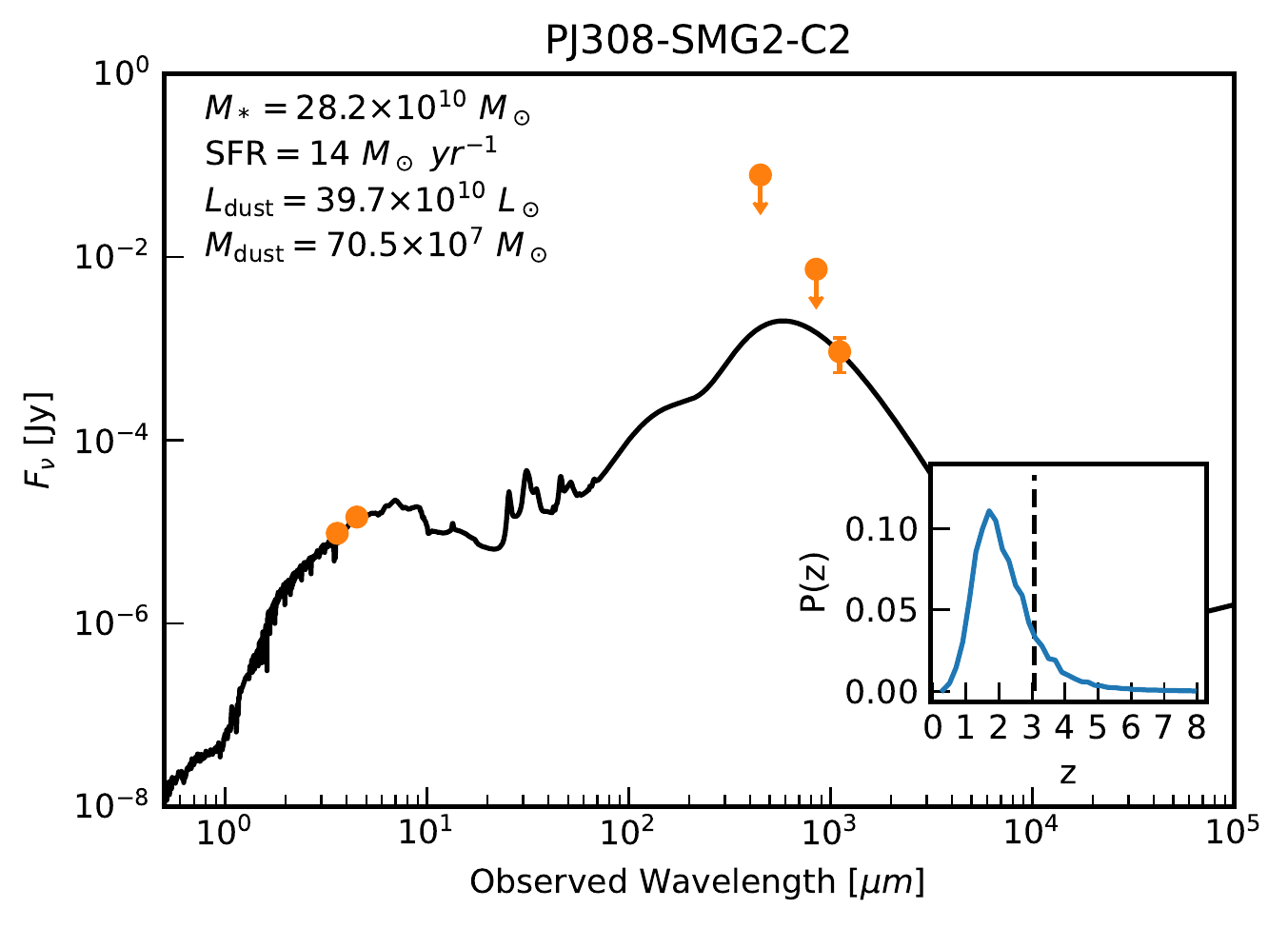} \\
    \includegraphics[width=0.45\textwidth]{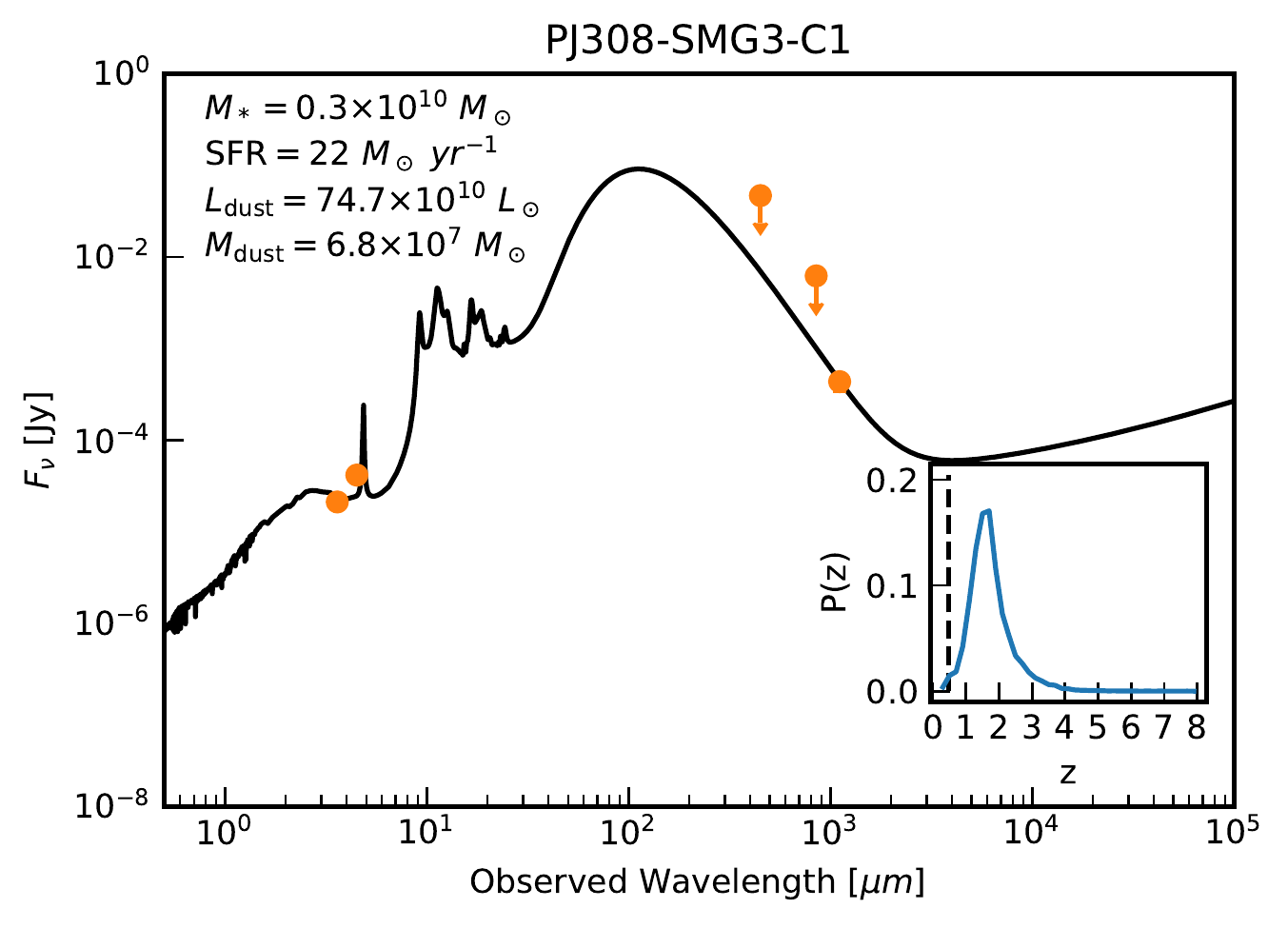}
    \includegraphics[width=0.45\textwidth]{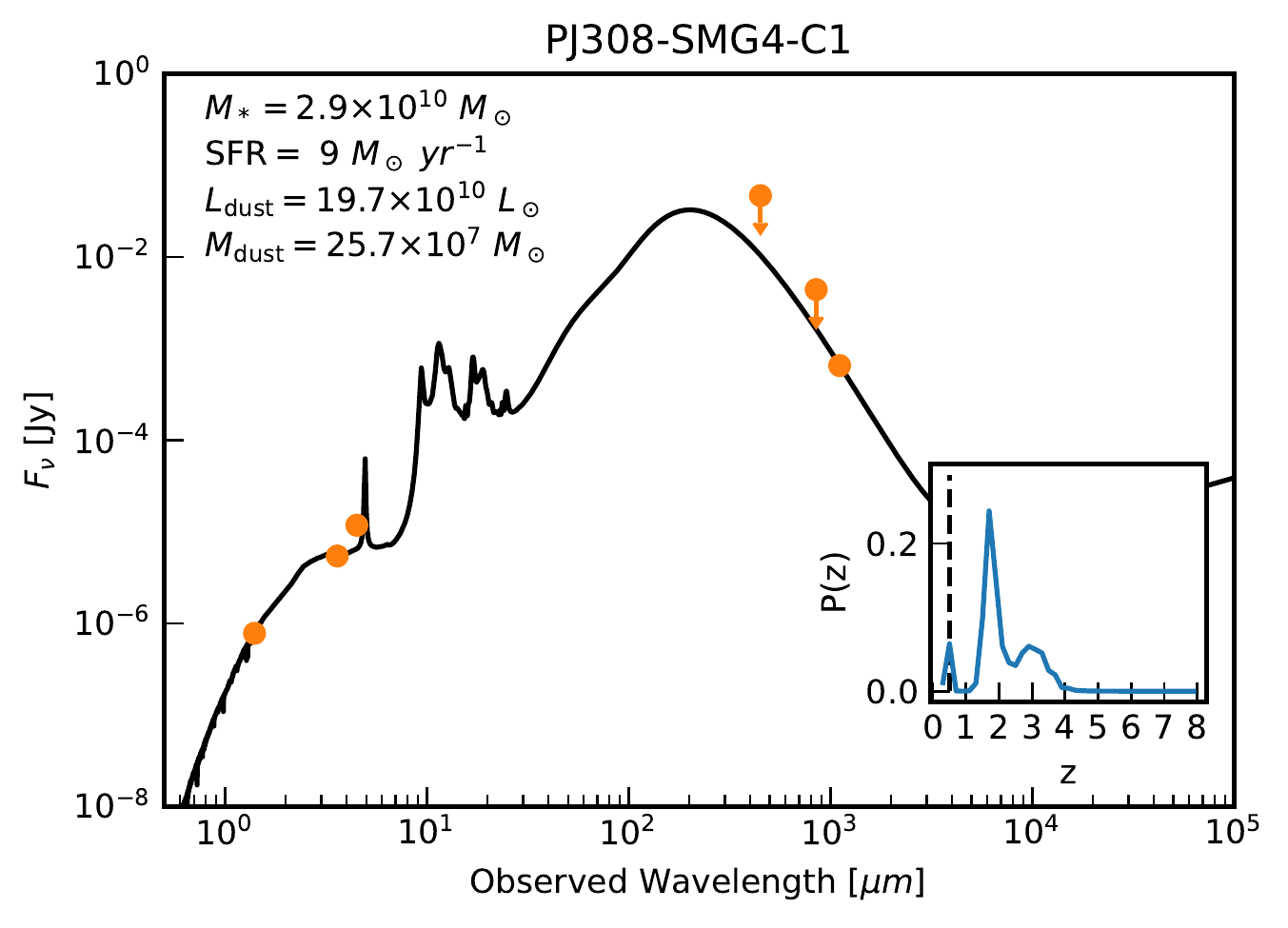} \\
    \includegraphics[width=0.45\textwidth]{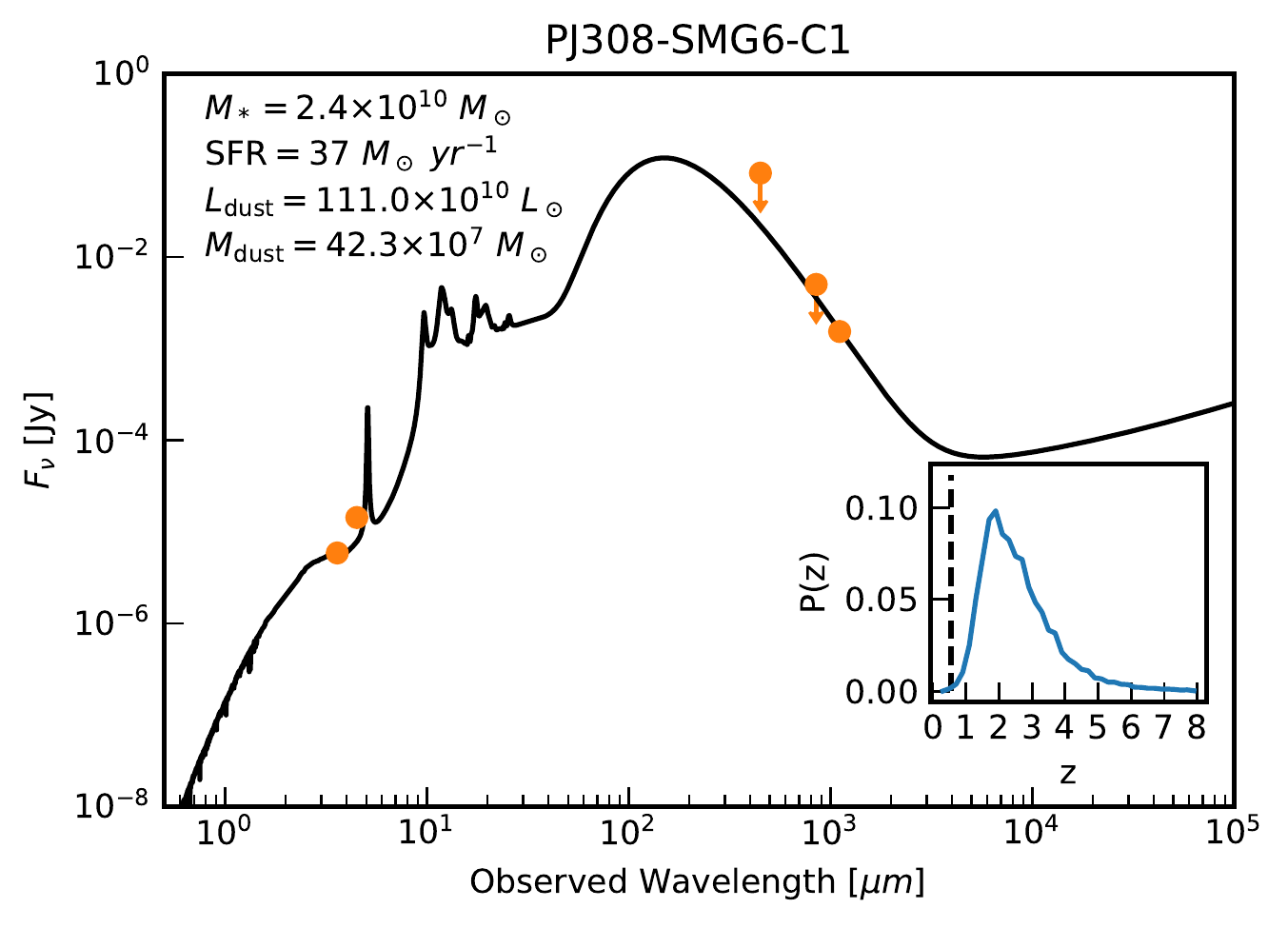}
    \includegraphics[width=0.45\textwidth]{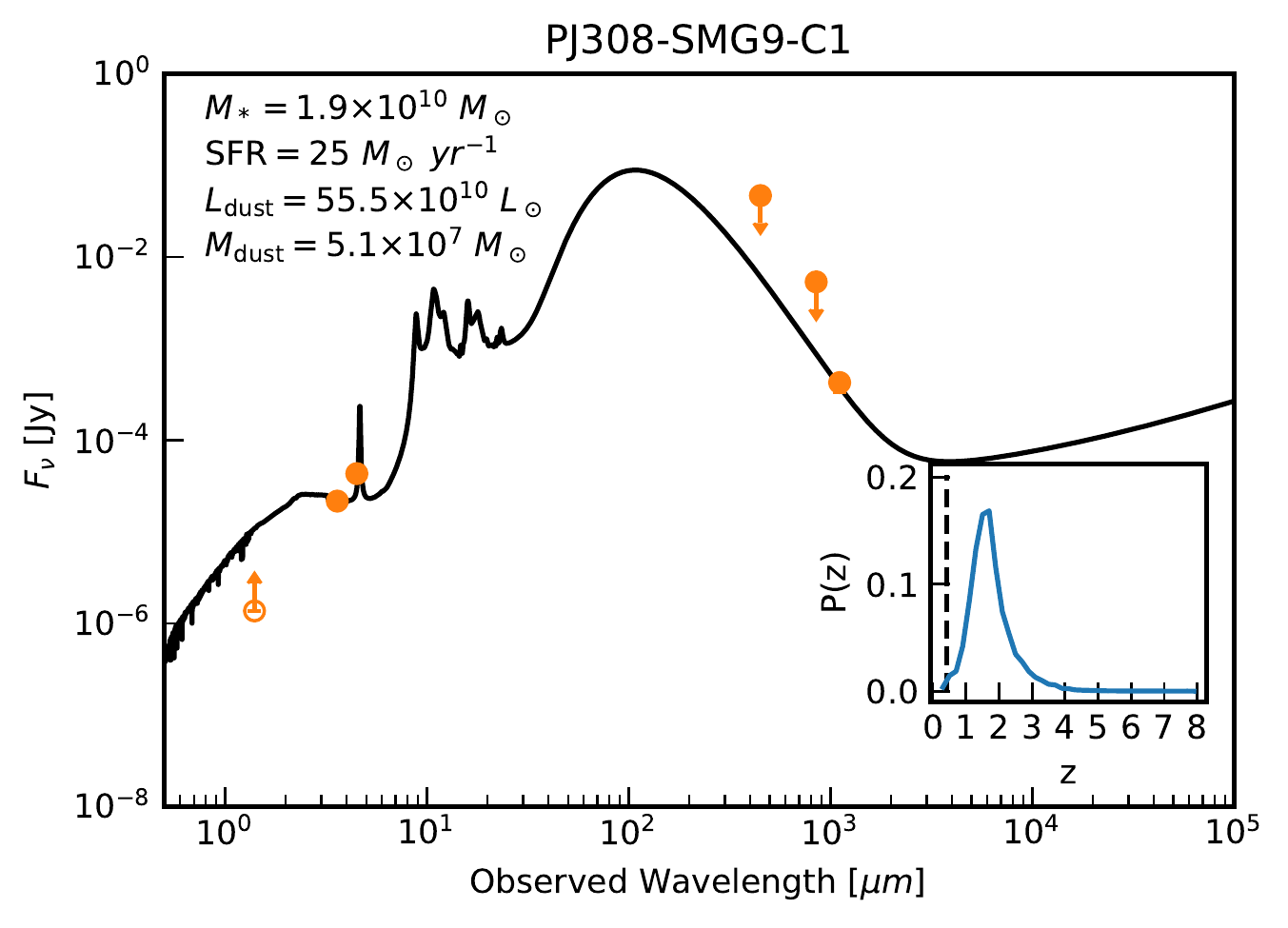}
    \caption{MAGPHYS best-fit SEDs, flux density measurements (orange) and photometric redshift posterior (inset, blue) for the continuum sources detected in our ALMA pointings (see Table \ref{tab:ancillary_photometry} for the ancillary photometry). The redshift of the best-fit SED is indicated as vertical dashed line in the inset and its main physical parameters are indicated in the upper left corner of the main plot.}
    \label{fig:magphys_SED_2}
\end{figure*}

We present in this appendix the best--fit \emph{MAGPHYS} SED resulting from our photometric redshift analysis in Section \ref{sec:photoz} in Figs. \ref{fig:magphys_SED_1} and \ref{fig:magphys_SED_2}. In each figure, the SED is presented in black with the observed fluxes in orange, and the inset gives the posterior distribution of the photometric redshift. The low number of datapoints (up to a maximum of three detections) leads to poorly constrained photometric redshifts with an SED modelling code sampling many more parameters. As such, the best-fit SED often has a different redshift than that of the photometric redshift posterior. The photometric redshift estimates of the SMGs in the field of PJ231-20 also differ from the spectroscopic redshifts by $\Delta z\sim 0.4$. Additional imaging data in the optical and sub-mm is needed to improve the SED modelling.

\begin{figure}
    \centering
    \includegraphics[width=\textwidth]{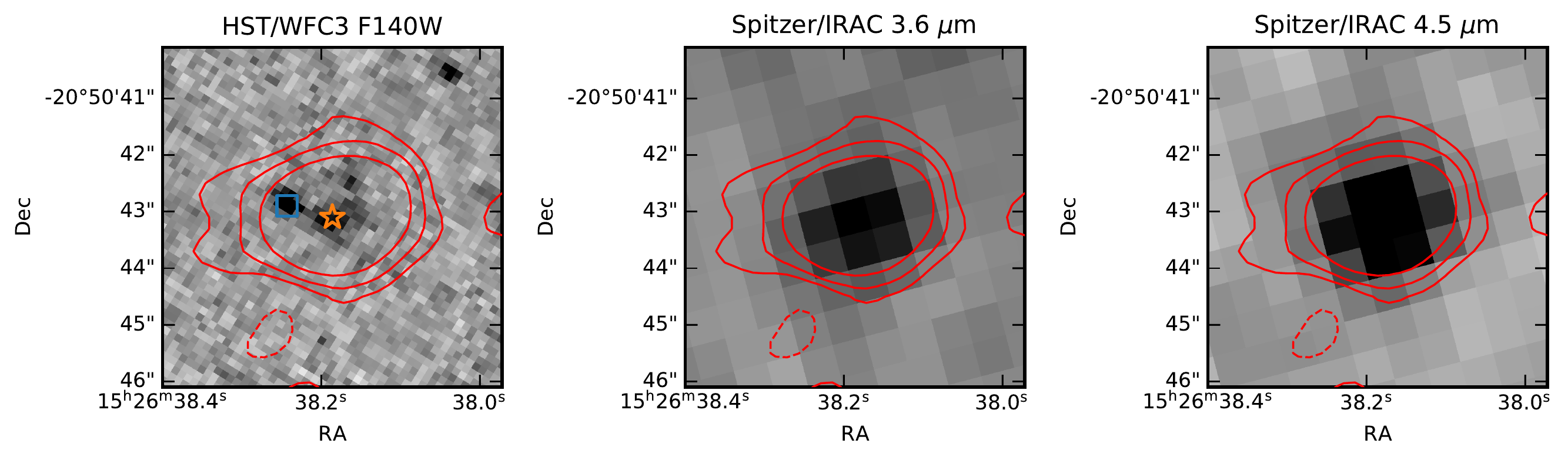}
    \caption{\textit{HST} and \textit{Spitzer} imaging of PJ231-SMG2-C1. The red contours show the (-4,-2,2,4)$\sigma$ levels of the dust continuum emission. For the purposes of the SED fitting, we assume the central source (orange star) is the ALMA/SCUBA2 SMG. The offset source indicated by a blue square is fainter by $\sim 0.5$ mag (F140W).}
    \label{fig:p231_smg2_c1_hst_spitzer}
\end{figure}

We also present the imaging data for PJ231-SMG2-C1 in Fig. \ref{fig:p231_smg2_c1_hst_spitzer}, which shows multiple sources in the \textit{HST} imaging. For the purposes of the SED fitting, we assume the brighter F140W central source is corresponding to the Spitzer, ALMA and SCUBA2 detections. We do not attempt to deblend the \textit{Spitzer} imaging given the lack of evidence for a second source.

\section{Candidate \cii\ line emitters}
\label{app:cii_line_search_candidates}
\begin{figure} 
    \begin{minipage}[t]{0.49\textwidth}
    \includegraphics[height=0.14\textheight]{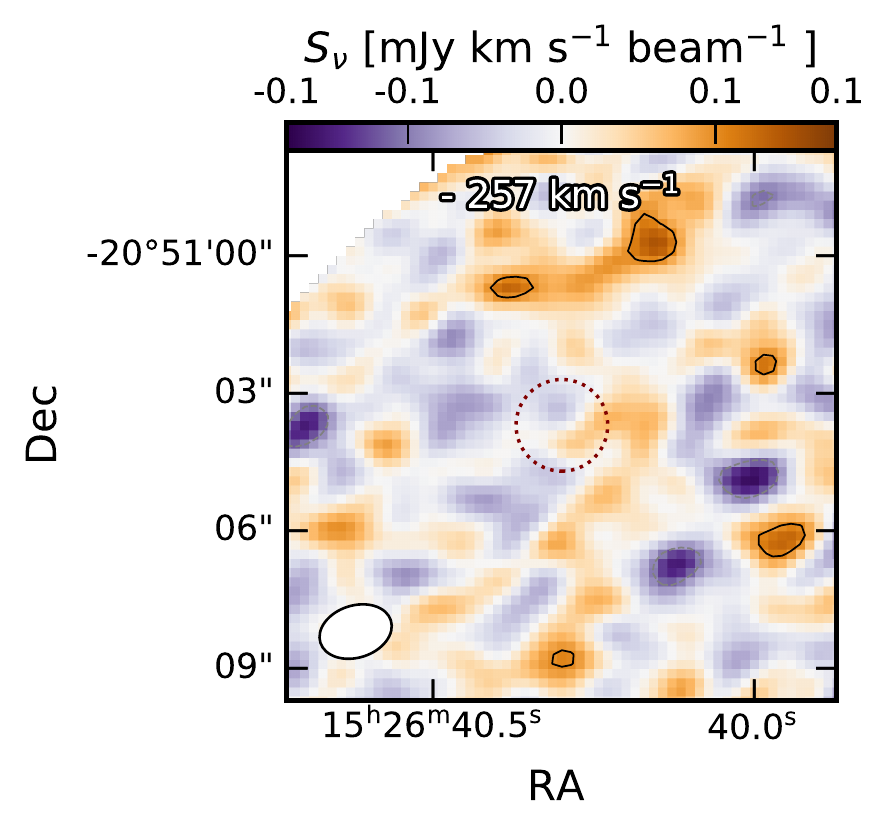}
    \includegraphics[height=0.14\textheight,trim={2.75cm 0 0 0},clip]{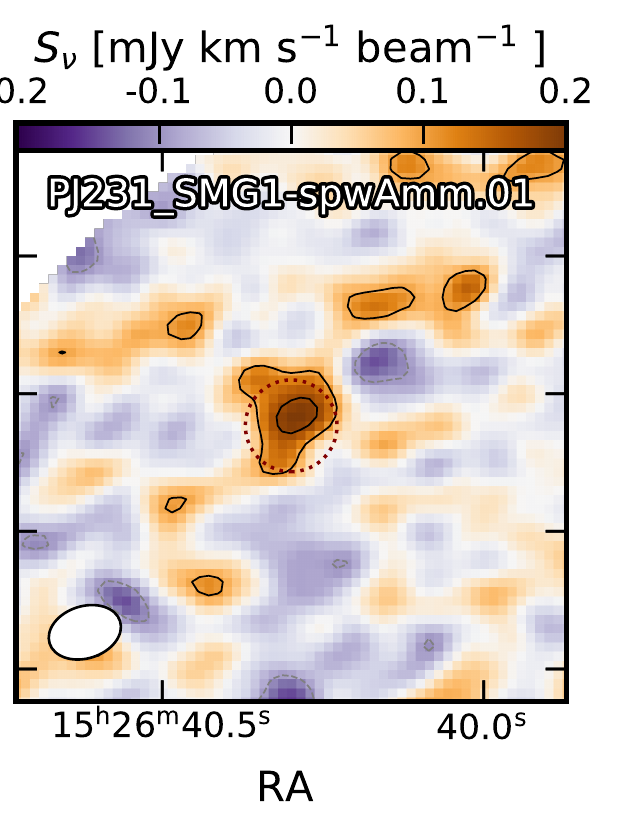} 
    \includegraphics[height=0.14\textheight,trim={2.75cm 0 0 0},clip]{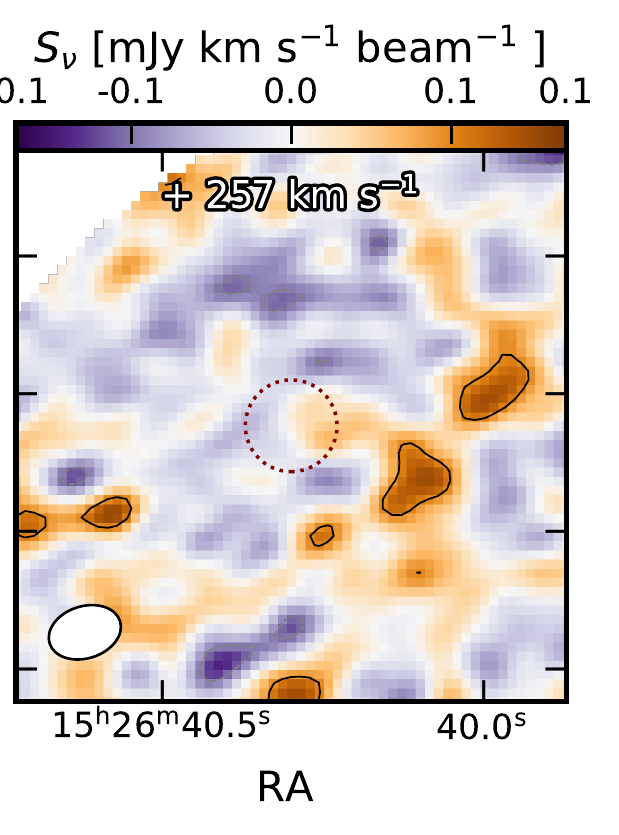} \\ 
    \includegraphics[height=0.14\textheight]{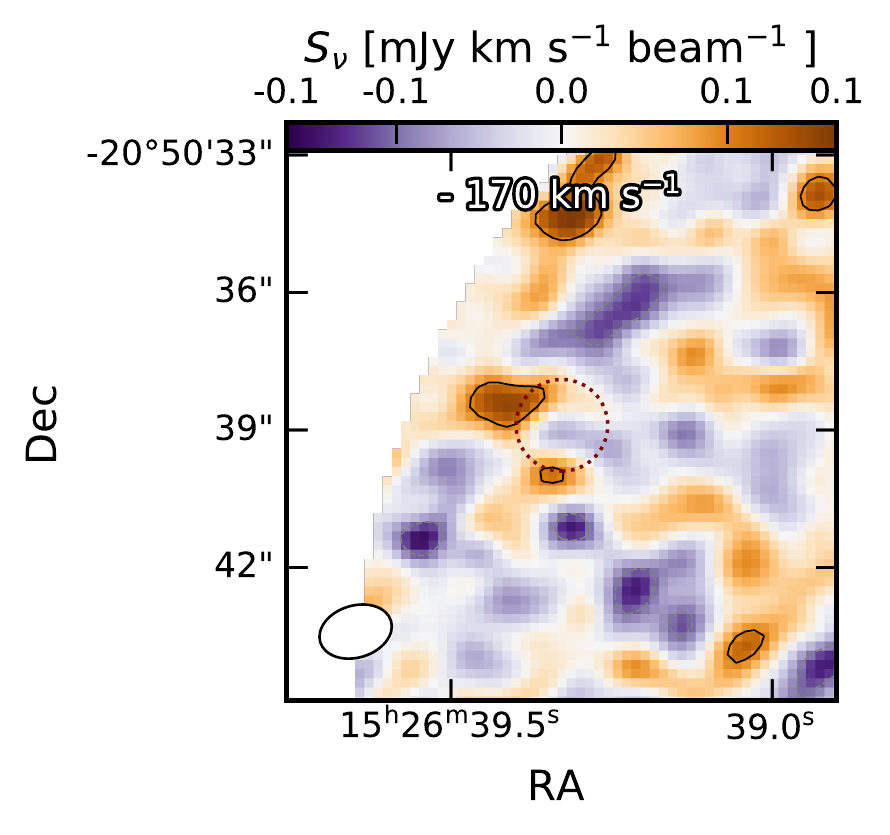}
    \includegraphics[height=0.14\textheight,trim={2.75cm 0 0 0},clip]{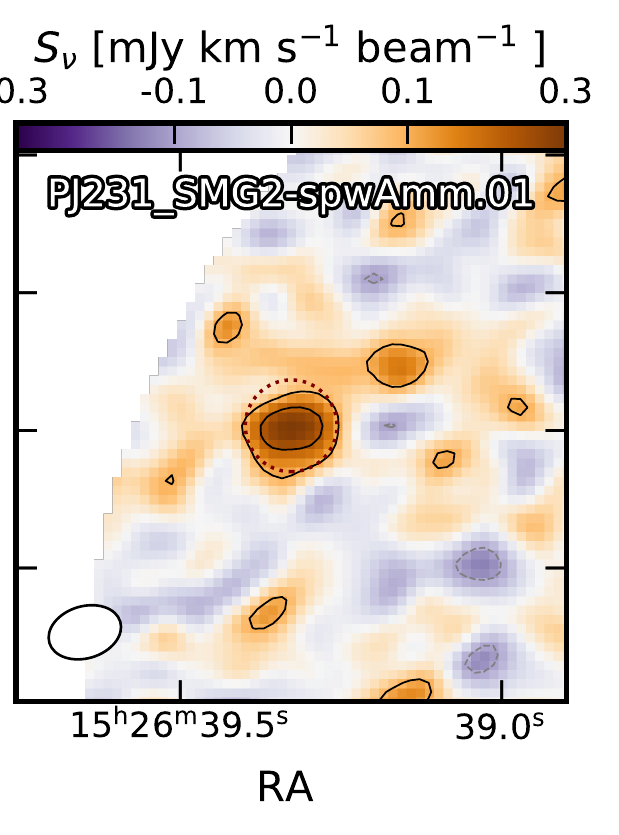} 
    \includegraphics[height=0.14\textheight,trim={2.75cm 0 0 0},clip]{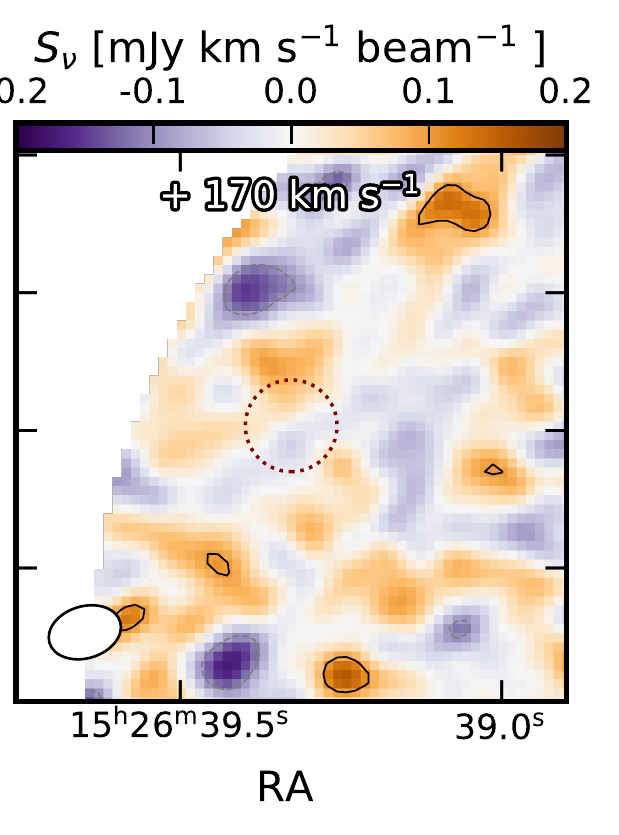} \\
    \raggedleft
    \includegraphics[height=0.14\textheight]{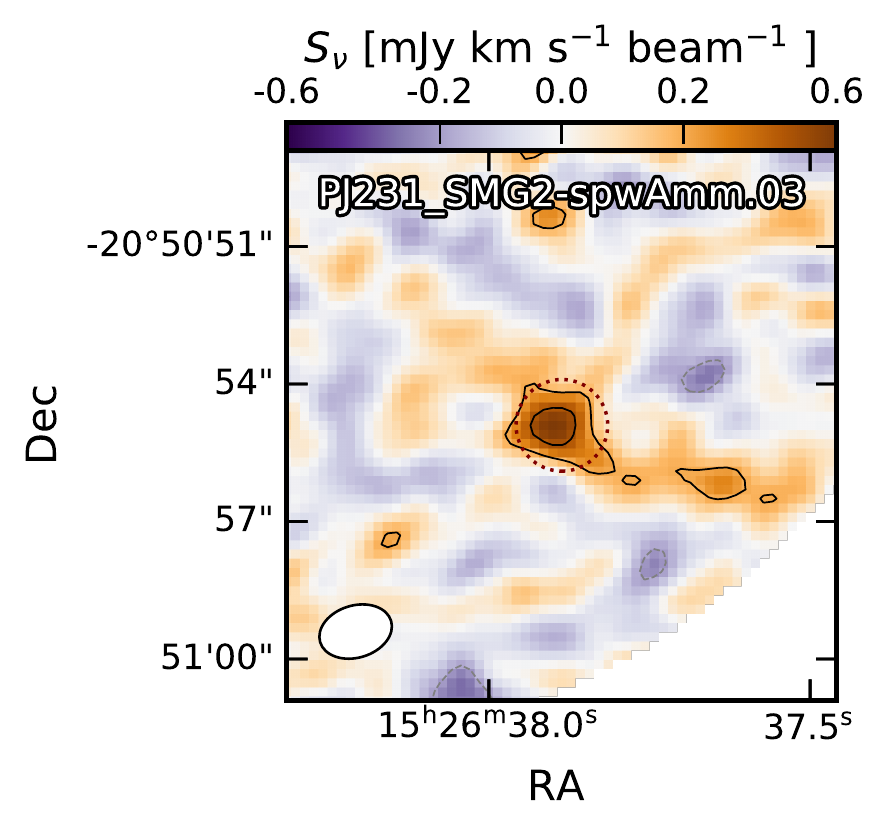} 
    \includegraphics[height=0.14\textheight,trim={2.75cm 0 0 0},clip]{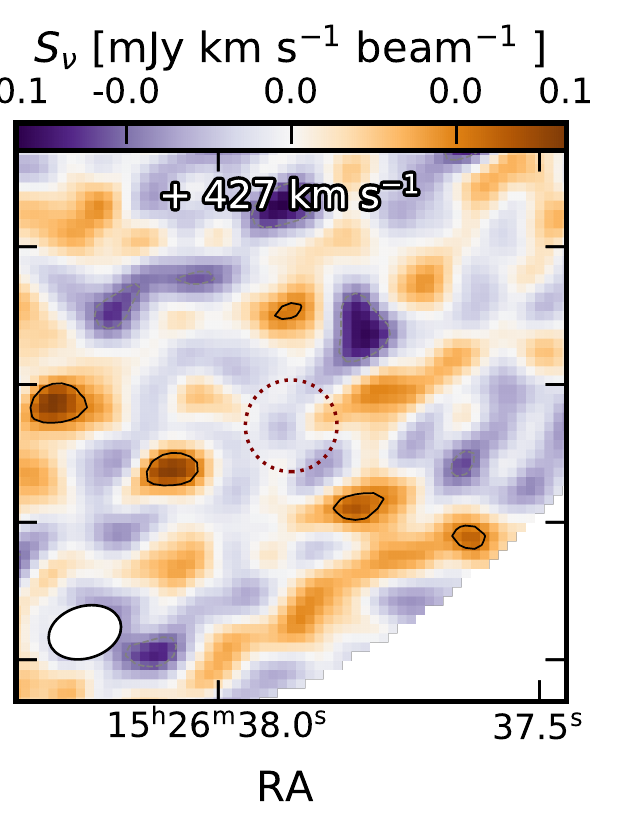} \\
    \includegraphics[height=0.14\textheight]{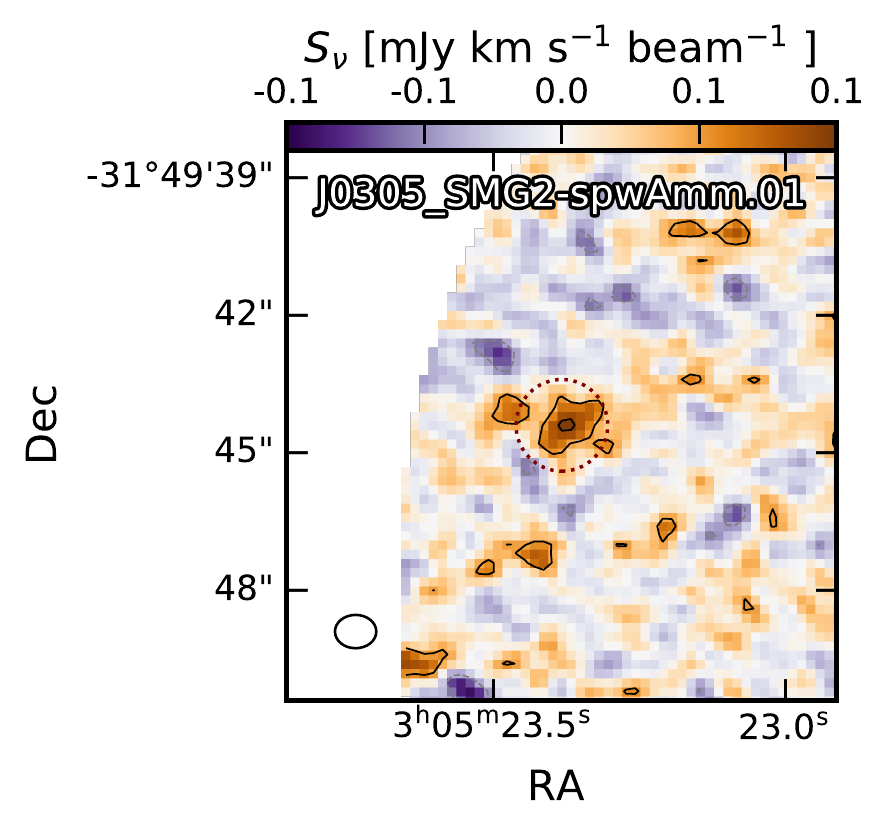}
    \includegraphics[height=0.14\textheight,trim={2.75cm 0 0 0},clip]{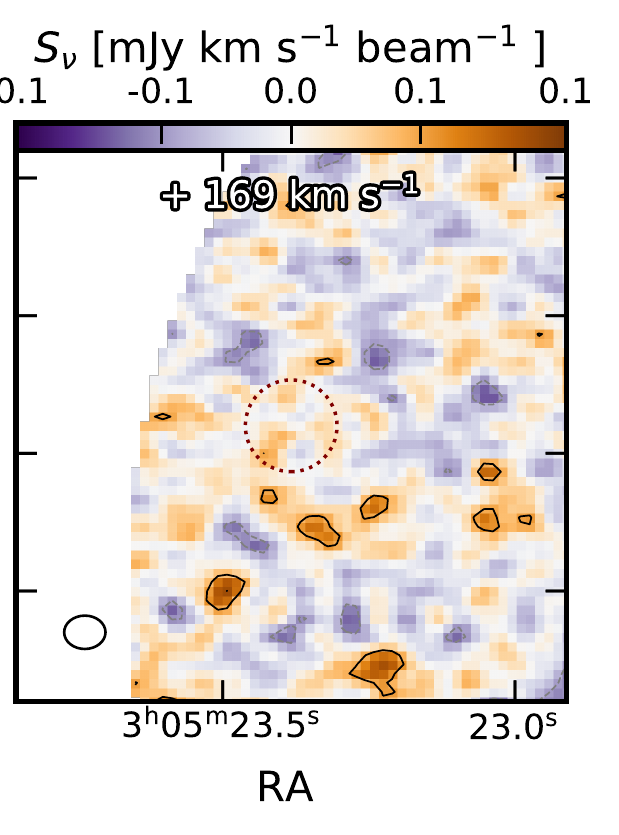} \\
    \raggedright \hspace{0.3cm}
    \includegraphics[height=0.14\textheight]{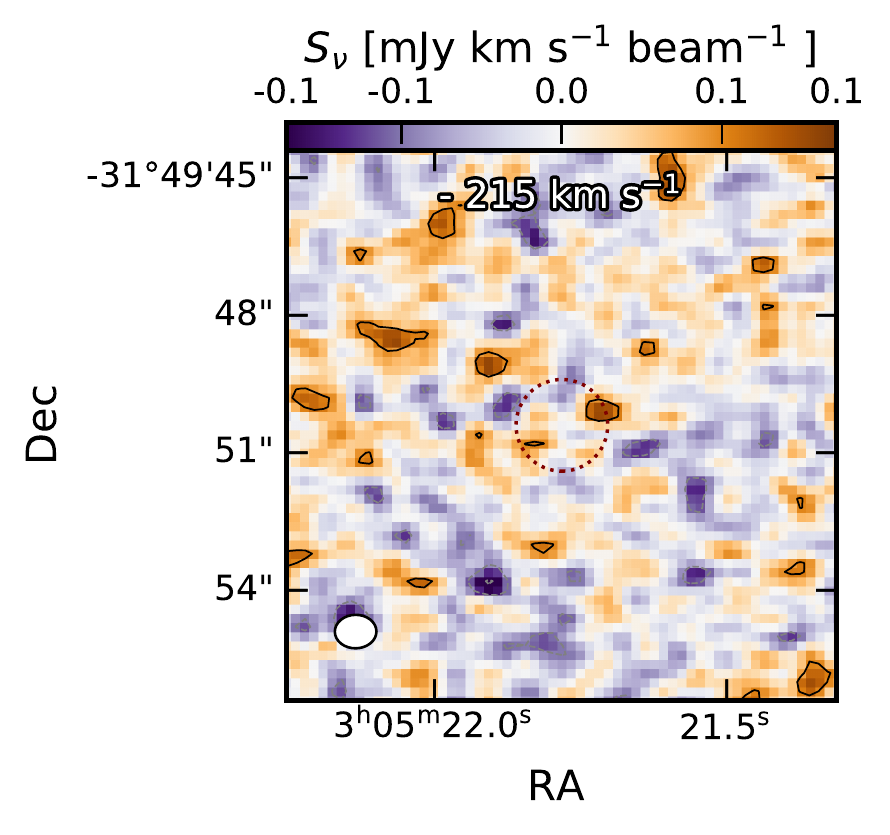}
    \includegraphics[height=0.14\textheight,trim={2.75cm 0 0 0},clip]{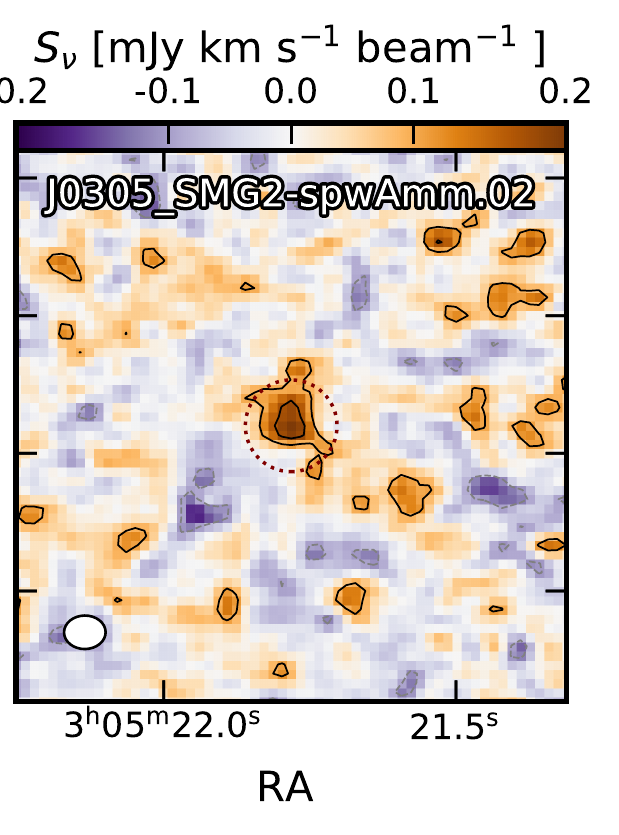} \\ 
    \raggedleft
    \includegraphics[height=0.14\textheight]{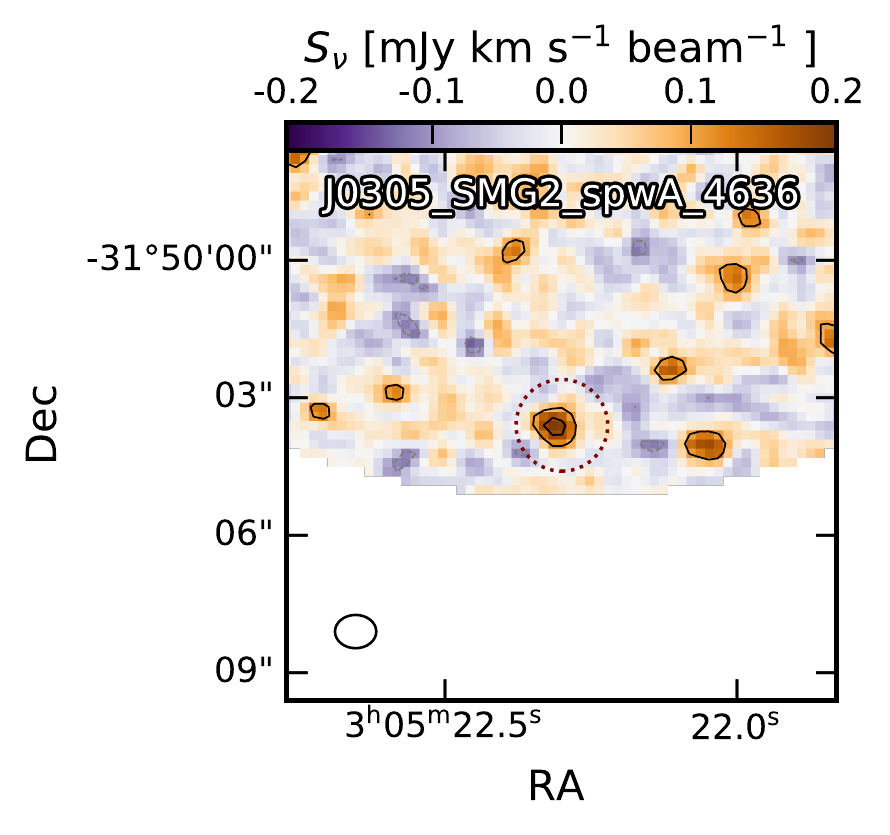}
    \includegraphics[height=0.14\textheight,trim={2.75cm 0 0 0},clip]{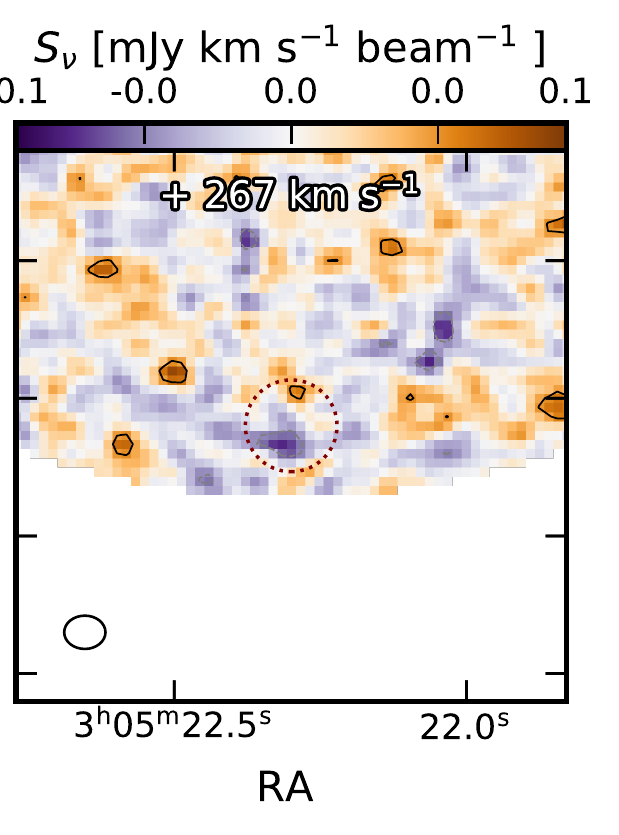}  \\
    \end{minipage}
    \begin{minipage}[t]{0.49\textwidth}
    \includegraphics[height=0.14\textheight]{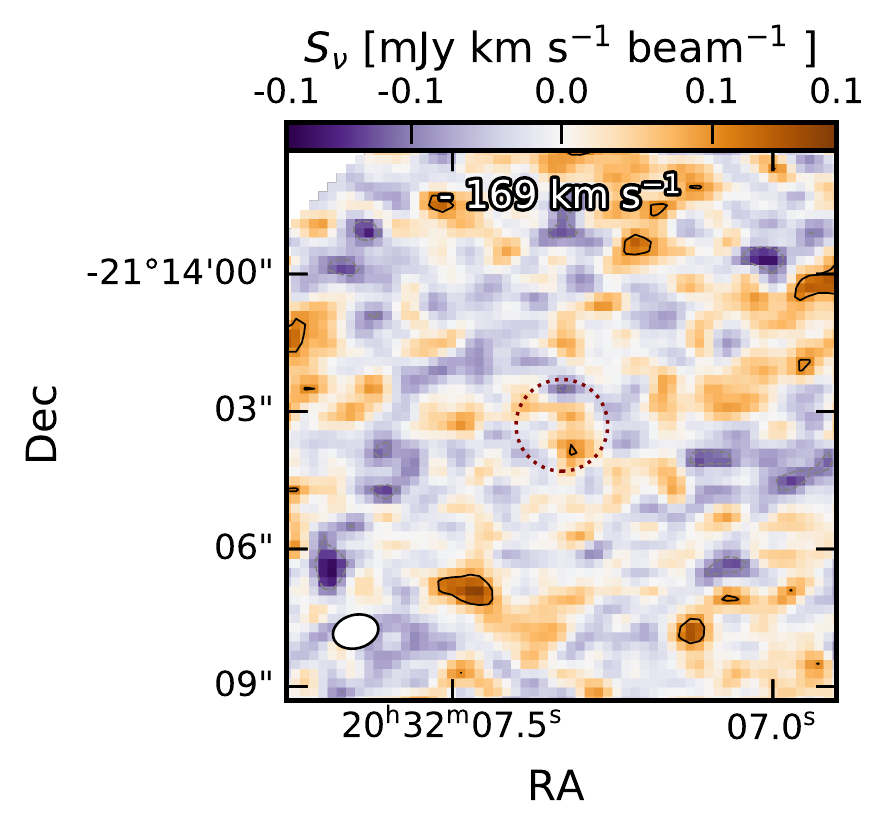}
    \includegraphics[height=0.14\textheight,trim={2.75cm 0 0 0},clip]{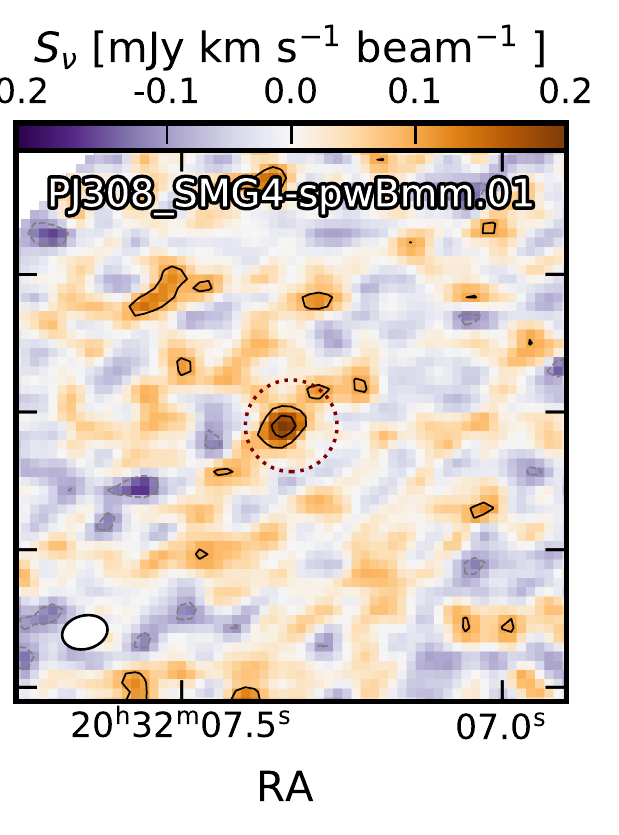} 
    \includegraphics[height=0.14\textheight,trim={2.75cm 0 0 0},clip]{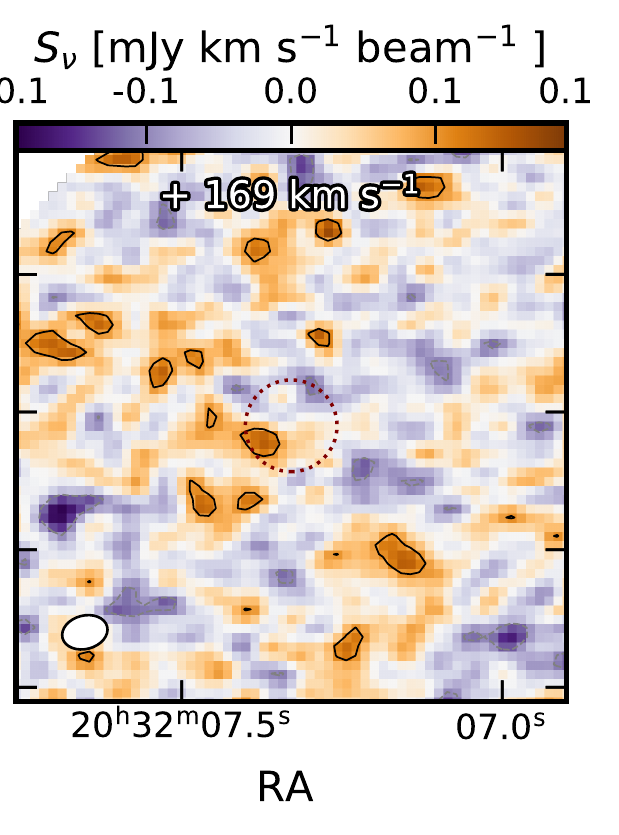} \\ 
    \includegraphics[height=0.14\textheight]{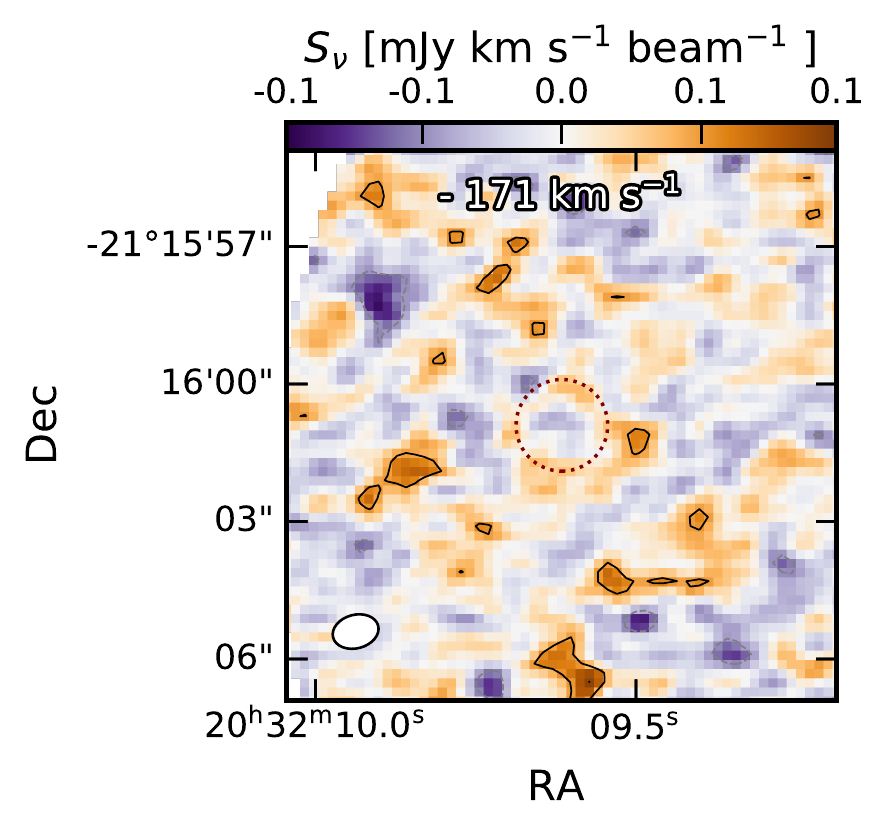}
    \includegraphics[height=0.14\textheight,trim={2.75cm 0 0 0},clip]{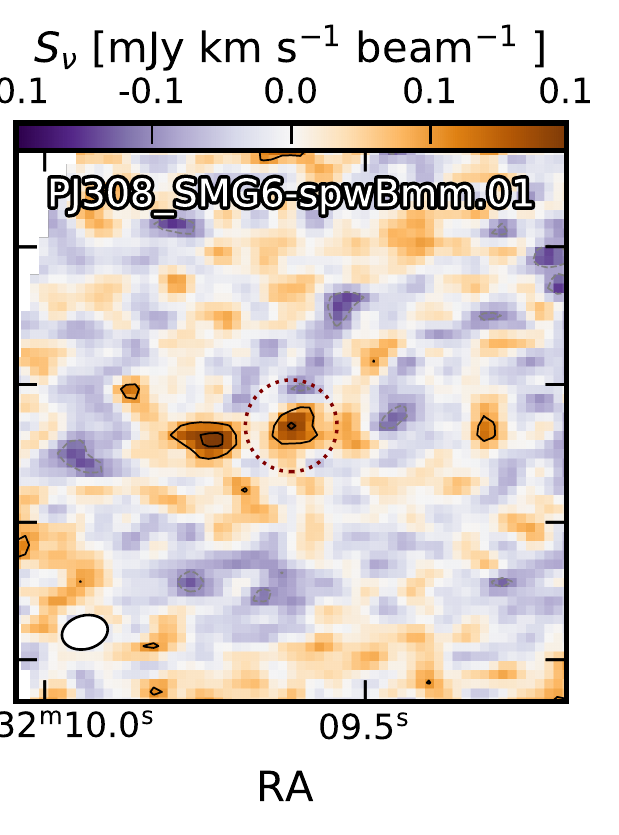} 
    \includegraphics[height=0.14\textheight,trim={2.75cm 0 0 0},clip]{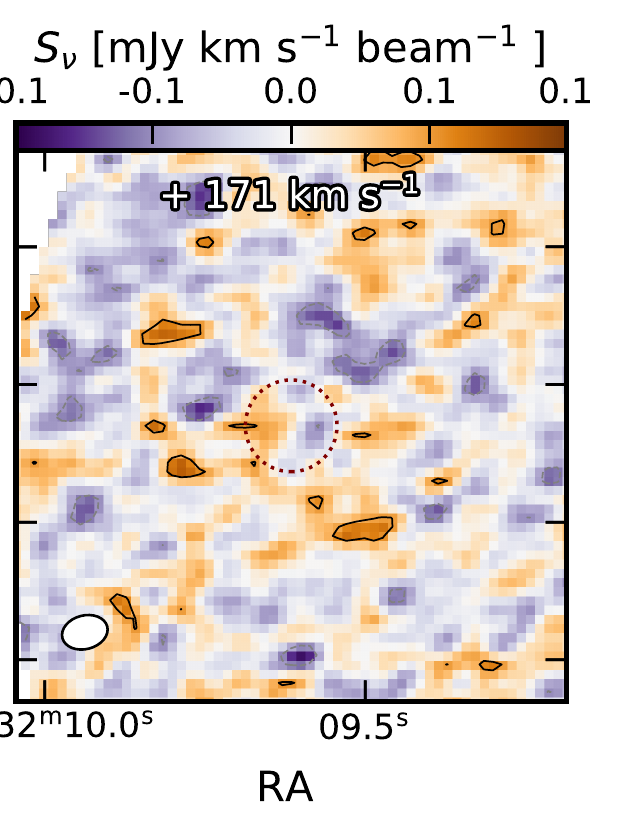} \\
    \includegraphics[height=0.14\textheight]{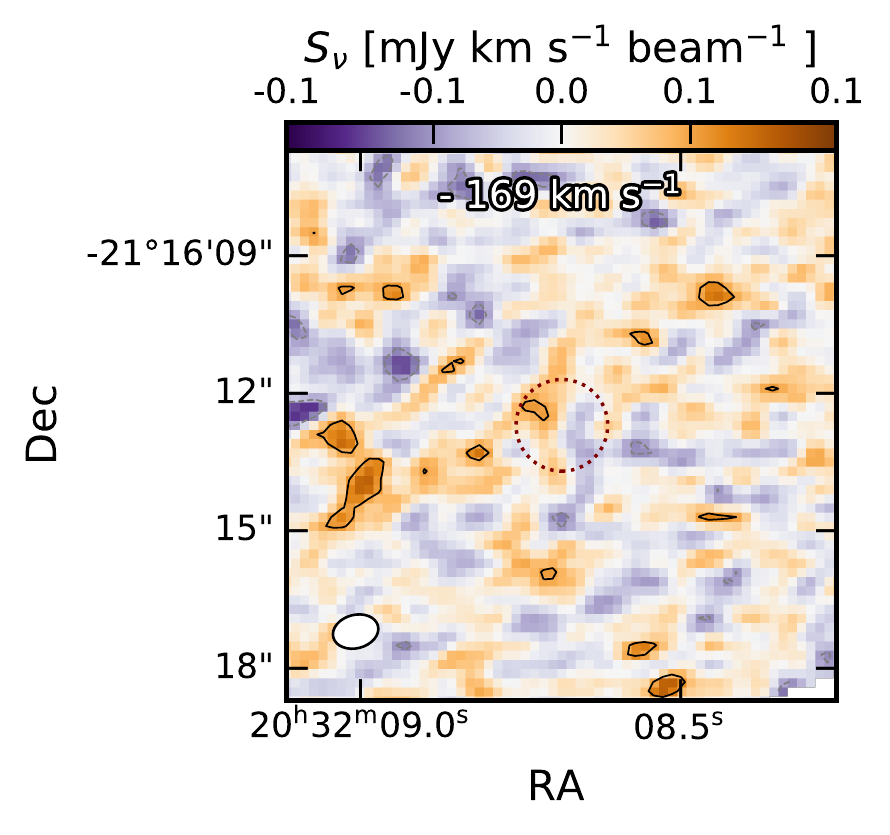} 
    \includegraphics[height=0.14\textheight,trim={2.75cm 0 0 0},clip]{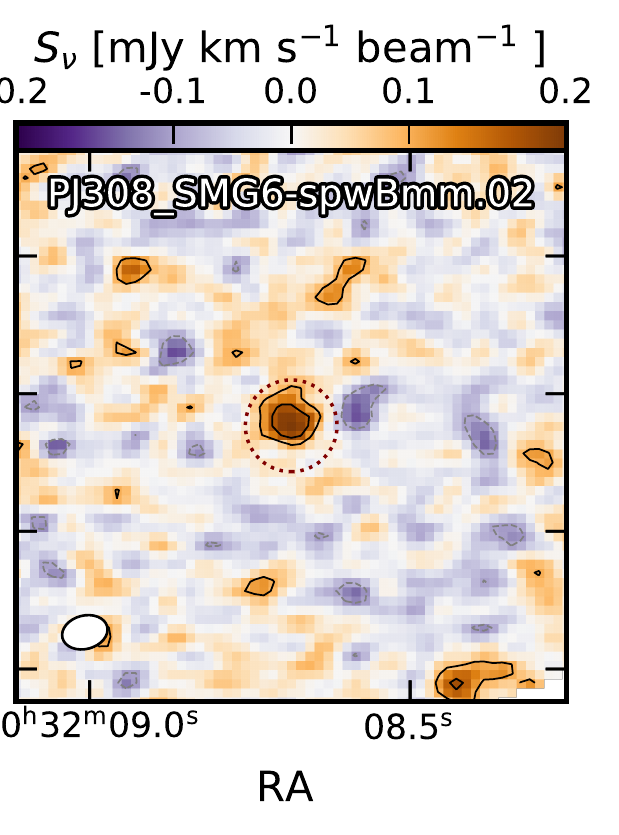}
    \includegraphics[height=0.14\textheight,trim={2.75cm 0 0 0},clip]{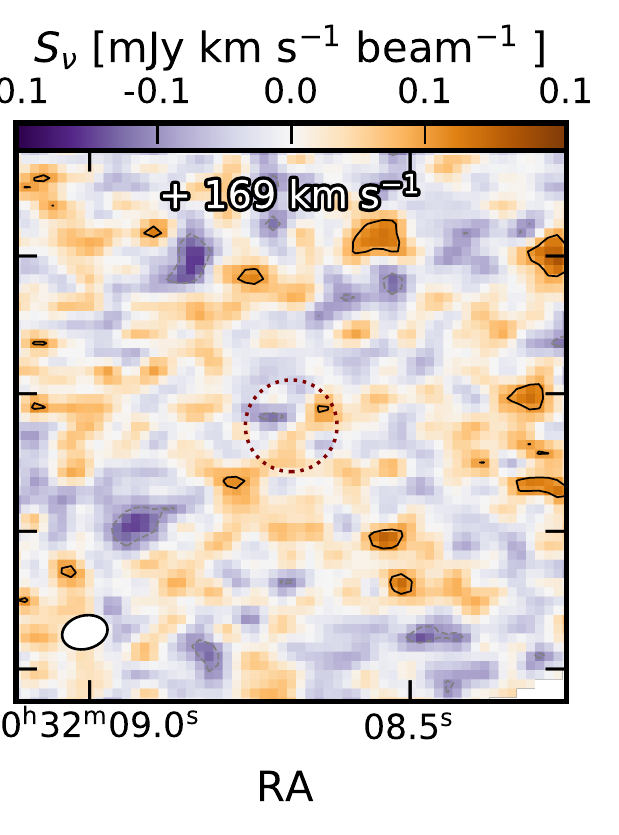} \\
    \includegraphics[height=0.14\textheight]{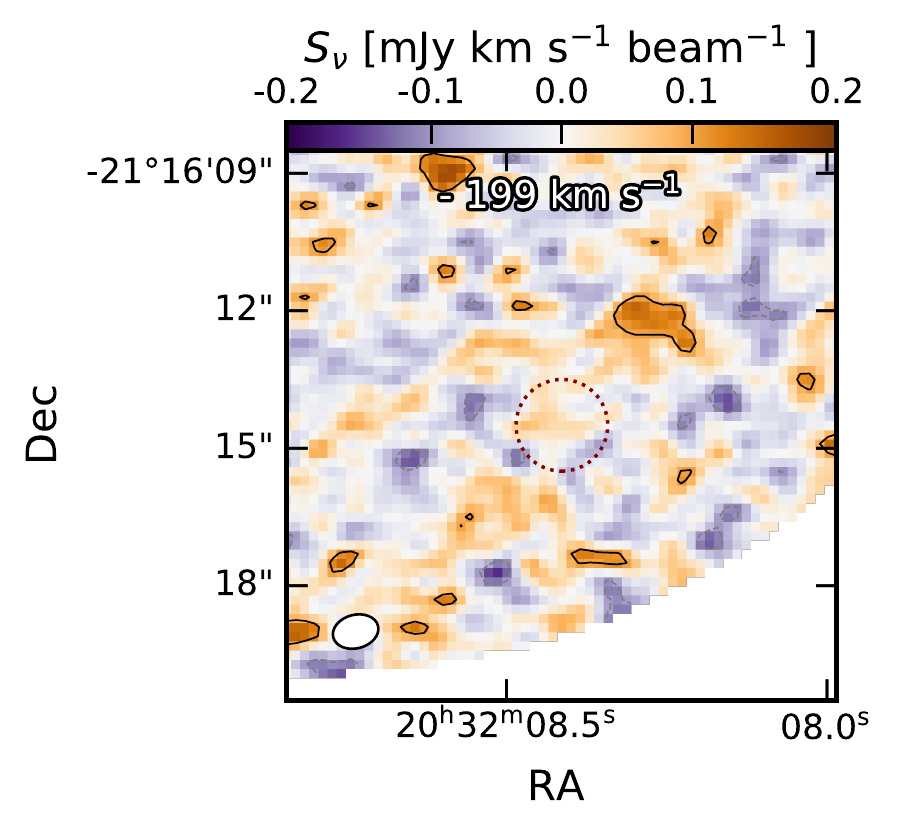} 
    \includegraphics[height=0.14\textheight,trim={2.75cm 0 0.2cm 0},clip]{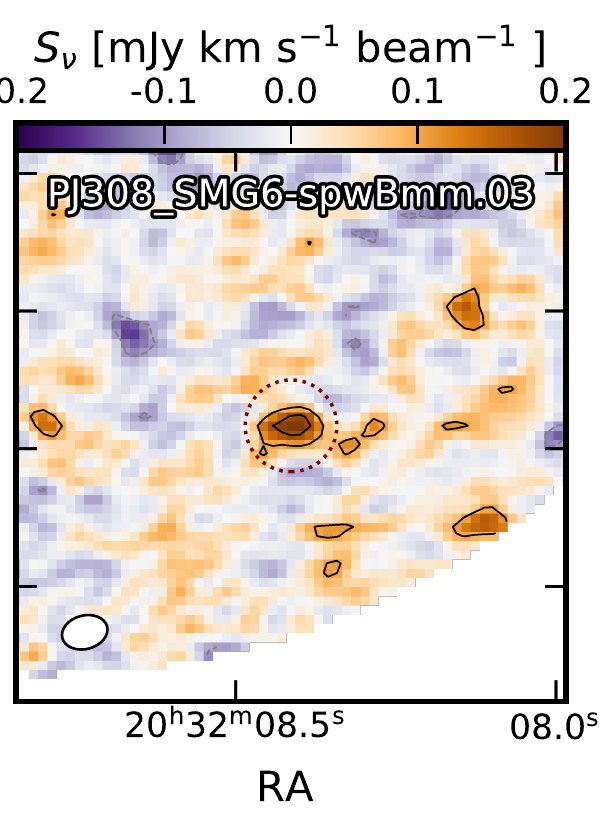}
    \includegraphics[height=0.14\textheight,trim={2.75cm 0 0.2cm 0},clip]{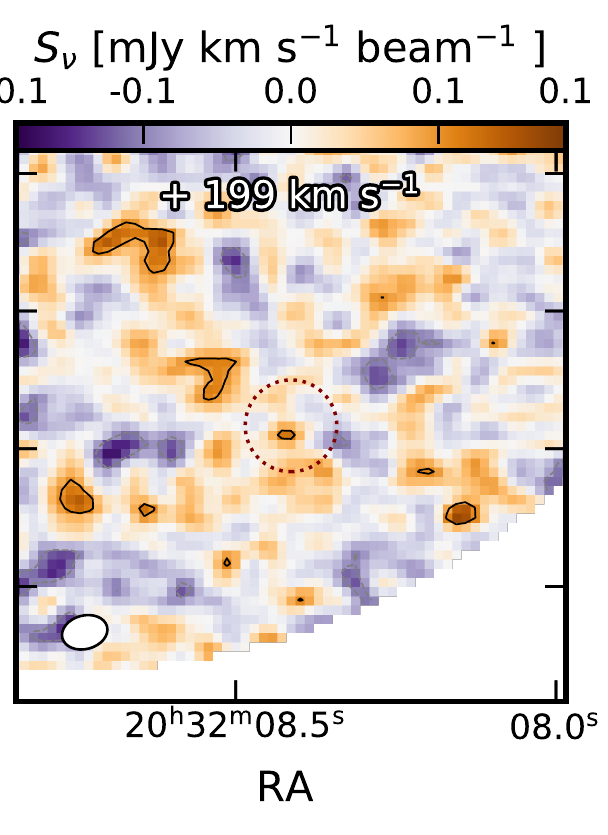} \\
    \includegraphics[height=0.14\textheight]{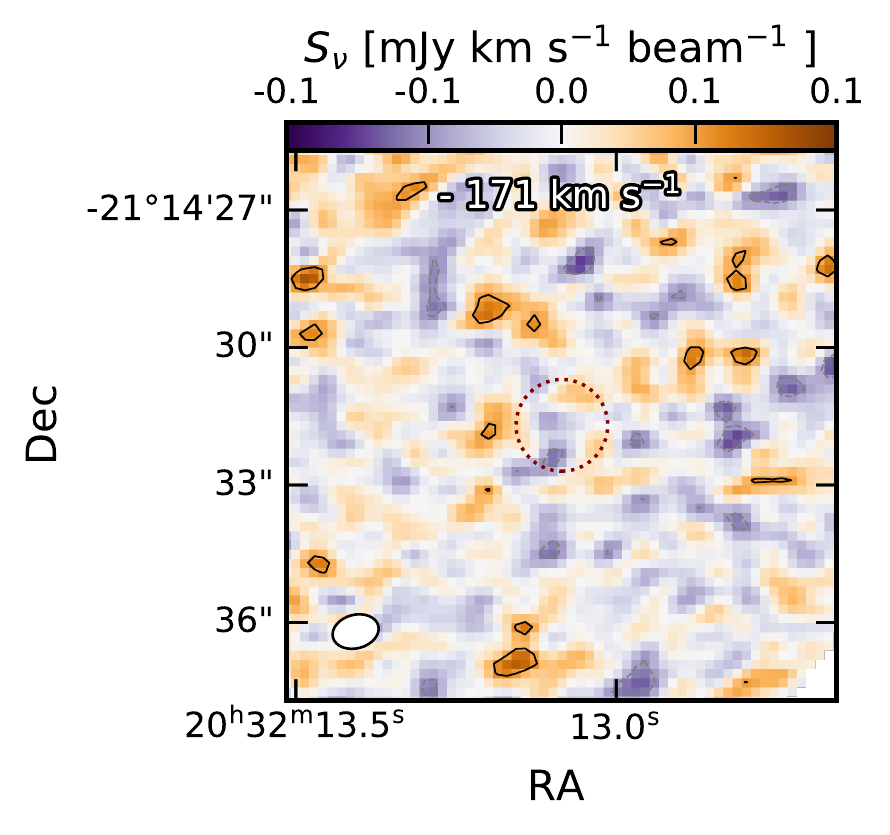} 
    \includegraphics[height=0.14\textheight,trim={2.75cm 0 0 0},clip]{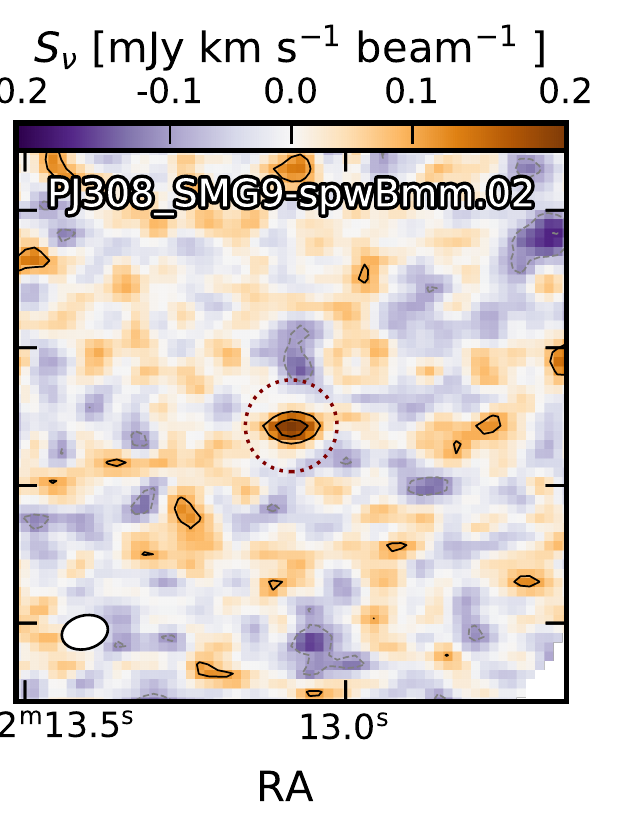}
    \includegraphics[height=0.14\textheight,trim={2.75cm 0 0 0},clip]{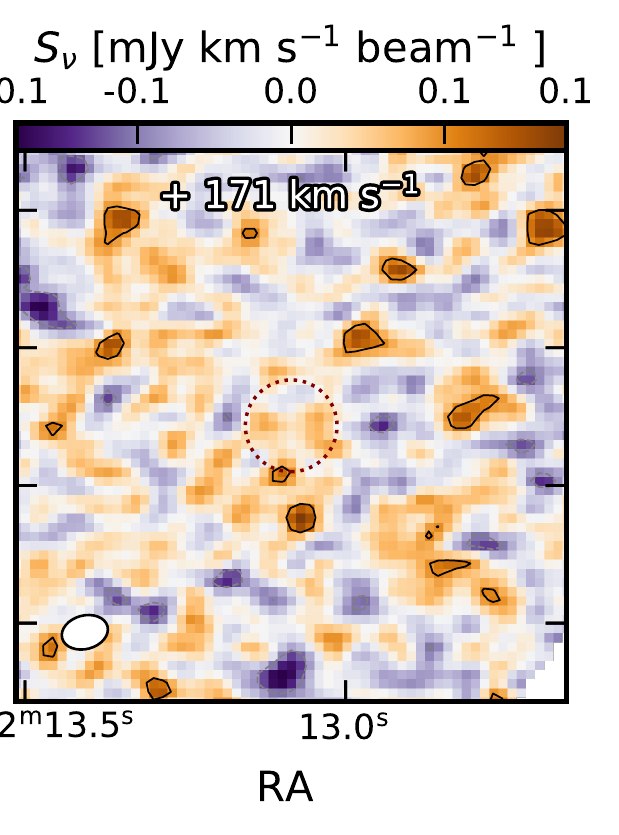}\\ \vfill
    \end{minipage}
    
    \caption{Emission line map for the secure line emitters (center, circled in dashed red) in the SMG fields. As in Fig. \ref{fig:line_dets_continuum_sources}, the emission line maps are integrated over $1.2\times$FWHM of the fitted Gaussian profile to the spectra, and control maps with similar width but offset in velocity are provided in each case. The color scaling is log--linear for better contrast, and the contours are logarithmic: $(-2,2,4,8,16,32) \sigma$, where $\sigma$ is the rms noise (see Table \ref{tab:quasar_obs_params}). The beam is plotted in the lower left corner. }
    \label{fig:cii_emitters_line_maps_1}
\end{figure}

In this appendix we present the candidate \cii\ line emitters recovered in our search detailed in Section \ref{sec:line_sources} in Fig. \ref{fig:cii_emitters_line_maps_1}. For each source, we show the emission line map as well as the two control maps velocity--integrated over $1.2\times$ the FWHM of the detected line in the $r=2"$ aperture--integrated spectrum.

\bibliography{bib}{}
\bibliographystyle{aasjournal}

\end{document}